\documentclass{aastex631}

\usepackage{newtxtext,newtxmath,amsmath,float,xspace,threeparttable,bm,amsmath,booktabs}
\usepackage{rotating}
\newcommand{\todo}{\ifmmode \text{\Huge{\(\bullet\)}} \else {\Huge$\bullet$}\fi}
\newcommand{\tido}{\ifmmode {\bullet} \else $\bullet$\fi}

\newcommand{\E        }[1]{\ifmmode 10^{#1} \else $10^{#1}$\fi}
\newcommand{\tE        }[1]{\ifmmode \times10^{#1} \else $\times10^{#1}$\fi}
\newcommand{\til}{\ifmmode \sim \else $\sim$\fi}
\renewcommand{\~} {\ifmmode \sim \else $\sim$\fi}

\newcommand{\pc}	{\ifmmode {\rm pc} \else pc\fi}
\newcommand{\ld}	{\ifmmode {\rm l.d.} \else l.d.\fi}
\newcommand{\kms}	{\ifmmode {\rm km\,s}^{-1} \else km\,s$^{-1}$\fi}
\newcommand{\cc}	{\ifmmode {\rm cm}^{-3}    \else cm$^{-3}$\fi}
\newcommand{\cmii}	{\ifmmode {\rm cm}^{-2}    \else cm$^{-2}$\fi}
\newcommand{\ergs}	{\ifmmode {\rm erg\,s}^{-1} \else erg s$^{-1}$\fi}
\newcommand{\ergcms}	{\ifmmode {\rm erg\,cm}^{-2}\,{\rm s}^{-1} \else erg\,cm$^{-2}$\,s$^{-1}$\fi}
\newcommand{\ergcmsA}	{\ifmmode {\rm erg\,cm}^{-2}\,{\rm s}^{-1}\,{\rm\AA}^{-1}
\else erg\,cm$^{-2}$\,s$^{-1}$\,\AA$^{-1}$\fi}
\newcommand{  \ergcmsHz  }{\ifmmode{\rm erg\,cm}^{-2}\,{\rm s}^{-1}\,{\rm Hz}^{-1}
                       \else ergs\,cm$^{-2}$\,s$^{-1}$\,Hz$^{-1}$\fi}
\newcommand{\kev}	{\ifmmode {\rm keV} \else keV\fi}

\newcommand{\mic}	{\ifmmode {\rm \mu m} \else $\mu$m\fi}
\newcommand{\vFWHM}	{\ifmmode v_{\mbox{\tiny FWHM}} \else $v_{\mbox{\tiny FWHM}}$\fi}
\newcommand{\vBLR}	{\ifmmode v_{\mbox{\tiny BLR}} \else $v_{\mbox{\tiny BLR}}$\fi}
\newcommand{\sigBLR}	{\ifmmode \sigma_{\mbox{\tiny BLR}} \else $\sigma_{\mbox{\tiny BLR}}$\fi}
\newcommand{\vNLR}	{\ifmmode v_{\mbox{\tiny NLR}} \else $v_{\mbox{\tiny NLR}}$\fi}
\newcommand{\tauBLR}	{\ifmmode \tau_{\mbox{\tiny BLR}} \else $\tau_{\mbox{\tiny BLR}}$\fi}

\newcommand{\Hubble}	{\ifmmode {\rm km\,s}^{-1}\,{\rm Mpc}^{-1} \else km\,s$^{-1}$\,Mpc$^{-1}$\fi}
\newcommand{\NDunit}	{\ifmmode {\rm Mpc}^{-3} \else Mpc$^{-3}$\fi}
\newcommand{\LFunit}	{\ifmmode {\rm Mpc}^{-3}\,{\rm mag}^{-1} \else Mpc$^{-3}$\,mag$^{-1}$\fi}
\newcommand{\MFunit}	{\ifmmode {\rm Mpc}^{-3}\,{\rm dex}^{-1} \else Mpc$^{-3}$\,dex$^{-1}$\fi}

\newcommand{\Msun}{\ifmmode M_{\odot} \else $M_{\odot}$\fi}
\newcommand{\Lsun}{\ifmmode L_{\odot} \else $L_{\odot}$\fi}
\newcommand{\Zsun}{\ifmmode Z_{\odot} \else $Z_{\odot}$\fi}
\newcommand{\mpyr}{\ifmmode \Msun\,{\rm yr}^{-1} \else $\Msun\,{\rm yr}^{-1}$\fi}

\newcommand{\logNH }{\ifmmode \log (N_{\rm H}/{\rm cm}^{-2}) \else $\log (N_{\rm H}/{\rm cm}^{-2})$\fi}

\newcommand{\qnote}{\ifmmode q_{0} \else $q_{0}$\fi}
\newcommand{\Hnote}{\ifmmode H_{0} \else $H_{0}$\fi}
\newcommand{\hnote}{\ifmmode h_{0} \else $h_{0}$\fi}
\newcommand{\anote}{\ifmmode a_{0} \else $a_{0}$\fi}

\newcommand{  \Halpha   }{\ifmmode {\rm H}\alpha \else H$\alpha$\fi}
\newcommand{  \ha   	}{\ifmmode {\rm H}\alpha \else H$\alpha$\fi}
\newcommand{  \Hbeta    }{\ifmmode {\rm H}\beta \else H$\beta$\fi}
\newcommand{  \hb    	}{\ifmmode {\rm H}\beta \else H$\beta$\fi}
\newcommand{  \Hgamma   }{\ifmmode {\rm H}\gamma \else H$\gamma$\fi}
\newcommand{  \Hdelta   }{\ifmmode {\rm H}\delta \else H$\delta$\fi}
\newcommand{  \Lya      }{\ifmmode {\rm Ly}\alpha \else Ly$\alpha$\fi}
\newcommand{  \Lyb      }{\ifmmode {\rm Ly}\beta \else Ly$\beta$\fi}
\newcommand{  \Pa       }{\ifmmode {\rm P}\alpha \else P$\alpha$\fi}
\newcommand{  \Pb       }{\ifmmode {\rm P}\beta \else P$\beta$\fi}
\newcommand{  \Bra      }{\ifmmode {\rm Br}\alpha \else Br$\alpha$\fi}
\newcommand{  \Brg      }{\ifmmode {\rm Br}\gamma \else Br$\gamma$\fi}
\newcommand{  \hii      }{\ifmmode {\rm H}\,\textsc{ii} \else H\,\textsc{ii}\fi}
\newcommand{  \hei      }{\ifmmode {\rm He}\,\textsc{i} \else He\,\textsc{i}\fi}
\newcommand{  \heii     }{\ifmmode {\rm He}\,\textsc{ii} \else He\,\textsc{ii}\fi}
\newcommand{  \HeIIuv   }{\ifmmode {\rm He}\,\textsc{ii}\,\lambda1640 \else He\,\textsc{ii}\,$\lambda1640$\fi}
\newcommand{  \HeIIop   }{\ifmmode {\rm He}\,\textsc{ii}\,\lambda4686 \else He\,\textsc{ii}\,$\lambda4686$\fi}
\newcommand{  \cii      }{\ifmmode {\rm C}\,\textsc{ii}  \else C\,\textsc{ii}\fi}
\newcommand{  \ciii     }{\ifmmode {\rm C}\,\textsc{iii}\right] \else C\,\textsc{iii}]\fi}
\newcommand{  \CIII     }{\ifmmode {\rm C}\,\textsc{iii}\right]\,\lambda1909 \else C\,\textsc{iii}]\,$\lambda1909$\fi}
\newcommand{  \civ      }{\ifmmode {\rm C}\,\textsc{iv}  \else C\,\textsc{iv}\fi}
\newcommand{  \CIV      }{\ifmmode {\rm C}\,\textsc{iv}\,\lambda1549 \else C\,\textsc{iv}\,$\lambda1549$\fi}
\newcommand{  \nii      }{\ifmmode [{\rm N}\,\textsc{ii}]  \else [N\,\textsc{ii}]\fi}
\newcommand{  \niii     }{\ifmmode {\rm N}\,\textsc{iii} \else N\,\textsc{iii}\fi}
\newcommand{  \niv      }{\ifmmode {\rm N}\,\textsc{iv}  \else N\,\textsc{iv}\fi}
\newcommand{  \NIVuv    }{\ifmmode {\rm N}\,\textsc{iv}\,\lambda1486 \else N\,\textsc{iv}\,$\lambda1486$\fi}
\newcommand{  \nv       }{\ifmmode {\rm N}\,\textsc{v}   \else N\,\textsc{v}\fi}
\newcommand{\oi}{\ifmmode \left[{\rm O}\,\textsc{i}\right] \else [O\,{\sc i}]\fi}
\newcommand{\OI}{\ifmmode \left[{\rm O}\,\textsc{i}\right]\,\lambda6300 \else [O\,{\sc i}]$\,\lambda6300$\fi}

\newcommand{\oii}{\ifmmode \left[{\rm O}\,\textsc{ii}\right] \else [O\,{\sc ii}]\fi}
\newcommand{\OII}{\ifmmode \left[{\rm O}\,\textsc{ii}\right]\,\lambda3727 \else [O\,{\sc ii}]\,$\lambda3727$\fi}
\newcommand{\oiii}{\ifmmode \left[{\rm O}\,\textsc{iii}\right] \else [O\,{\sc iii}]\fi}
\newcommand{\OIII}{\ifmmode \left[{\rm O}\,\textsc{iii}\right]\,\lambda5007 \else [O\,{\sc iii}]\,$\lambda5007$\fi}

\newcommand{\NII}{\ifmmode \left[{\rm N}\,\textsc{ii}\right]\,\lambda6583 \else [N\,{\sc ii}]$\,\lambda6583$\fi}

\newcommand{\NeIII}{\ifmmode \left[{\rm Ne}\,\textsc{iii}\right]\,\lambda3968 \else [Ne\,{\sc iii}]$\,\lambda3968$\fi}

\newcommand{\NeV}{\ifmmode \left[{\rm Ne}\,\textsc{v}\right]\,\lambda3426 \else [Ne\,{\sc v}]$\,\lambda3426$\fi}

\newcommand{\HeII}{\ifmmode {\rm He}\,\textsc{ii}\,\lambda4686 \else He\,{\sc ii}$\,\lambda4686$\fi}

\newcommand{\sii}{\ifmmode \left[{\rm S}\,\textsc{ii}\right] \else [S\,{\sc ii}]\fi}

\newcommand{\SII}{\ifmmode \left[{\rm S}\,\textsc{ii}\right]\,\lambda6717,6731 \else [S\,{\sc ii}]$\,\lambda6717,6731$\fi}

\newcommand{  \OIIIuv   }{\ifmmode {\rm O}\,\textsc{iii}\,\lambda1663 \else O\,\textsc{iii}\,$\lambda1663$\fi}
\newcommand{  \oiv      }{\ifmmode {\rm O}\,\textsc{iv}  \else O\,\textsc{iv}\fi}
\newcommand{  \OIVuv    }{\ifmmode {\rm O}\,\textsc{iv}\,\lambda1402  \else O\,\textsc{iv}\,$\lambda1402$\fi}
\newcommand{  \OIVIR    }{\ifmmode {\rm O}\,\textsc{iv}\,25.9\,\mu {\rm m} \else O\,\textsc{iv}\,$25.9\,\mu$m\fi}
\newcommand{  \ovi      }{\ifmmode {\rm O}\,\textsc{vi}   \else O\,\textsc{vi}\fi}
\newcommand{  \Ovi      }{\ifmmode {\rm O}\,\textsc{vi}\,\lambda1035 \else O\,\textsc{vi}\,$\lambda1035$\fi}
\newcommand{  \nei      }{\ifmmode {\rm Ne}\,\textsc{i}   \else Ne\,\textsc{i}\fi}
\newcommand{  \neii     }{\ifmmode {\rm Ne}\,\textsc{ii}  \else Ne\,\textsc{ii}\fi}
\newcommand{  \NeiiIR   }{\ifmmode {\rm Ne}\,\textsc{ii}\,12.8\,\mu {\rm m} \else Ne\,\textsc{ii}\,$12.8\,\mu$m\fi}
\newcommand{  \neiii    }{\ifmmode {\rm Ne}\,\textsc{iii} \else Ne\,\textsc{iii}\fi}
\newcommand{  \neiv     }{\ifmmode {\rm Ne}\,\textsc{iv}  \else Ne\,\textsc{iv}\fi}
\newcommand{  \nev      }{\ifmmode {\rm Ne}\,\textsc{v}   \else Ne\,\textsc{v}\fi}
\newcommand{  \NevIR    }{\ifmmode {\rm Ne}\,\textsc{v}\,24.3\,\mu {\rm m} \else Ne\,\textsc{v}\,$24.3\,\mu$m\fi}
\newcommand{  \nevi     }{\ifmmode {\rm Ne}\,\textsc{vi}  \else Ne\,\textsc{vi}\fi}
\newcommand{  \mgi      }{\ifmmode {\rm Mg}\,\textsc{i}   \else Mg\,\textsc{i}\fi}
\newcommand{  \mgii     }{\ifmmode {\rm Mg}\,\textsc{ii}  \else Mg\,\textsc{ii}\fi}
\newcommand{  \MgII     }{\ifmmode {\rm Mg}\,\textsc{ii}\,\lambda2798 \else Mg\,\textsc{ii}\,$\lambda2798$\fi}

\newcommand{  \siii     }{\ifmmode {\rm S}\,\textsc{iii} \else S\,\textsc{iii}\fi}
\newcommand{  \siv      }{\ifmmode {\rm S}\,\textsc{iv}  \else S\,\textsc{iv}\fi}
\newcommand{  \sili     }{\ifmmode {\rm Si}\,\textsc{i}   \else Si\,\textsc{i}\fi}
\newcommand{  \silii    }{\ifmmode {\rm Si}\,\textsc{ii}  \else Si\,\textsc{ii}\fi}
\newcommand{  \Siliv    }{\ifmmode {\rm Si}\,\textsc{iv}  \else Si\,\textsc{iv}\fi}
\newcommand{  \SilIVuv  }{\ifmmode {\rm Si}\,\textsc{iv}\,\lambda1400  \else Si\,\textsc{iv}\,$\lambda1400$\fi}

\newcommand{  \caii     }{\ifmmode {\rm Ca}\,\textsc{ii}   \else Ca\,\textsc{ii}\fi}
 \newcommand{\Mgb}{\ifmmode \left{\rm Mg}\,\textsc{i}\right\,\lambda5175 \else Mg\,{\sc i}\,$\lambda5175$\fi}
\newcommand{\Cahk}{\ifmmode \left[{\rm Ca H+K}\,\textsc{ii}\right\,\lambda3935,3968 \else Ca H+K$\,\lambda3935,3968$\fi}
\newcommand{  \feii     }{\ifmmode {\rm Fe}\,\textsc{ii}  \else Fe\,\textsc{ii}\fi}
\newcommand{  \feiii    }{\ifmmode {\rm Fe}\,\textsc{iii} \else Fe\,\textsc{iii}\fi}

\newcommand{ \Lhb   }{\ifmmode L\left(\hb\right) \else $L\left(\hb\right)$\fi}
\newcommand{ \fwhb  }{\ifmmode {\rm FWHM}\left(\hb\right) \else FWHM(\hb)\fi}
\newcommand{ \Lha   }{\ifmmode L\left(\ha\right) \else $L\left(\ha\right)$\fi}
\newcommand{ \fwha  }{\ifmmode {\rm FWHM}\left(\ha\right) \else FWHM(\ha)\fi}
\newcommand{ \Lmg   }{\ifmmode L\left(\mgii\right) \else $L\left(\mgii\right)$\fi}
\newcommand{ \fwmg  }{\ifmmode {\rm FWHM}\left(\mgii\right) \else FWHM(\mgii)\fi}
\newcommand{ \Lciv  }{\ifmmode L\left(\civ\right) \else $L\left(\civ\right)$\fi}
\newcommand{ \fwciv }{\ifmmode {\rm FWHM}\left(\civ\right) \else FWHM(\civ)\fi}
\newcommand{ \fwhm  }{\ifmmode {\rm FWHM} \else FWHM\fi} 
\newcommand{ \voff  }{\ifmmode v_{\rm off} \else $v_{\rm off}$\fi} 

\newcommand{ \mumg  }{\ifmmode \mu\left(\mgii\right) \else $\mu\left(\mgii\right)$\fi}
\newcommand{ \fmg   }{\ifmmode f\left(\mgii\right) \else $f\left(\mgii\right)$\fi}
\newcommand{ \muciv }{\ifmmode \mu\left(\civ\right) \else $\mu\left(\civ\right)$\fi}
\newcommand{ \fciv  }{\ifmmode f\left(\civ\right) \else $f\left(\civ\right)$\fi}

\newcommand{  \auvo     }{\ifmmode \alpha_{\nu,{\rm UVO}} \else $\alpha_{\nu,{\rm UVO}}$\fi}
\newcommand{  \Ledd     }{\ifmmode L_{\rm Edd} \else $L_{\rm Edd}$\fi}
\newcommand{  \lamLlam  }{\ifmmode \lambda L_{\lambda} \else $\lambda L_{\lambda}$\fi}
\newcommand{  \lLl      }{\ifmmode \lambda L_{\lambda} \else $\lambda L_{\lambda}$\fi}
\newcommand{  \nuLnu    }{\ifmmode \nu L_{\nu} \else $\nu L_{\nu}$\fi}
\newcommand{  \nLn      }{\ifmmode \nu L_{\nu} \else $\nu L_{\nu}$\fi}
\newcommand{  \Luv      }{\ifmmode L_{1450} \else $L_{1450}$\fi}
\newcommand{  \Lop      }{\ifmmode L_{5100} \else $L_{5100}$\fi}
\newcommand{  \lLop     }{\ifmmode \log\left(\Lop/\ergs\right) \else $\log\left(\Lop/\ergs\right)$\fi}
\newcommand{  \Lthree   }{\ifmmode L_{3000} \else $L_{3000}$\fi}
\newcommand{  \lLthree  }{\ifmmode \log\left(\Lthree/\ergs\right) \else $\log\left(\Lthree/\ergs\right)$\fi}

\newcommand{\Fthree}{\ifmmode F_{3000} \else $F_{3000}$\fi}
\newcommand{\fuv}{\ifmmode f_{\lambda}\left(1450{\rm \AA}\right) \else $f_{\lambda}\left(1450 {\rm \AA}\right)$\fi}
\newcommand{\fthree}{\ifmmode f_{\lambda}\left(3000{\rm \AA}\right) \else $f_{\lambda}\left(3000{\rm \AA}\right)$\fi}
\newcommand{\fH}{\ifmmode f_{\lambda}\left(1.65\micron\right) \else
$f_{\lambda}\left(1.65\micron\right)$\fi}

\newcommand{\fbol}{\ifmmode f_{\rm bol} \else $f_{\rm bol}$\fi}
\newcommand{\fbolwv}{\ifmmode f_{\rm bol}\left(\lambda\right) \else $f_{\rm bol}\left(\lambda\right)$\fi}
\newcommand{\fbolopt}{\ifmmode f_{\rm bol}\left(5100{\rm \AA}\right) \else $f_{\rm bol}\left(5100{\rm \AA}\right)$\fi}
\newcommand{\fbolthree}{\ifmmode f_{\rm bol}\left(3000{\rm \AA}\right) \else $f_{\rm bol}\left(3000{\rm \AA}\right)$\fi}
\newcommand{\fboluv}{\ifmmode f_{\rm bol}\left(1450{\rm \AA}\right) \else $f_{\rm bol}\left(1450{\rm \AA}\right)$\fi}

\newcommand{  \mbh      }{\ifmmode M_{\rm BH} \else $M_{\rm BH}$\fi}
\newcommand{  \lmbh     }{\ifmmode \log\left(\mbh/\Msun\right) \else $\log\left(\mbh/\Msun\right)$\fi} 
\newcommand{  \lledd    }{\ifmmode L/L_{\rm Edd} \else $L/L_{\rm Edd}$\fi}
\newcommand{  \Lbol     }{\ifmmode L_{\rm bol} \else $L_{\rm bol}$\fi}
\newcommand{  \lbol     }{\ifmmode L_{\rm bol} \else $L_{\rm bol}$\fi}
\newcommand{  \lLbol    }{\ifmmode \log\left(\Lbol/\ergs\right) \else $\log\left(\Lbol/\ergs\right)$\fi} 
\newcommand{  \Lagn     }{\ifmmode L_{\rm AGN} \else $L_{\rm AGN}$\fi}
\newcommand{  \lagn     }{\ifmmode L_{\rm AGN} \else $L_{\rm AGN}$\fi}

\newcommand{  \tgrow     }{\ifmmode t_{\rm growth} \else $t_{\rm growth}$\fi}
\newcommand{  \tUni      }{\ifmmode t_{\rm Universe} \else $t_{\rm Universe}$\fi}

\newcommand{  \Mindot	}{\ifmmode \dot{M}_{\rm infall} \else $\dot{M}_{\rm infall}$\fi}
\newcommand{  \Mbhdot	}{\ifmmode \dot{M}_{\rm BH} \else $\dot{M}_{\rm BH}$\fi}
\newcommand{  \Maddot	}{\ifmmode \dot{M}_{\rm AD} \else $\dot{M}_{\rm AD}$\fi}

\newcommand{  \as	}{\ifmmode a_{\rm *} 		\else $a_{\rm *}$\fi}
\newcommand{  \avec	}{\ifmmode \vec{a}_{\rm *} 	\else $\vec{a}_{\rm *}$\fi}
\newcommand{  \re	}{\ifmmode \eta      	\else $\eta$\fi}

\newcommand{  \mseed    }{\ifmmode M_{\rm seed} \else $M_{\rm seed}$\fi}
\newcommand{  \mbul     }{\ifmmode M_{\rm Bulge} \else $M_{\rm Bulge}$\fi} 
\newcommand{  \mstar    }{\ifmmode M_{*} \else $M_{*}$\fi} 
\newcommand{  \mgal     }{\ifmmode M_{*} \else $M_{*}$\fi} 
\newcommand{  \mhost    }{\ifmmode M_{\rm Host} \else $M_{\rm Host}$\fi}
\newcommand{  \mm       }{\ifmmode M_{*}/M_{\rm BH} \else $M_{*}/M_{\rm BH}$\fi}
\newcommand{  \mmsmall  }{\ifmmode M_{\rm BH}/M_{*} \else $M_{\rm BH}/M_{*}$\fi}
\newcommand{  \mmlarge  }{\ifmmode M_{*}/M_{\rm BH} \else $M_{*}/M_{\rm BH}$\fi}
\newcommand{  \mmwp     }{\ifmmode \left(M_{*}/M_{\rm BH}\right) \else $\left(M_{*}/M_{\rm BH}\right)$\fi}
\newcommand{  \ml       }{\ifmmode M_{*}/L_{*} \else $M_{*}/L_{*}$\fi}
\newcommand{  \mlwp     }{\ifmmode \left(M_{*}/L\right) \else $\left(M_{*}/L\right)$\fi}
\newcommand{  \mlk      }{\ifmmode \left(M_{*}/L_{K}\right) \else $\left(M_{*}/L_{K}\right)$\fi}
\newcommand{  \sigs     }{\ifmmode \sigma_{*} \else $\sigma_{*}$\fi}
\newcommand{  \Reff     }{\ifmmode R_{\rm e} \else $R_{\rm e}$\fi}

\newcommand{\bj}{\ifmmode b_{\rm J} \else $b_{\rm J}$\fi}

\newcommand{\iab}{\ifmmode i_{\rm AB} \else $i_{\rm AB}$\fi}

\newcommand{\jab}{\ifmmode J_{\rm AB} \else $J_{\rm AB}$\fi}
\newcommand{\hab}{\ifmmode H_{\rm AB} \else $H_{\rm AB}$\fi}
\newcommand{\kab}{\ifmmode K_{\rm AB} \else $K_{\rm AB}$\fi}

\newcommand{\jveg}{\ifmmode J_{\rm Vega} \else $J_{\rm Vega}$\fi}
\newcommand{\hveg}{\ifmmode H_{\rm Vega} \else $H_{\rm Vega}$\fi}
\newcommand{\kveg}{\ifmmode K_{\rm Vega} \else $K_{\rm Vega}$\fi}

\newcommand {\Fsoftobs} {$F^{\rm obs}_{\mathrm{2-10\ keV}}$\xspace}
\def\ergpcmsqps{\hbox{$\erg\cm^{-2}\s^{-1}\,$}}

\newcommand {\Fbatobs} {$F^{\rm obs}_{\mathrm{14-195\ keV}}$\xspace}
\newcommand {\Fbatricci} {$F^{\rm obs}_{\mathrm{14-150\ keV}}$\xspace}
\def\ergpcmsqps{\hbox{$\erg\cm^{-2}\s^{-1}\,$}}
\newcommand {\Fsoftint} {$F^{\rm in}_{\mathrm{2-10\ keV}}$\xspace}

\def\arcmin{\hbox{$^\prime$}}
\def\arcsec{\hbox{$^{\prime\prime}$}}

\newcommand{  \Chisq    }{\ifmmode \chi^{2} \else $\chi^{2}$}
\newcommand{  \nelec    }{\ifmmode n_{e} \else $n_{e}$\fi}     
\newcommand{  \nh       }{\ifmmode n_{H} \else $n_{H}$\fi}     
\newcommand{  \Ncol     }{\ifmmode N_{col} \else $N_{col}$\fi} 
\newcommand{  \NH       }{\ifmmode N_{H} \else $N_{\rm H}$\fi}     

\def\deg{\hbox{$^\circ$}}
\def\sun{\hbox{$\odot$}}
\def\arcmin{\hbox{$^\prime$}}
\def\arcsec{\hbox{$^{\prime\prime}$}}

\def\ion#1#2{#1$\;${\small\rm\@Roman{#2}}\relax}

\newcommand{\OIIIa}{\ifmmode \left[{\rm O}\,\textsc{iii}\right]\,\lambda4959 \else [O\,{\sc iii}]\,$\lambda4959$\fi}
\newcommand{\NIIa}{\ifmmode \left[{\rm N}\,\textsc{ii}\right]\,\lambda6548 \else [N\,{\sc ii}]\,$\lambda6548$\fi}
\newcommand{\SIIa}{\ifmmode \left[{\rm S}\,\textsc{ii}\right]\,\lambda6716 \else [S\,{\sc ii}]\,$\lambda6716$\fi}
\newcommand{\SIIb}{\ifmmode \left[{\rm S}\,\textsc{ii}\right]\,\lambda6732 \else [S\,{\sc ii}]\,$\lambda6731$\fi}
\newcommand{\NeVa}{\ifmmode \left[{\rm Ne}\,\textsc{v}\right]\,\lambda3346 \else [Ne\,{\sc v}]\,$\lambda3346$\fi}
\newcommand{\NeVb}{\ifmmode \left[{\rm Ne}\,\textsc{v}\right]\,\lambda3426 \else [Ne\,{\sc v}]\,$\lambda3426$\fi}
\newcommand{\NeIIIa}{\ifmmode \left[{\rm Ne}\,\textsc{iii}\right]\,\lambda3869 \else [Ne\,{\sc iii}]\,$\lambda3869$\fi}
\newcommand{\NeIIIb}{\ifmmode \left[{\rm Ne}\,\textsc{iii}\right]\,\lambda3968 \else [Ne\,{\sc iii}]\,$\lambda3968$\fi}

\newcommand{\mgb}{\ifmmode \left{\rm Mg}\,\textsc{i}\right \else Mg\,{\sc i}\fi}

\def\arcmin{{\mbox{$^{\prime}$}}}
\def\degree{{\mbox{$^{\circ}$}}}
\def\arcsec{{\mbox{$^{\prime \prime}$}}}
\def\cm{{\rm\thinspace cm}}

\def\erg{{\rm\thinspace erg}}

\def\g{{\rm\thinspace g}}

\def\K{{\rm\thinspace K}}

\def\km{{\rm\thinspace km}}

\def\Lsun{\hbox{$\rm\thinspace L_{\odot}$}}
\def\m{{\rm\thinspace m}}

\def\pc{{\rm\thinspace pc}}

\def\s{{\rm\thinspace s}}

\newcommand{\halpha}{\Halpha}

\newcommand{\hbeta}{\Hbeta}

\newcommand{\HeIIir}{\ifmmode {\rm He}\,\textsc{ii}\,\lambda8237 \else He\,{\sc ii}$\,\lambda8237$\fi}
\newcommand{\HeIir}{\ifmmode {\rm He}\,\textsc{ii}\,\lambda10830 \else He\,{\sc ii}$\,\lambda10830$\fi}

\newcommand{\SIII}{\ifmmode \left[{\rm S}\,\textsc{iii}\right]\,\lambda9531 \else [S\,\textsc{ii}]\,$\lambda9531$\fi}

\newcommand {\Lsoftint} {\ifmmode L^{\rm in}_{\mathrm{2-10\ keV}} \else $L^{\rm in}_{\mathrm{2-10\ keV}}$\fi}

\newcommand {\ergpersec} {\ifmmode {\rm erg~s}^{-1} \else erg~s$^{-1}$ \fi}

\newcommand {\nhunit} {cm$^{-2}$\xspace}
\newcommand {\degrees} {$^{\circ}$\xspace}

\def\micron{{\mbox{$\mu{\rm m}$}}}
\def\arcsec{{\mbox{$^{\prime \prime}$}}}
\def\arcmin{{\mbox{$^{\prime}$}}}
\def\degree{{\mbox{$^{\circ}$}}}

\def \wise{{\em WISE\ }}

\def\arcsec{{\mbox{$^{\prime \prime}$}}}
\def\cm{{\rm\thinspace cm}}

\def\erg{{\rm\thinspace erg}}

\def\g{{\rm\thinspace g}}

\def\K{{\rm\thinspace K}}

\def\km{{\rm\thinspace km}}

\def\Lsun{\hbox{$\rm\thinspace L_{\odot}$}}
\def\m{{\rm\thinspace m}}

\def\pc{{\rm\thinspace pc}}

\def\s{{\rm\thinspace s}}

\def\ergpcmsqps{\hbox{$\erg\cm^{-2}\s^{-1}\,$}}

\def\ergps{\hbox{$\erg\s^{-1}\,$}}

\def\kmps{\hbox{$\km\s^{-1}\,$}}

\def\apjl{\hbox{ApJL}}
\def\apjs{\hbox{ApJS}}

\def\aap{\hbox{AAP}}
\def\micron{{\mbox{$\mu{\rm m}$}}}
\def\arcsec{{\mbox{$^{\prime \prime}$}}}
\def\arcmin{{\mbox{$^{\prime}$}}}
\def\degree{{\mbox{$^{\circ}$}}}

\newcommand {\iraf}{{\sc IRAF}\xspace}
\newcommand {\molecfit}{\texttt{molecfit}}
\newcommand {\ppxf}{\texttt{pPXF}}

\newcommand{\nuvr}{\ifmmode {\rm NUV}-r \else NUV-$r$\fi}
\newcommand{\mh}{\ifmmode M_{\rm H_2} \else $M_{\rm H_2}$\fi}
\newcommand{\mhi}{\ifmmode M_{\rm HI} \else $M_{\rm HI}$\fi}

\newcommand{\must}{\ifmmode \mu_{\ast} \else $\mu_{\ast}$\fi}
\newcommand{\hmol}{\ifmmode H_2 \else $H_2$\fi}
\newcommand{\rmol}{\ifmmode R_{\rm mol} \else $R_{\rm mol}$\fi}
\newcommand{\tdep}{\ifmmode t_{\rm dep}({\rm H_2}) \else $t_{\rm dep}({\rm H_2})$\fi}
\newcommand{\tdepHI}{\ifmmode t_{\rm dep}({\rm HI}) \else $t_{\rm dep}({\rm HI})$\fi}
\newcommand{\fgas}{\ifmmode f_{\rm H_2} \else $f_{\rm H_2}$\fi}
\newcommand{\fhi}{\ifmmode f_{\rm HI} \else $f_{\rm HI}$\fi}
\newcommand{\xco}{\ifmmode \alpha_{\rm CO} \else $\alpha_{\rm CO}$\fi}

\newcommand{\Ndrnew}{22}
\newcommand{\Ngalacticnew}{17}
\newcommand{\Nspec}{1449} 
\newcommand{\Nspecnew}{1182} 

\newcommand{\Nnewz}{47}

\newcommand{\NAGN}{858}  
\newcommand{\Ndrtwospec}{816} 
\newcommand{\Ndrtwospecper}{95.1\%} 
\newcommand{\Nmbh}{790}
\newcommand{\Nbeamed}{105}
\newcommand{\Nunbeamed}{752}

\newcommand{\NPalomarAGN}{402}
\newcommand{\NXSHOOTERAGN}{211}
\newcommand{\NllamaTAGN}{19}
\newcommand{\NSDSSAGN}{151}

\newcommand{\NPalomarhigh}{66}
\newcommand{\NSOAR}{153}
\newcommand{\NduPont}{119}
\newcommand{\Narchival}{90}
\newcommand{\NFORS}{69}
\newcommand{\Nkecklris}{21}
\newcommand{\NMage}{12}
\newcommand{\NDRoneonly}{42}
\newcommand{\Nvltmuse}{6}

\newcommand{\Nrever  }{48}
\newcommand{\Nmegamaser }{10}
\newcommand{\NIFU  }{12}

\accepted{April 10, 2022}

\submitjournal{ApJS}

\shorttitle{BASS XXII: DR2 Catalog}
\shortauthors{Koss et al.}

\graphicspath{{./}{./figures/}}

\begin{document}

\title{BASS XXII: The BASS DR2 AGN Catalog and Data}

\correspondingauthor{Michael Koss}
\email{mike.koss@eurekasci.com}

\author[0000-0002-7998-9581]{Michael J. Koss}
\affiliation{Eureka Scientific, 2452 Delmer Street, Suite 100, Oakland, CA 94602-3017, USA}
\affiliation{Space Science Institute, 4750 Walnut Street, Suite 205, Boulder, CO 80301, USA}

\author[0000-0001-5231-2645]{Claudio Ricci}
\affiliation{N\'ucleo de Astronom\'ia de la Facultad de Ingenier\'ia, Universidad Diego Portales, Av. Ej\'ercito Libertador 441, Santiago 22, Chile}
\affiliation{Kavli Institute for Astronomy and Astrophysics, Peking University, Beijing 100871, People's Republic of China}

\author[0000-0002-3683-7297]{Benny Trakhtenbrot}
\affiliation{School of Physics and Astronomy, Tel Aviv University, Tel Aviv 69978, Israel}

\author[0000-0002-5037-951X]{Kyuseok Oh}
\affiliation{Korea Astronomy \& Space Science institute, 776, Daedeokdae-ro, Yuseong-gu, Daejeon 34055, Republic of Korea}
\affiliation{Department of Astronomy, Kyoto University, Kitashirakawa-Oiwake-cho, Sakyo-ku, Kyoto 606-8502, Japan}
\affiliation{JSPS Fellow}

\author[0000-0002-8760-6157]{Jakob S. den Brok}
\affiliation{Institute for Particle Physics and Astrophysics, ETH Z{\"u}rich, Wolfgang-Pauli-Strasse 27, CH-8093 Z{\"u}rich, Switzerland}
\affiliation{Argelander Institute for Astronomy, Auf dem H{\"u}gel 71, D-53231, Bonn, Germany}

\author[0000-0001-8450-7463]{Julian E. Mej\'ia-Restrepo}
\affiliation{European Southern Observatory, Casilla 19001, Santiago 19, Chile}

\author[0000-0003-2686-9241]{Daniel Stern}
\affiliation{Jet Propulsion Laboratory, California Institute of Technology, 4800 Oak Grove Drive, MS 169-224, Pasadena, CA 91109, USA}

\author[0000-0003-3474-1125]{George C. Privon}
\affiliation{National Radio Astronomy Observatory, 520 Edgemont Road, Charlottesville, VA 22903, USA}
\affiliation{Department of Astronomy, University of Florida, P.O. Box 112055, Gainesville, FL 32611, USA}

\author[0000-0001-7568-6412]{Ezequiel Treister}
\affiliation{Instituto de Astrof{\'i}sica, Facultad de F{\'i}sica, Pontificia Universidad Cat{\'o}lica de Chile, Casilla 306, Santiago 22, Chile}

\author[0000-0003-2284-8603]{Meredith C. Powell}
\affiliation{Kavli Institute of Particle Astrophysics and Cosmology, Stanford University, 452 Lomita Mall, Stanford, CA 94305, USA}

\author[0000-0002-7962-5446]{Richard Mushotzky}
\affiliation{Department of Astronomy, University of Maryland, College Park, MD 20742, USA}

\author[0000-0002-8686-8737]{Franz E. Bauer}
\affiliation{Instituto de Astrof\'{\i}sica  and Centro de Astroingenier{\'{\i}}a, Facultad de F\'{i}sica, Pontificia Universidad Cat\'{o}lica de Chile, Casilla 306, Santiago 22, Chile}
\affiliation{Millennium Institute of Astrophysics (MAS), Nuncio Monse{\~{n}}or S{\'{o}}tero Sanz 100, Providencia, Santiago, Chile}
\affiliation{Space Science Institute, 4750 Walnut Street, Suite 205, Boulder, Colorado 80301, USA}

\author[0000-0001-8211-3807]{Tonima T. Ananna}
\affiliation{ Department of Physics and Astronomy, Dartmouth College, 6127 Wilder Laboratory, Hanover, NH 03755, USA}

\author[0000-0003-0476-6647]{Mislav Balokovi\'c}
\affiliation{Yale Center for Astronomy \& Astrophysics and Department of Physics, Yale University, P.O. Box 208120, New Haven, CT 06520-8120, USA}

\author[0000-0001-5481-8607]{Rudolf E. B\"{a}r}
\affiliation{Institute for Particle Physics and Astrophysics, ETH Z{\"u}rich, Wolfgang-Pauli-Strasse 27, CH-8093 Z{\"u}rich, Switzerland}

\author[0000-0003-2344-263X]{George Becker}
\affiliation{Department of Physics \& Astronomy, University of California, Riverside, CA 92521, USA}

\author[0000-0002-0205-5940]{Patricia Bessiere}
\affiliation{Instituto de Astrof{\'i}sica, Facultad de F{\'i}sica, Pontificia Universidad Cat{\'o}lica de Chile, Casilla 306, Santiago 22, Chile}

\author[0000-0003-1014-043X]{Leonard Burtscher}
\affiliation{Leiden Observatory, P.O. Box 9513, 2300 RA, Leiden, the Netherlands}

\author[0000-0002-9144-2255]{Turgay Caglar}
\affiliation{Leiden Observatory, PO Box 9513, 2300 RA, Leiden, the Netherlands}

\author[0000-0002-8549-4083]{Enrico Congiu}
\affiliation{Departamento de Astronom\'ia, Universidad de Chile, Camino del Observatorio 1515, Las Condes, Santiago, Chile}

\author[0000-0002-8465-3353]{Phil Evans}
\affiliation{Department of Physics and Astronomy, University of Leicester,  Leicester, LE1 7RH, UK}

\author{Fiona Harrison}
\affiliation{Cahill Center for Astronomy and Astrophysics, California Institute of Technology, Pasadena, CA 91125, USA}

\author[0000-0002-1082-7496]{Marianne Heida}
\affiliation{European Southern Observatory, Karl-Schwarzschild-Str. 2, D-85748 Garching bei M{\"u}nchen, Germany}

\author[0000-0002-4377-903X]{Kohei Ichikawa}
\affil{Frontier Research Institute for Interdisciplinary Sciences, Tohoku University, Sendai 980-8578, Japan}
\affil{Astronomical Institute, Graduate School of Science
Tohoku University, 6-3 Aramaki, Aoba-ku, Sendai 980-8578, Japan}
\affil{Max-Planck-Institut f{\"u}r extraterrestrische Physik (MPE), Giessenbachstrasse 1, D-85748 Garching bei M{\"u}unchen, Germany
}

\author[0000-0002-3233-2451]{Nikita Kamraj}
\affiliation{Cahill Center for Astronomy and Astrophysics, California Institute of Technology, Pasadena, CA 91125, USA}

\author[0000-0003-3336-5498]{Isabella Lamperti}
\affiliation{Centro de Astrobiología (CAB), CSIC–INTA, Cra. de Ajalvir Km. 4, E-28850 Torrejón de Ardoz, Madrid, Spain}

\author[0000-0001-9879-7780]{Fabio Pacucci}
\affil{Center for Astrophysics, Harvard \& Smithsonian,
Cambridge, MA 02138, USA}
\affil{Black Hole Initiative, Harvard University,
Cambridge, MA 02138, USA}

\author[0000-0001-5742-5980]{Federica Ricci}
\affiliation{Instituto de Astrof{\'i}sica, Facultad de F{\'i}sica, Pontificia Universidad Cat{\'o}lica de Chile, Casilla 306, Santiago 22, Chile}
\affiliation{Dipartimento di Fisica e Astronomia, Università di Bologna, via Gobetti 93/2, 40129 Bologna, Italy}

\author[0000-0002-1321-1320]{Rog\'erio Riffel}
\affiliation{Departamento de Astronomia, Instituto de F\'\i sica, Universidade Federal do Rio Grande do Sul, CP 15051, 91501-970, Porto Alegre, RS, Brazil}

\author[0000-0003-0006-8681]{Alejandra F. Rojas}
\affiliation{Centro de Astronomía (CITEVA), Universidad de Antofagasta, Avenida Angamos 601, Antofagasta, Chile}

\author[0000-0001-5464-0888]{Kevin Schawinski}
\affiliation{Modulos AG, Technoparkstrasse 1, CH-8005 Zurich, Switzerland}

\author[0000-0001-8433-550X]{Matthew Temple}
\affiliation{Universidad Diego Portales, Av. Ejército Libertador 441, Santiago 8370109, Chile}

\author[0000-0002-0745-9792]{C. Megan Urry}
\affiliation{Yale Center for Astronomy \& Astrophysics and Department of Physics, Yale University, P.O. Box 208120, New Haven, CT 06520-8120, USA}

\author[0000-0002-3158-6820]{Sylvain Veilleux}
\affiliation{Department of Astronomy, University of Maryland, College Park, MD 20742, USA}

\author[0000-0002-0441-3502]{Jonathan Williams}
\affiliation{Department of Astronomy, University of Maryland, College Park, MD 20742, USA}
 
\begin{abstract}
We present the AGN catalog and optical spectroscopy for the second data release of the Swift BAT AGN Spectroscopic Survey (BASS DR2). With this DR2 release we provide \Nspec\ optical spectra,  of which \Nspecnew\ are released for the first time, for the \NAGN\ hard-X-ray-selected AGN in the Swift BAT 70-month sample.  The majority of the spectra (801/\Nspec, 55\%) are newly obtained from Very Large Telescope (VLT)/X-shooter or Palomar/Doublespec.  Many of the spectra have both higher resolution (R${>}$2500, N$\sim$450) and/or very wide wavelength coverage (3200-10000~\AA, $N\sim$600) that are important for a variety of AGN and host galaxy studies.    We include newly revised AGN counterparts for the full sample and review important issues for population studies, with \Nnewz\ AGN redshifts determined for the first time and \Nmbh\ black hole mass and accretion rate estimates.  This release is spectroscopically complete for all AGN (100\%, \NAGN/\NAGN) with 99.8\% having redshift measurements  (857/\NAGN) and 96\% completion in black hole mass estimates of unbeamed AGN (722/752).  This AGN sample represents a unique census of the brightest hard-X-ray-selected AGN in the sky, spanning many orders of magnitude in Eddington ratio (\lledd=$10^{-5}$--100), black hole mass (\mbh=$10^{5}{-}10^{10}$\Msun), and AGN bolometric luminosity (\lbol=$10^{40}{-}10^{47}$\ergps).

\end{abstract}


\section{Introduction} \label{sec:intro}

Many different optical spectroscopic surveys have been done of AGN in X-ray survey fields.  Early work focused on bright sources from the Einstein \citep[e.g.][]{Stocke:1991:813} and ROSAT surveys \citep[e.g.,][]{Voges:1999:389} typically focused on obtaining basic redshift and counterpart information for tens to hundreds of sources.    This was later extended to deeper and fainter sources in Chandra fields \citep[e.g.,][]{Green:2004:43,Szokoly:2004:271, Eckart:2006:19, Treister:2009:1713, Silverman:2010:124} or XMM-Newton surveys \citep[e.g.,][]{Menzel:2016:110}.  Full spectroscopic completeness was difficult  in the deepest surveys owing to the optical faintness of distant targets \citep{Brandt:2005:827}.     More recently, these efforts have focused on direct estimates of supermassive black hole (SMBH) masses for X-ray-selected AGN with broad emission lines using virial relations \citep[e.g.,][]{Shen:2011:45}.   


Larger optical spectroscopic samples of X-ray-selected active galactic nuclei (AGN) now exist within the Sloan Digital Sky Survey (SDSS) footprint, though typically focused on unobscured AGN.  A total of 7005 ROSAT sources were cross-matched with the SDSS spectroscopic footprint \citep{Anderson:2007:313}.  Due to the soft X-ray sensitivity  of ROSAT (e.g. 0.1--2.4\,keV), the majority, 89\% (6224/7005), were broad-line AGN.   Similarly, a study by \citet{Mahony:2010:1151} in the southern hemisphere using spectra from the 6-degree-Field Galaxy Survey \citep{Jones:2004:747,Jones:2009:683}, a near-infrared-selected redshift survey covering a large area in the Southern hemisphere ($\sim$17000\degree$^2$), had spectroscopic identifications for 1715 ROSAT sources in the Southern hemisphere.  More recently SPIDERS \citep[SPectroscopic IDentification of eROSITA Sources,][]{Dwelly:2017:1065,Comparat:2020:A97} is currently collecting 40,000 spectra of X-ray-selected AGN from ROSAT as well as the XMM slew survey (0.5--12\,keV), which is more sensitive to obscured AGN.

Hard X-ray emission ($>$10\,keV) from the corona of the AGN can probe the innermost parts of the central engine of AGN with the advantage over UV/optical/soft X-rays that it can even find AGN in highly obscured (e.g.~\NH$>10^{22}-10^{25}$ \nhunit) systems.  The Swift-BAT survey \citep{Barthelmy:2005:143}, with its all-sky coverage that is insensitive to obscuration up to Compton-thick levels \citep{Koss:2016:85}, provides the largest, most complete sample of bright, local ($z<0.1$), powerful AGN.  The spectroscopic coverage in the SDSS for BAT AGN, is, however, only $\sim$15\% \citep{Koss:2017:74}.  Therefore, a complete optical spectroscopic sample of BAT AGN, including the unobscured to the highly obscured AGN which are largely absent from ROSAT surveys, provides a unique way to fully understand BH growth and its relation to the host galaxy.  The BAT AGN survey also provides a bright complement that is more sensitive to obscured AGN compared  to the currently ongoing eROSITA satellite mission and its all-sky survey \citep{2021A&A...647A...1P}.  

The goal of the BAT AGN Spectroscopic Survey (BASS) is to provide the largest available spectroscopic sample of Swift BAT ultrahard X-ray (14--195\,keV) detected AGN.  In the BASS DR1 \citep{Koss:2017:74}, mostly archival optical telescope data was used for 641 BAT AGN from the 70-month BAT catalog \citep{Baumgartner:2013:19} and 102 AGN comosing the NIR DR1 \citep{Lamperti:2017:540}. These data were then used in a variety of scientific studies, such as between X-ray emission and high-ionization optical lines \citep{Berney:2015:3622}, ionized gas outflows \citep{Rojas:2020:5867}, and radio emission \citep{Baek:2019:4317,Smith:2020:4216}.  Several works identified the importance of the Eddington ratio in various scaling relations \citep[e.g.,][]{Oh:2017:1466,Ricci:2017:488,Ricci:2018:1819} and links to host galaxy properties such as molecular gas \citep{Koss:2021:29}.  

In the BASS data release 2 (DR2), we have identified all AGN among the 1210 sources in the BAT 70-month survey in order to obtain a 100\% spectroscopically complete sample of high-quality optical spectra and BH mass estimates for a large fraction of AGN across the entire sky.  High signal-to-noise ratio (S/N) and spectral resolution optical spectroscopy with measurements of continuum, emission and absorption lines, over the full optical range (3200-10000\AA) provides a large number of important diagnostics.  To name a few, these include star formation rates, stellar masses, stellar population ages, dust reddening, metallicities, AGN-driven outflows, and dynamical properties of the galaxy such as velocity dispersions of stellar populations \citep[e.g.,][]{Tremonti:2004:898,Vazdekis:2012:157,Yates:2012:215, Shimizu:2018:154, Rojas:2020:5867}.  Repeat optical spectroscopy can also probe the time-variable nature of these emission components such as in changing optical type AGN \citep[e.g.,][]{Collin-Souffrin:1973:343,Shappee:2014:48}.


In this first catalog paper of the DR2 release series, we provide an updated list of counterparts and the \Nspec\ optical spectra of \NAGN\ AGN among the 1210 sources in the BAT 70-month survey \citep{Baumgartner:2013:19}.  We provide an explanation of all the optical spectra obtained, their reductions, and general derived measurements.  An overview of the DR2 release and scientific results and a comparison with other surveys are provided in \citet{Koss_DR2_overview}.  Further catalogs of derived measurements will be provided in subsequent papers such as broad-line measurements \citep{Mejia_Broadlines}, narrow emission-line measurements from the best available optical spectra \citep{Oh_DR2_NLR}, velocity dispersion measurements from stellar absorption lines \citep{Koss_DR2_sigs}, and the NIR spectroscopic measurements \citep{denBrok_DR2_NIR,Ricci_DR2_NIR_Mbh}.  We will also provide scientific investigations using DR2 data such as the BH mass and Eddington ratio distribution function for obscured and unobscured AGN \cite{Ananna_DR2_XLF_BHMF_ERDF}, the ability of the MIR to recover obscured AGN \citep{Pfeifle:2022:3}, and the $M_{BH}$ - $\sigma_{\star}$ relation of type 1 unobscured AGNs \citep{Caglar_DR2_Msigma}.    Throughout this work, we adopt $\Omega_{\rm M} {=} 0.3$, $\Omega_\Lambda {=} 0.7$, and $H_0 {=} 70 \,\kms\,{\rm Mpc}^{-1}$.  To determine extinction due to Milky Way foreground dust, we use the maps of \citet{Schlegel:1998:525} and the extinction law derived by \citet{Cardelli:1989:245} with $R_v$=3.1.  


\section{Revised AGN Catalog}
The initial 70-month catalog \citep{Baumgartner:2013:19} was composed of 1210 sources, including 822 classified as AGN or associated with a galaxy and likely an AGN, 287 Galactic sources (e.g., high/low-mass X-ray binary, cataclysmic variable, pulsar), 19 clusters, and 82 unknown sources.  The counterpart positions and AGN classifications were updated based on WISE and X-ray data for 838 AGN in the BASS DR1 \citet[see Appendix A,][]{Ricci:2017:17}, which included three dual AGN systems.  However, even after the DR1, 44 unknown BAT sources, typically near the Galactic plane ($\lvert b\rvert{<}10^{\circ}$), had not been associated with counterparts.  Here we discuss the BASS DR2 AGN counterparts after an extensive examination of the remaining unknown 70-month catalog sources and important issues for surveys of these AGN.

Given the large (FWHM=19.5\arcmin) Swift BAT point spread function (PSF),  it is important to consider cases such as chance alignment of multiple AGN, AGN clustering, and dual AGN for population studies. Due to the scarcity of Swift BAT sources in the sky, the likelihood of chance alignment of any two unassociated sources within the BAT beam is small.  Specifically, there are $\sim$850 sources at $\lvert b\rvert{>}5^{\circ}$, (91\% of the sky), so the likelihood of chance alignment of one unassociated source within the BAT beam is very small, $\sim$0.2\%.  However, there are some unique cases involving galaxy mergers \citep[e.g.,][]{Koss:2018:214a} and galaxy clustering where multiple AGN systems occur.  Additionally, at lower fluxes, where AGN are more numerous, there may be some cases of flux boosting where two sources below the sky sensitivity both contribute to be above the detection sensitivity.  

Throughout this work, we refer to AGN as Sy1 (with optical broad \hbeta), Sy1.9 (narrow \hbeta\ and broad \halpha), and Sy 2 (with narrow \hbeta\ and narrow \halpha, including small numbers of LINERs and AGN in H$_2$-dominated regions).  This nomenclature is used for the sake of simplicity and consistency with previous work, despite the fact that many of our BASS DR2 AGNs may not be considered as
Seyfert galaxies, given their high (X-ray) luminosities.  Additionally, a small number of Sy2 sources would be better classified as HII or LINERs or composites based on their position in the BPT diagram \citep{Koss:2017:74} or even elusive AGN \citep{Smith:2014:112}, due to a lack of prominent lines.  These classifications are discussed in detail in the BASS DR2 paper on narrow emission-line measurements \citep{Oh_DR2_NLR}.  Finally, some Sy2 sources with narrow optical lines are known to have polarized broad-lines or NIR broad-lines \citep[e.g.,][]{Lamperti:2017:540}.

\subsection{Newly Identified AGN and Galactic Sources}
There were still some unidentified sources listed in the 70-month catalog, typically within the Galactic plane, that were not part of the BASS DR1 sample.  We examined them all to ensure that our sample provided a complete census of all AGN detected with Swift BAT across the sky.  Further optical spectroscopy found many of them to be stellar in nature.  


We uncovered an additional \Ndrnew\ AGN among the remaining unidentified sources (\autoref{tab:newagn}).  These sources were the brightest 2--10\,keV sources within the 5\arcmin\ Swift BAT position error circle, and optical spectroscopy confirmed their AGN nature.       The total sample then increases from 838 to \NAGN\ AGN.\footnote{838 AGN+\Ndrnew\ newly identified-2 DR1 AGN found to be stellar}  Further details on the X-ray modeling of these newly detected AGN will be provided in C. Ricci,  et al., (2022 in preparation).

\begin{deluxetable}{lllllllrr}
\tabletypesize{\scriptsize}
\tablewidth{0pt}
\tablecaption{New 70-month AGN Counterparts in BASS DR2 \label{tab:newagn}}
\tablehead{
\colhead{BAT ID}& \colhead{Swift Name}& \colhead{Counterpart}& \colhead{R.A.}& \colhead{Decl.}& \colhead{DR2 Type}& \colhead{$z$}&\colhead{$b$}& \colhead{$A_V$}\\
\colhead{(1)}& \colhead{(2)}& \colhead{(3)}& \colhead{(4)}&\colhead{(5)}& \colhead{(6)}& \colhead{(7)}&
\colhead{(8)}&\colhead{(9)}
}
\startdata
25&SWIFT J0041.0+2444&SWIFT J004039.9+244539&10.1661783&24.7609374&Sy1.9&0.078365&-38.0&0.11\\
323&SWIFT J0612.2-4645&PMN J0612-4647&93.1121548&-46.788457&BZQ&0.317767&-25.8&0.17\\
343&SWIFT J0640.0-4737&SWIFT J064013.50-474132.9&100.056201&-47.692985&Sy2&0.057242&-21.5&0.36\\
359&SWIFT J0709.3-1527&PKS 0706-15&107.302151&-15.450997&BZB&0.142277&-3.2&2.00\\
364&SWIFT J0714.7-2521&SWIFT J0714.7-2521&108.654098&-25.290303&Sy1&0.042503&-6.5&1.62\\
367&SWIFT J0723.8-0804&1RXS J072352.4-080623&110.971135&-8.1039597&Sy1.9&0.144926&3.4&0.94\\
396&SWIFT J0755.4+8402&2MASS J07581638+8356362&119.571766&83.9435807&Sy1&0.133952&28.7&0.15\\
410&SWIFT J0812.3-4004&1RXS J081215.2-400336&123.058378&-40.05667&Sy1&0.074934&-3.3&5.05\\
433&SWIFT J0854.3-0827&SWIFT J085429.35-082428.6&133.621953&-8.4076316&Sy2&0.188435&22.6&0.10\\
494&SWIFT J1020.5-0237A&SDSS J102103.08-023642.6&155.262884&-2.6118136&Sy2&0.293645&43.1&0.13\\
516&SWIFT J1045.3-6024&2MASS J10445192-6025115&161.216286&-60.419879&Sy2&0.047 \tablenotemark{a}&-1.3&9.54\\
761&SWIFT J1512.2-1053A&NVSS J151148-105023&227.952883&-10.840131&BZQ&0.94672&39.0&0.34\\
780&SWIFT J1548.1-6406&SWIFT J1548.1-6406&237.1265671&-64.0263441&BZQ&1.693124&-7.6&0.64\\
897&SWIFT J1737.7-5956A&1RXS J173751.2-600408&264.466913&-60.066598&BZQ&3.656025&-14.8&0.23\\
1000&SWIFT J1852.2+8424A&SWIFT J185024.2+842240&282.60455&84.3790556&Sy1&0.183122&27.0&0.30\\
1001&SWIFT J1852.2+8424B&1RXS J184642.2+842506&281.707105&84.4181331&Sy1&0.225381&27.1&0.30\\
1007&SWIFT J1852.8+3002&GALEXASC J185249.68+300425.8&283.206294&30.0741661&Sy1.9&0.057301&12.9&0.62\\
1075&SWIFT J2024.0-0246&1RXS J202400.8-024527&306.008863&-2.7590708&Sy1.9&0.137523&-21.8&0.21\\
1083&SWIFT J2034.0-0943&2MASX J20341926-0945586&308.5803195&-9.7664501&Sy2&0.081551&-27.2&0.19\\
1091&SWIFT J2048.4+3815&1RXS J204826.8+381120&312.112497&38.1903626&Sy1&0.105394&-3.4&2.76\\
1096&SWIFT J2059.6+4301B&SWIFT J210001.06+430209.6&315.004153&43.036367&Sy2&0.066023&-2.0&4.09\\
1164&SWIFT J2243.2-4539&2MASX J22422135-4539093&340.588956&-45.652581&Sy2&0.120675&-58.4&0.03\\
\enddata
\tablecomments{A detailed description of this table's contents is given in \autoref{sec:surveytable}.  Columns (1--2): BAT 70-month survey catalog ID and Swift name (\url{https://swift.gsfc.nasa.gov/results/bs70mon/}).  Column (3): corresponding galaxy counterpart name in NED or SIMBAD based on the WISE positions. Columns (4--5): right ascension and decl. of the IR counterpart of the BAT AGN, in decimal degrees, based on \wise positions.  Column (6): AGN type based on optical spectroscopy--Sy1 (with broad \hbeta), Sy1.9 (narrow \hbeta\ and broad \halpha), and Sy 2 (with narrow \hbeta\ and \halpha).  For beamed AGN, the types include those with the presence of broad lines (BZQ), only host galaxy features lacking broad lines (BZG), or traditional continuum-dominated blazars with no emission lines or host galaxy features (BZB).  Column (7):  best DR2 redshift measurement and the line or method used for the measurement. Measurements are from a broad-line fitting code \citep{Mejia_Broadlines}, when available for all Sy1 and BZQ sources with broad-line \hbeta.  For narrow-line sources, the redshift is based on emission-line fitting of \OIII, when possible.  For some high Galactic extinction sources or high-redshift sources $z{>}1$, other lines are used.  Finally, host galaxy templates are used for some continuum-dominated blazars (BZB) with no emission lines. Column (8): Galactic latitude, in decimal degrees. (9) visual extinction due to Milky Way foreground dust, using maps of \citet{Schlegel:1998:525} and the extinction law derived by \citet{Cardelli:1989:245}.}
\tablenotetext{a}{This particular AGN resides in a region of very high Galactic extinction ($A_{V}$=9.5), and the redshift is based on an \HeIir\ line measurement from the literature \citep{Fortin:2018:A150}.}
\end{deluxetable}


A summary of the \Ngalacticnew\ sources that are newly classified as Galactic is provided in \autoref{tab:newgalactic}.  For 10/\Ngalacticnew\ sources, we determine the stellar nature based on optical spectroscopy, with the remaining sources being classified in recent publications.

\begin{figure*}
\centering
\includegraphics[width=16cm]{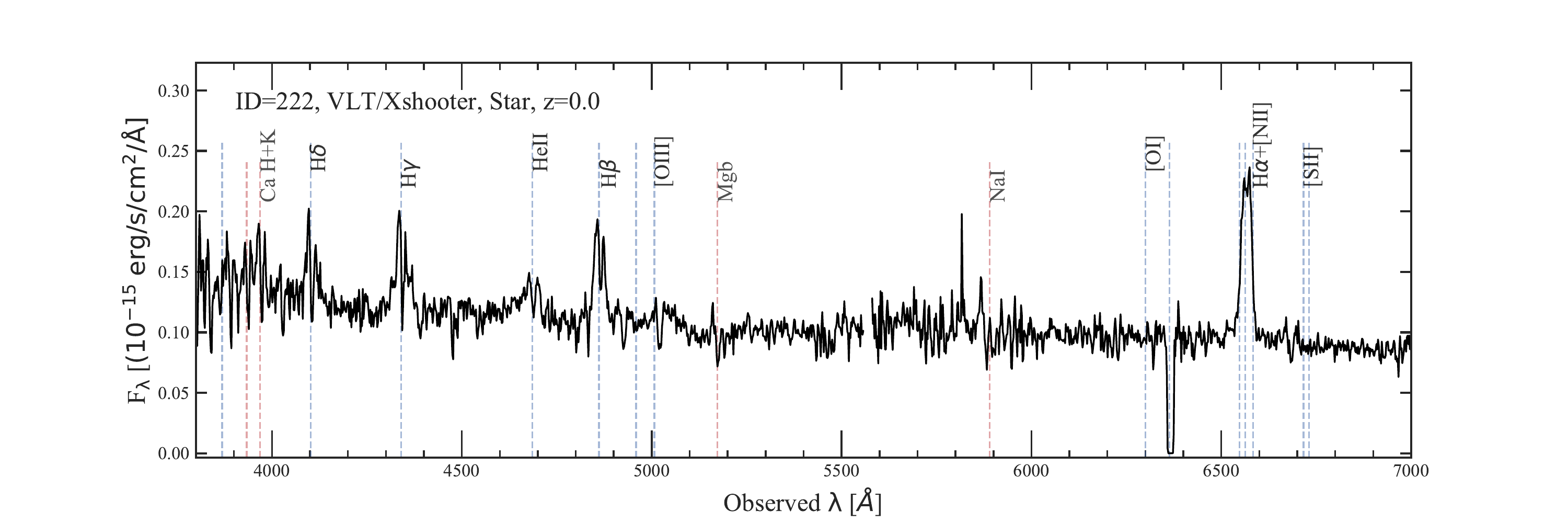}
\includegraphics[width=16cm]{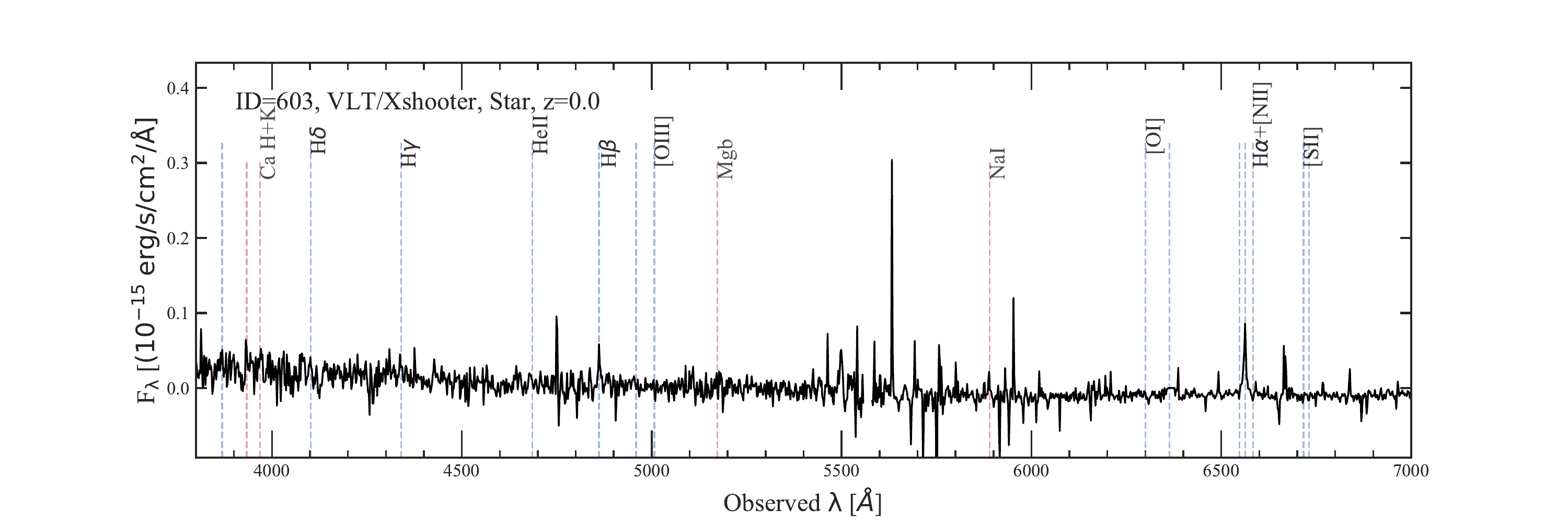}
\includegraphics[width=16cm]{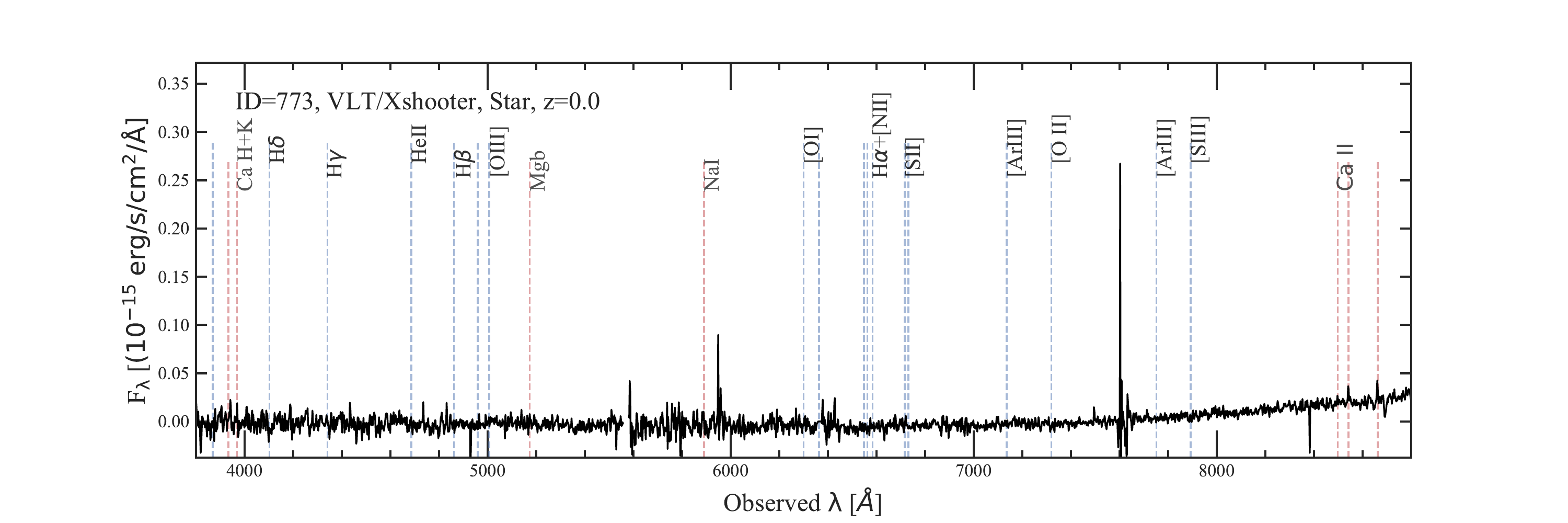}
\includegraphics[width=16cm]{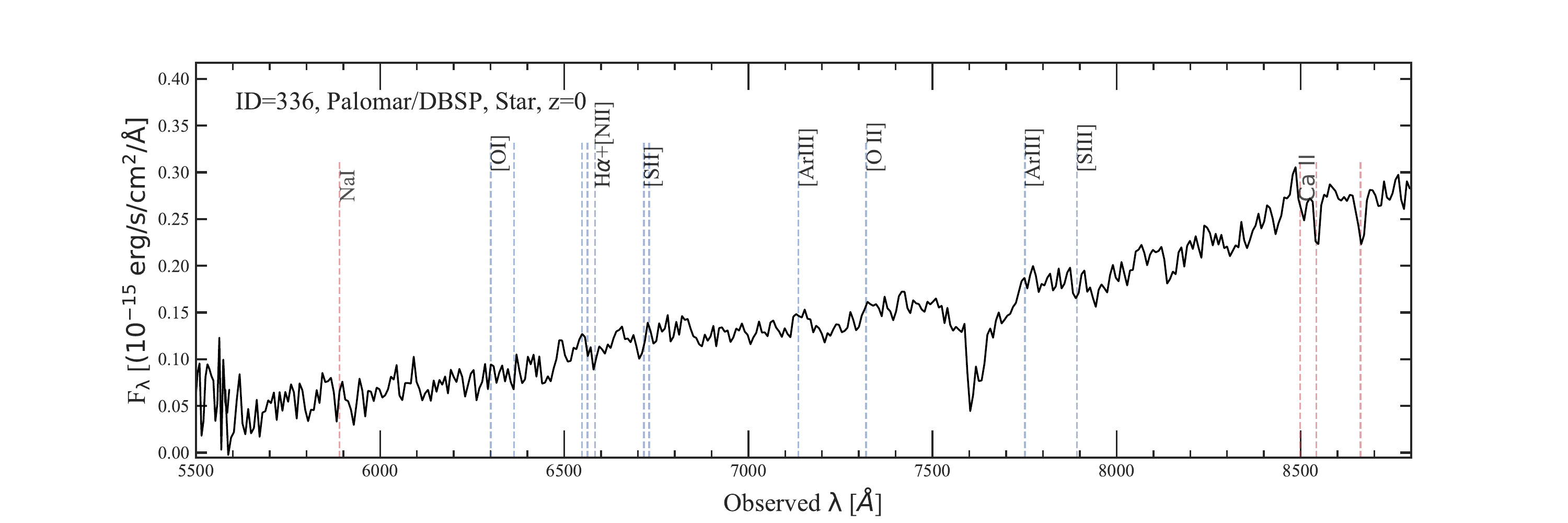}
\caption{Figure showing spectra of sources identified as Galactic based on emission or absorption lines at redshift zero.  }
\label{fig:Galactic_spec}
\end{figure*}

There are two cases that were thought to be AGN based on only an X-ray and WISE detection in the BASS DR1, but optical follow-up (\autoref{fig:Galactic_spec}) found the source to be Galactic.   SWIFT J0428.2-6704A, was found to be an eclipsing X-ray binary \citep{Kennedy:2020:3912} which follow-up optical spectra confirmed. We observed SWIFT J1535.8-5749 (aka. IGR J15360-5750), but found the source consistent with a very red star based on the \caii\ triplet spectral region (8450--8700\,\AA) and CO-band heads (2.29-2.51\,\micron).

In the other 15 cases, the 70-month X-ray classification was uncertain \citep{Baumgartner:2013:19}, but the sources were found to be Galactic.  For instance, SWIFT J0630.9+1129, showed a very red star with \halpha\ emission at $z{=}0$.  SWIFT J1213.2-6020 (aka IGR J12134-6015) was found to be located within 0\farcs3 of 2MASS J12132397-6015169 based on Chandra \citep{Karasev:2012:629}.  The WISE and X-ray counterpart has a GAIA source within 1\arcsec\ with $1.4\pm0.055$ mas parallax, and a spectrum consistent with a star.

\begin{deluxetable}{ccccccccccc}
\tabletypesize{\scriptsize}
\tablewidth{0pt}
\tablecaption{New Galactic Sources in DR2 \label{tab:newgalactic}}
\tablehead{
\colhead{BATID}& \colhead{Counterpart}& \colhead{R.A.}& \colhead{Decl.}& \colhead{Telescope}& \colhead{$b$}& \colhead{New Class}& \colhead{70-month}& \colhead{Feature}& \colhead{Ref.}
\\\colhead{(1)}& \colhead{(2)}& \colhead{(3)}& \colhead{(4)}&\colhead{(5)}& \colhead{(6)}& \colhead{(7)}&
\colhead{(8)}&\colhead{(9)}&\colhead{(10)}
}
\startdata
27&CXOM31 J004232.0+411314&10.6336579&41.2206639&&-21.6&Pulsar&in M31&&Y17\\
222&SWIFT J042749.42-670436.1&66.9571685&-67.076319&VLT/X-shooter&-38.6&Star&&Broad \Halpha&Ke20\\
289&SWIFT J053457.91+282837.9&83.7412468&28.4770753&&-2.2&CV&on plane&&H18\\
336&WISEA J063033.56+113414.1&97.63994&11.57063&&0.8&Star&&Broad \Halpha&\\
365&1RXS J071748.9-215306&109.4510999&-21.88376786&&-4.3&CV&in plane&&H18\\
462&SWIFT J092752.58-694438.8&141.979246&-69.745714&&-13.5&CV&SRC/X-RAY&&H18\\
603&1RXS J121324.5-601458&183.3497138&-60.2546853&VLT/X-shooter&2.3&Star&&Broad \Halpha&Ka12\\
618&XSSJ12270-4859&186.994503&-48.895133&&13.8&Pulsar&CV?&&D15\\
620&1RXS J123113.2-423524&187.80036&-42.588943&VLT/X-shooter&20.1&Star&X-ray src&Broad \Halpha&\\
773&IRAS15318-5740&234.011699&-57.814826&VLT/X-shooter&-1.7&Star&&Ca II Emission&\\
879&1RXS J172032.0-514414&260.135722&-51.736956&VLT/X-shooter&-8.3&Star&SRC/X-RAY&Ca II Emission&\\
933&IGR J17507-2856&267.6024&-29.037&&-0.9&LMXB&transient&&L01\\
989&SWIFT J183905.82-571507.6&279.774816&-57.251419&VLT/X-shooter&-21&Star&SRC/X-RAY&Ca II Absorption&\\
993&SWIFT J1842.2-1124&280.5729&-11.4172&Palomar/DBSP&-3.2&LMXB&transient&Ca II Absorption&C16\\
1061&SWIFT J200622.36+364140.9&301.591132&36.6982992&Palomar/DBSP&2.5&CV&New src&Ca II Absorption&H18\\
1095&SWIFT J205915.84+430109.5&314.815403&43.0189524&Palomar/DBSP&-1.9&HMXB&SRC/X-RAY&Ca II Absorption&\\
1159&SWIFT J223703.9+632338&339.155838&63.492667&Palomar/DBSP&4.5&CV&X-ray transient&Broad \Halpha&H18\\
\enddata
\tablecomments{Column (1): BAT 70-month survey catalog ID. Column (2): corresponding galaxy counterpart name in NED or SIMBAD based on the WISE positions. Column (3--4): right ascension and decl. of the IR counterpart of the BAT AGN, in decimal degrees, based on \wise positions. Column (5): telescope and instrument used to observe the Galactic source. Column (6): Galactic latitude, in decimal degrees.  Column (7): Galactic classification based on reference or optical spectral features. Star denotes stellar features in optical spectrum at redshift zero, but uncertain classification. Column (8): initial classification in 70-month catalog \citep{Baumgartner:2013:19}. Column (9): optical spectral feature used in redshift measurement. (10) References.}
\tablerefs{C16: \citet{Corral-Santana:2016:A61}; D15: \citet{deMartino:2015:2190}; H18: \citet{Halpern:2018:247}; Ka12: \citet{Karasev:2012:629}; Ke20: \citet{Kennedy:2020:3912}; L01: \citet{Liu:2001:1021}; Y17: \citet{Yukita:2017:47}.}
\end{deluxetable}

\subsection{Excluded Unknown Sources in the Galactic Plane} \label{appen_excluded}

The DR2 sample defined here is fully complete for all BAT 70-month BAT AGN and unknown sources above ($\lvert b\rvert{>}3^{\circ}$) or below $A_{V}$=5 mag. Beyond these limits an additional seven sources were excluded (\autoref{tab:excluded}) in the analysis owing to their very high extinction values (5-43 $A_V$ mag). 

 Many of these sources have been surveyed extensively as part of INTEGRAL surveys \citep[e.g.,][]{Tomsick:2008:1143}.   These sources are very close to the Galactic plane, with many foreground stars and very high extinction levels that make any accurate optical targeting impractical,  though they may host AGN in some cases. A follow-up Chandra observation SWIFT J1848.5-0046 (aka IGR J18485-0047) showed a source coincident with strong radio emission consistent with an AGN \citep{Tomsick:2009:811}.  SWIFT J1403.6-6146 (aka IGR J14044-6146), was observed with Chandra, but no detection was found after an earlier Swift XRT detection, suggesting variability that may be associated with an AGN \citep{Bodaghee:2012:113}.  In one case (SWIFT J2037.2+4151), optical spectral follow-up showed a very red spectrum, but there was no obvious counterpart.  SWIFT J2056.8+4939 (aka 4C 49.35), has both radio emission and a likely Fermi detection consistent with a blazar-like beamed AGN.   None of the remaining sources have been observed with Chandra, XMM-Newton, or NuSTAR. Further source localization and characterization with Chandra, deeper X-ray observations with XMM-Newton or NuSTAR, radio, and finally NIR spectroscopy will likely be required to identify these sources and their possible AGN nature.

There was also one remaining 70-month transient source outside the Galactic plane, SWIFT J0325.6-0907, which we exclude because it is likely a transient.  The source shows a declining significance in the 105-month light curve \citep{Oh:2018:4}, with a drop to zero XMM-Newton flux after 2010 January.  A 10 ks Swift XRT exposure finds no sources above \Fsoftobs${=}10^{-13}$\ergpcmsqps in flux that could be a likely counterpart.  No observations of this source have been performed with Chandra, XMM-Newton, or NuSTAR.

\begin{deluxetable}{ccccccc}
\tabletypesize{\footnotesize}
\tablewidth{0pt}
\tablecaption{Unknown Sources at $\lvert b\rvert <3\degree$ and $A_V$ Excluded from DR2\label{tab:excluded}}
\tablehead{
\colhead{BAT ID}& \colhead{Swift Name}& \colhead{Counterpart}& \colhead{RA}& \colhead{DEC}& \colhead{$b$}& \colhead{$A_V$}}
\startdata
464&SWIFT J0931.5-5105&SWIFT J093118.1-510845&142.825292&-51.146017&0.20&5.63\\
706&SWIFT J1403.6-6146&[CG2001] G311.45-0.13&211.12571&-61.789482&-0.13&43.08\\
742&SWIFT J1447.0-5814&SWIFT J144719.3-581616&221.831011&-58.271689&1.24&7.42\\
1005&SWIFT J1848.5-0046&IGR J18485-0047&282.106051&-0.7764179&0.32&35.98\\
1011&SWIFT J1855.5+0516&SWIFT J1855.5+0516&283.92069&5.220055&1.54&21.85\\
1086&SWIFT J2037.2+4151&SSTSL2 J203705.58+415005.3&309.273359&41.8347872&0.52&20.65\\
1094&SWIFT J2056.8+4939&RX J2056.6+4940&314.174892&49.6715329&2.76&9.25\\
\enddata
\tablecomments{Column descriptions are the same as in \autoref{tab:newagn}, unless otherwise noted.}
\end{deluxetable}

\subsection{Dual AGN} \label{dual_appen}
In the case of bright dual AGN, both may contribute significantly to the BAT flux \citep[e.g.][]{Koss:2016:L4}.  Dual AGN are typically close together ($\sim<30$ kpc), show small offsets in redshift (${<}500$\,\kmps), and often show signs of interaction in imaging.  For these sources, we report them as if they would have been individually detected based on the sky sensitivity \citep[][]{Baumgartner:2013:19}, and report the fluxes based on the soft X-ray emission \citep{Ricci:2017:17}.    Additional spectra of merging companions \citep[e.g.,][]{Koss:2010:L125, Koss:2012:L22} will be presented in a separate release (Koss, M., et al. 2022, in preparation). 

A summary of bright dual AGN in the sample can be found in \autoref{tab:duals}.   This only includes dual AGN that are both X-ray detected, not lower-luminosity AGN detected only in optical spectroscopy.   The galaxy group Arp 318 (NGC 835/NGC 833) and the close dual AGN Mrk 463 are unique in that the sources are individually below the median BAT sky sensitivity, but the combined flux is above the median sky sensitivity.   To be complete, we set the limit at \Fbatobs${>} 5{\times}10^{-12}$ \ergpcmsqps which corresponds to the faintest detected sources in the sky by BAT rather than the median sensitivity \Fbatobs${>} 1.03{\times}10^{-11}$ \ergpcmsqps. We assume \Fbatobs$\propto$ \Fsoftint for this calculation, and $\Gamma$=1.8, which is consistent at \NH${<}10^{23.5}$\,\nhunit \citep{Koss:2016:85,Ricci:2018:1819}.   This leads to three sets of  bright dual BAT-detected AGN in the DR2. See \citet[][]{Ricci:2017:17} and \citet{Koss:2012:L22} for a further discussion of these sources.

\begin{deluxetable}{ccccccccc}
\tabletypesize{\scriptsize}
\tablewidth{0pt}
\tablecaption{Dual AGN in the DR2 \label{tab:duals}}
\tablehead{
\colhead{BAT ID}& \colhead{Swift Name}& \colhead{Primary }& \colhead{Sec.}& \colhead{BAT Total}& \colhead{Ratio}& \colhead{\Fbatobs Sec.}& \colhead{Incl. }& \colhead{Ref.}\\
\colhead{}& \colhead{}& \colhead{}& \colhead{}& \colhead{$10^{-12}$\ergpcmsqps}& \colhead{}& \colhead{$10^{-12}$\ergpcmsqps}& \colhead{}& \colhead{}\\\colhead{(1)}& \colhead{(2)}& \colhead{(3)}& \colhead{(4)}&\colhead{(5)}& \colhead{(6)}& \colhead{(7)}&
\colhead{(8)}&\colhead{(9)}}
\startdata
841&SWIFT J1652.9+0223&NGC6240S&NGC6240N&72&0.32&23.0&Y&P16\\
1077&SWIFT J2028.5+2543&MCG+04-48-002&NGC 6921&78&0.20&15.6&Y&K16\\
112&SWIFT J0209.5-1010&NGC 835&NGC 833&15&0.46&7.1&Y&O18\\
\hline
471&SWIFT J0945.6-1420&NGC 2992&NGC 2993&27&0.14&3.8&N&K12\\
703&SWIFT J1355.9+1822&Mrk 463E&Mrk 463W&11&0.25&2.7&N&K12\\
678&SWIFT J1334.8-2328&ESO 509-IG066W&ESO 509-IG066E&18&0.10&1.8&N&Ko17\\
176&SWIFT J0324.9+4044&IRAS 03219+4031&2MASXJ03251221+4042021&19&0.08&1.6&N&K12\\
1198&SWIFT J2328.9+0328&NGC 7679&NGC 7682&15&0.10&1.5&N&K12\\
497&SWIFT J1023.5+1952&NGC 3227&NGC 3226&109&0.01&1.1&N&K12\\
552&SWIFT J1136.0+2132&Mrk 739E&Mrk 739W&13&0.09&1.2&N&K11\\
\enddata
\tablecomments{Columns (1--2): BAT 70-month survey catalog ID and Swift name (\url{https://swift.gsfc.nasa.gov/results/bs70mon/}).  Columns (3--4): primary and secondary galaxy where the primary galaxy is the one with brighter soft X-ray emission (\Fsoftint). Column (5): total hard X-ray emission detected from BAT for both sources.  Column (6): The ratio of the soft X-ray emission of the secondary to the total emission based on intrinsic soft X-ray emission (\Fsoftint). Column (7): the predicted \Fbatobs for the secondary given the soft X-ray emission. Column (8): whether or not the secondary galaxy optical spectra is included in the DR2 release. Column (9): references.} 
\tablerefs{Ko17: \citet{Kosec:2017:168}; K11: \citet{Koss:2011:L42}; K12: \citet{Koss:2012:L22}. O18: \citet{Oda:2018:79}; P16: \citet{Puccetti:2016:A157}.}
\end{deluxetable}

\subsection{Weakly Associated AGN} \label{tab:AGNcluster_appen}

Previous studies have also found that galaxy clustering is higher around BAT AGN \citep[e.g.,][]{Cappelluti:2010:L209,Koss:2010:L125} and this extends to secondary nearby AGN on scales of 70-1000 kpc.  These AGN will be separated by significant distances on the sky of several arcminutes, but show small offsets in redshift (${<}500$\,\kmps) and can often be found with several other galaxies clustering at the same redshift.   These two AGN will reside in the BAT beam and may both be bright enough to be individually detected.   In other cases, the clustered AGN would not be individually detected, but they are detected by BAT because of flux boosting. See \autoref{tab:AGNcluster_appen} for further details on these systems.

Here we provide a list of multiple soft X-ray AGN that are detected within the BAT beam but are weakly associated ($\sim$70-700 kpc; \autoref{tab:associatedagn}).  This includes seven pairs of AGN.  In three cases (SWIFT J0202.4+6824B, SWIFT J0359.0-3015A, SWIFT J1051.2-1704B), one of the AGN is below the median BAT sky sensitivity and should be excluded from population studies.

\begin{deluxetable}{cccccccccccc}
\tabletypesize{\scriptsize}
\tablewidth{0pt}
\tablecaption{Weakly Associated Counterparts in DR2 \label{tab:associatedagn}}
\tablehead{
\colhead{BAT ID}&\colhead{Swift ID}& \colhead{R.A.}& \colhead{Decl.}& \colhead{Counterpart}& \colhead{DR2 Type}& \colhead{$z$}& \colhead{\Fbatobs}& \colhead{Offset\tablenotemark{a}}& \colhead{}& \colhead{}\\
\colhead{}&\colhead{}& \colhead{}& \colhead{}& \colhead{}& \colhead{}& \colhead{}& \colhead{($10^{-12}$\ergpcmsqps})& \colhead{(\kmps)}& \colhead{(\arcsec)}& \colhead{(kpc)}}
\startdata
103&SWIFT J0202.4+6824A&30.5723165&68.3626549&LEDA89913&Sy2&0.011836&11.53&58&294&72\\
104&SWIFT J0202.4+6824B&30.384773&68.4060751&LEDA137972&Sy1.9&0.012028&7.68&&&\\
202&SWIFT J0359.0-3015A&59.8354003&-30.20269&SARS059.33488-30.34397&Sy1.9&0.097452&8.42&1086&391&681\\
203&SWIFT J0359.0-3015B&59.7867976&-30.302943&SARS059.28692-30.44439&Sy2&0.093833&10.98&&&\\
520&SWIFT J1051.2-1704A&162.812665&-17.008078&NGC3431&Sy2&0.01744&22.43&337&529&188\\
521&SWIFT J1051.2-1704B&162.9061&-17.124721&LEDA32573&Sy2&0.018563&3.72&&&\\
657&SWIFT J1306.4-4025A&196.608846&-40.41461&ESO323-77&Sy1.5&0.01563&33.06&171&525&168\\
658&SWIFT J1306.4-4025B&196.800383&-40.407563&ESO323-81&Sy1&0.0162&16.91&&&\\
746&SWIFT J1451.0-5540A&222.888099&-55.677311&LEDA3079667&Sy1.9&0.018091&40.84&172&1216&461\\
747&SWIFT J1451.0-5540B&222.303112&-55.605703&LEDA3085605&Sy2&0.018663&26.5&&&\\
754&SWIFT J1506.7+0353A&226.485635&3.7073107&Mrk1392&Sy1.5&0.036009&19.01&375&905&648\\
755&SWIFT J1506.7+0353B&226.684&3.862&2MASXJ15064412+0351444&Sy2&0.037259&15.67&&&\\
1173&SWIFT J2254.2+1147A&343.681327&11.7141072&UGC12243&Sy2&0.028508&12.55&88&428&245\\
1174&SWIFT J2254.2+1147B&343.581956&11.7825716&UGC12237&Sy2&0.028215&17.11&&&\\
\enddata
\tablecomments{Column descriptions are the same as in \autoref{tab:newagn}, unless otherwise noted.}
\tablenotetext{a}{Offset between two BAT AGN in measured redshift (\kmps) and \wise position (\arcsec\ and kpc) at the redshift of the first AGN listed.}
\end{deluxetable}

\subsection{Multiple Unassociated Faint X-Ray Counterparts} \label{AGNmult_appen}
Some sources in the initial 70-month catalog were reported with multiple likely counterparts within the error circle of the BAT beam position \citep{Baumgartner:2013:19}.  The details of many of these sources and their fluxes were described further in \citet{Ricci:2017:17} as part of the BASS DR1. Sources below the median sky sensitivity (\Fbatobs${<} 10.3{\times}10^{-12}$ \ergpcmsqps) should be excluded from population studies based on the all-sky sensitivity because they were likely detected only because of X-ray follow-up (\autoref{tab:faintagn}).  Among these, 10 pairs of sources 76\% (20/26) are cases of flux boosting that  individually are below the median sensitivity.  Another six cases are fainter counterparts to a brighter BAT source that is above the median detection sensitivity. There is only one case of a pair of AGN with both AGN above the detection sensitivity (SWIFT J1652.0-5915) which includes NGC 6221 ($z{=}0.0041$) and the background galaxy ESO 138-1 ($z{=}0.0091$).




\begin{deluxetable}{cccccccc}
\tabletypesize{\scriptsize}
\tablewidth{0pt}
\tablecaption{Faint Soft X-Ray Counterparts in DR2 \label{tab:faintagn}}
\tablehead{
\colhead{BAT ID}& \colhead{Swift ID}& \colhead{R.A.}& \colhead{Decl.}& \colhead{Counterpart}& \colhead{DR2 Type}& \colhead{$z$}& \colhead{\Fbatricci} ($10^{-12}$\ergps)}
\startdata
29&SWIFT J0042.9+3016A&10.7578514&30.2887771&2MASXJ00430184+3017195&Sy2&0.04894&7.72\\
92&SWIFT J0149.2+2153A&27.2487096&21.7594329&LEDA1656658&Sy1.9&0.069397&8.69\\
93&SWIFT J0149.2+2153B&27.353505&21.9973509&NGC678&Sy2&0.009485&8.17\\
223&SWIFT J0428.2-6704B&67.4475784&-67.055521&2MASXJ04294735-6703205&Sy1.2&0.064848&3.21\\
320&SWIFT J0609.5-6245A&92.5272683&-62.720088&2MASXJ06100652-6243125&Sy1&0.157475&9.45\\
321&SWIFT J0609.5-6245B&92.1613084&-62.787855&LEDA2816519&Sy1&0.099173&3.88\\
494&SWIFT J1020.5-0237A&155.262884&-2.6118136&SDSSJ102103.08-023642.6&Sy2&0.293645&6.88\\
495&SWIFT J1020.5-0237B&154.994136&-2.5767385&SDSSJ101958.58-023436.2&Sy1&0.059739&5.93\\
528&SWIFT J1105.7+5854A&166.495981&58.9460474&Z291-28&Sy2&0.047752&5.89\\
529&SWIFT J1105.7+5854B&166.406764&58.8557915&2MASXJ11053754+5851206&Sy1.2&0.191213&5.34\\
550&SWIFT J1132.9+1019B&173.247877&10.395067&[HB89]1130+106&BZQ&0.539603&8.78\\
554&SWIFT J1138.9+2529A&174.640356&25.3981165&LEDA1735060&Sy2&0.025363&8.94\\
555&SWIFT J1138.9+2529B&174.813041&25.5993964&SDSSJ113915.13+253557.9&Sy2&0.21925&5.63\\
601&SWIFT J1213.1+3239B&183.265867&32.7935448&B21210+33&BZQ&2.50706&8.12\\
632&SWIFT J1240.2+3457A&189.96528&34.9749458&Mrk653&Sy2&0.042812&9.00\\
633&SWIFT J1240.2+3457B&190.435624&35.0627157&NGC4619&Sy1.9&0.022946&7.02\\
664&SWIFT J1313.6+3650A&198.453989&36.8994443&RXJ1313.8+3653&Sy1.5&0.066945&7.12\\
665&SWIFT J1313.6+3650B&198.364486&36.5938771&NGC5033&Sy1.9&0.002763&5.52\\
761&SWIFT J1512.2-1053A&227.952883&-10.840131&NVSS J151148-105023&BZQ&0.94672&5.97\\
762&SWIFT J1512.2-1053B&228.021112&-10.776578&2MASXJ15120505-1046356&Sy2&0.165799&8.31\\
897&SWIFT J1737.7-5956A&264.466913&-60.066598&1RXSJ173751.2-600408&BZQ&3.656025&2.18\\
924&SWIFT J1747.8+6837A&267.159564&68.7045251&Mrk507&Sy1.2&0.055038&4.83\\
925&SWIFT J1747.8+6837B&266.748141&68.6085193&2MASS J17465953+6836303&Sy1.2&0.063785&5.23\\
1000&SWIFT J1852.2+8424A&282.60455&84.3790556&SWIFT J185024.2+842240&Sy1&0.183122&5.51\\
1001&SWIFT J1852.2+8424B&281.707105&84.4181331&1RXSJ184642.2+842506&Sy1&0.225381&4.19\\
1096&SWIFT J2059.6+4301B&315.004153&43.036367&SWIFT J210001.06+430209.6&Sy2&0.066023&5.84\\
\enddata
\tablecomments{Column descriptions are the same as in \autoref{tab:newagn}.}
\end{deluxetable}

\subsection{Beamed and Lensed AGN}
The Swift BAT survey includes beamed, lensed, and unbeamed AGN and it is important to separate them in most scientific analyses.  The original DR1 included 105 beamed sources \citep[see, e.g., Table 1 of][]{Ricci:2017:17}, based on the Roma Blazar Catalog (BZCAT) catalog \citep{Massaro:2009:691} and DR1 optical spectra \citep{Koss:2017:74}.  Since this release, a further study by \citet{Paliya:2019:154} used recent Fermi LAT data and SED fitting to identify all the blazars in the 105-month catalog, which includes all DR2 70-month AGN. 

For continuity, we provide all of the new beamed AGN, or changes of beamed AGN from the DR1 classifications to unbeamed in DR2 or vice versa, in  \autoref{tab:beamedchange} in \autoref{appen_beamed}. In most cases, recent radio observations or SED fitting \citep{Paliya:2019:154} revealed beamed AGN to be unbeamed, or conversely, a recent Fermi detection revealed a previously categorized unbeamed AGN to be beamed.  In DR2, 13/105 (12\%) changed from beamed to unbeamed AGN classification.  Conversely, eight unbeamed AGN in the DR1 list were found to be beamed.  Finally,  five beamed AGN were included because they are part of the expanded list of AGN, that were not in the original list of 838 DR1 AGN.   

For DR2, we reviewed all the beamed classifications and found three AGN that were classified as beamed AGN in \citep{Paliya:2019:154}, but we now classify them as unbeamed AGN based on further analysis.  SWIFT J0312.9+4121 (aka QSO B0309+411), while detected in Fermi, is a broad-line radio galaxy, with double-lobe morphology within a compact structure \citep{Tzanetakis:1978:63P}.  For SWIFT J0519.5-3140 (aka PKS 0521-365), \citet{Angioni:2019:A148} found with VLBI imaging that the jet of PKS 0521-36 is not highly beamed, with viewing angles larger than 10\degrees.  Finally, SWIFT J1742.1-6054 (aka PKS1737-60), was classified as an FRII radio galaxy \citep{Bassani:2016:3165}.  Further polarimetric observations or detections of compact cores and superluminal motions using high-resolution radio imaging would be needed to further classify these sources.  

The beamed DR2 AGN then total \Nbeamed, the same as in the DR1 despite the 26 changes.  There is also a lensed AGN, SWIFT J1131.9-1233 (aka QSO J1131-1231 at $z{=}0.654$) by a galaxy at $z{=}0.295$ \citep{Berghea:2017:90, Sluse:2017:4838}.   Finally, SWIFT J1833.7-2105 (aka PKS 1830-211 at $z{=}2.5$) is both beamed and also lensed \citep{Lidman:1999:L57} by a foreground spiral at $z{=}0.89$ \citep{Winn:2002:103}.


\begin{deluxetable}{cccccccccc}
\tabletypesize{\footnotesize}
\tablewidth{0pt}
\tablecaption{DR2 Instrument Setups \label{tab:setups}}
\tablehead{
\colhead{Telescope}&\colhead{Instrument}&\colhead{Total}&\colhead{Grating}&\colhead{Range}&\colhead{Dispersion}&\colhead{Slit Width}&\colhead{Res.}&\colhead{$R$}&\colhead{LSF}\\
\colhead{}&\colhead{}&\colhead{}&\colhead{(lines/mm)}&\colhead{(\AA)}&\colhead{(Ang/pix)}&\colhead{(\arcsec)}&\colhead{FWHM (\AA)}&\colhead{}&\colhead{Type}\\\colhead{(1)}& \colhead{(2)}& \colhead{(3)}& \colhead{(4)}&\colhead{(5)}& \colhead{(6)}& \colhead{(7)}&
\colhead{(8)}&\colhead{(9)}&\colhead{(10)}}
\startdata
Palomar&Doublespec&440&600/316&3150-5598/5200-10500&1.07/1.54&1.5&4.1/4.9&1220/1730&Stars/\molecfit\\
&&33&&&&2&4.8/6.5&1040/1290&Sky lines/\molecfit\\
&&2&&&&1&3/3.6&1670/2360&Sky lines/\molecfit\\\\
&&1&&&&0.5&2.6/3.2&1920/2660&Sky lines/\molecfit\\\\
&&10&300/316&3413-5500:4760-10500&1.07/1.54&1.5&8.1/4.9&620/1730&Sky lines/\molecfit\\\\
&&10&600/600&3150-5598/5570-8930&1.07/0.82&2&4.8/3.4&1040/2500&Sky lines/\molecfit\\\\
&&6&&&&1.5&4.1/2.8&1220/3030&Sky lines/\molecfit\\\\
&&51&1200/1200&3970-5499/8050-9600&0.55/0.4&2&2.3/1.8&2170/4720&Sky lines/\molecfit\\
&&15&&&&1.5&2/1.3&2500/6000&Sky lines/\molecfit\\
VLT&X-shooter&179&Echelle&2990-5560/5337-10200&0.2&1.6/1.5&1.3/1.4&3850/6000&Stars/\molecfit\\
&&11&&&&1/0.9&0.9/1.0&5500/8900&Arc lines\\
&&4&&&&1.3/1.2&1.2/1.3&4120/6500&Arc lines\\
&&2&&&&1/1.2&0.9/1.3&5500/6500&Arc lines\\
&&1&&&&0.8/0.7&0.8/0.7&6650/11570&Arc lines/Stars\\
&&36&IFU&&&1.8&1/0.6&8600/13330&Sky lines\\
&FORS2&61&600&3400-6100&1.32&1&6&830&Sky lines\\
&&9&300&6100-11000&2.8&&6.8&1250&Sky lines\\
&MUSE&4&&4800-9300&1.25&2&2.7/2.7&1850/3150&Instrument\\
&&2&&&&1&2.7/2.7&1850/3150&Instrument\\
SOAR&Goodman&58&400&4560-8690&0.99/1.98&1.2&5.6/5.2&890/1630&Stars\\
&&9&600&5290-7200&1.29&1.2&3.8&1450&Sky lines\\
&&34&931&8000-9600&0.39/0.78&1.2&2.7&3150&\molecfit\\
&&50&1200&7900-9070&0.28&1.2&1.8&4720&\molecfit\\
&&2&&&&0.45&0.7&12100&\molecfit\\
APO&SDSS&146&Fiber&3830-9180&1.15/1.96&3&3.0/4.1&1760/2490&Survey\\
&&31&&3600-10330&&2&2.9/3.4&1650/2070&Survey\\
Dupont&BC&119&300&3000-9070&&1&10.4&480&Stars\\
Keck&LRIS&15&600/400&3200-5600/5450-10280&1.23/1.19&1&3.9/4.7&1280/1810&Sky lines/\molecfit\\
&&4&&&&1.5&4.6/6.3&1090/1490&Sky lines/\molecfit\\
&&1&600/600&&1.23/0.8&1.5&4.6/5.4&1090/1670&Sky lines/\molecfit\\
&&1&600/1200&&0.61/0.4&1&3.9/1.6&1280/5310&Sky lines/\molecfit\\
Magellan&MAGE&12&Echelle&3300-10010&0.25&1&1.3&3850&Sky lines\\
\enddata
\tablecomments{See \autoref{sec:surveytable} for a detailed description of this table's contents. Column (1): telescope. Column (2): instrument. Column (3): total number of DR2 spectra observed with this setup. Column (4): grating listing the lines/mm, if applicable.  For instruments with a blue and red side two numbers are listed associated with each grating.  Columns (5--9):  wavelength range (Range), pixel dispersion, resolution (Res.), and resolving power ($R$).  These represent typical values for this setup and may have small differences within individual spectra depending on observing conditions.  These quantities may also be wavelength dependent in some cases, and so the values are given at 5000\,\AA\ and 8500\,\AA\ depending on the spectral range.  Two values are listed when the instrument had both a blue and red arm with different settings. Column (10): method used to determine the instrumental resolution, either with sky lines, telluric features with \molecfit, with arc lines, fitting stellar templates to stars, or based on resolutions provided within the survey (e.g. SDSS).  The measurements are for 5000 or 8500\,\AA\ or both depending on the spectral range.  See \autoref{sec:instres} for further details.}
\end{deluxetable}

\section{Survey, Observations, and Data Reduction}
Here we provide an overview of DR2 survey and observations.  The DR2 targeting criteria goals were to provide the largest possible sample of BH mass measurements from either broad Balmer lines or stellar velocity dispersion measurements, as well as the broadest possible spectral coverage (e.g. 3000--10000\,\AA) for emission-line measurements for the entire catalog of \NAGN\ AGN.  In practice, outside of echelle instruments, this required multiple spectra with broad wavelength coverage with lower-resolution or higher-resolution gratings with narrower wavelength coverage.  Repeat observations were done if the S/N of the broad Balmer lines (\hbeta\ or \halpha) was too low for measurements \citep[][]{Mejia_Broadlines} or the low S/N and/or spectral resolution of the stellar absorption features resulted in a failed measurement of velocity dispersion.  We did not reobserve targets with acceptable spectra and measurements from the SDSS.  The requirement for high S/N, high spectral resolution, and broad wavelength coverage, combined with queue mode observing approved months in advance, sometimes resulted in duplicate (or more) observations of the same source.  Example spectra for different instrumental setups are provided in \autoref{spec_example}.  All spectra associated with DR2 for each AGN will be provided at the BASS website.{\footnote{\href{https://www.bass-survey.com/}{https://www.bass-survey.com/}}}

A summary of all the observational setups used is shown in \autoref{tab:setups}.  We did not specifically exclude sources with high Galactic extinction in spectroscopic targeting if there was an obvious optical counterpart to the WISE counterpart of the soft X-ray emission.  This resulted in observations of 12 AGN with very high extinctions (between $A_{V}$=5 and 10 mag) close to the Galactic plane ($0<b<3^{\circ}$), primarily to determine the first redshift and AGN type.

The data reduction and analysis of DR2 spectra maintain the uniform approach described in the initial DR1 paper \citep{Koss:2017:74}.  All new spectra are processed using the standard tasks for cosmic-ray removal, 1D spectral extraction, wavelength, and flux calibrations, in either \iraf or the ESO/{\tt Reflex} environment for the Very Large Telescope (VLT) instruments. The spectra are flux-calibrated using standard stars, which were typically observed two to three times per night.  The spectra are corrected for Galactic reddening.   Finally, a telluric absorption correction is applied to the spectra with the software \molecfit.      

\subsection{Master Observing Table \label{sec:spectable}} 

\begin{deluxetable}{rll}
\label{tab:specall}
\tablecaption{Column Description for the Master Observing Table.}
\tablehead{
\colhead{Column} & \colhead{Name} & \colhead{Units}}
\startdata
1& BAT\_ID                            &\\
2&Telescope&\\
3&Diameter&m\\
4&Instrument&\\
5&File&\\
6&File Red&\\
7&Flags&\\
8&Date&\\
9&Range&\AA\\
10&Grating&\\
11&Grating Red&\\
12&$R$&\\
13&$R$ Red&\\
14&Res&\AA\\
15&Res Red&\AA\\
16&Slit Width&\arcsec\\
17&Slit Width Red&\arcsec\\
18&Slit Length&\arcsec\\
19&Slit Length Red&\arcsec\\
20&Angle&\degrees\\
21&Seeing&\arcsec\\
22&Air mass&\\
23&Exposure&s\\
24&CDELT&\AA/pix\\
25&CDELT Red&\AA/pix\\
26&BC&\kmps\\
27&EBV&mag\\
\enddata
\tablecomments{See \autoref{sec:spectable} for detailed descriptions of each field.}
\end{deluxetable}

We provide the following key parameters when possible for each individual spectrum (\autoref{tab:specall}):

\begin{enumerate}

\item \texttt{BAT ID:} Catalog ID in the BAT survey.\footnote{\url{https://swift.gsfc.nasa.gov/results/bs70mon/}}


\item \texttt{Telescope, diameter, instrument}: Name of observatory, its diameter, and instrument used.


\item \texttt{File and  File red}: Name of associated fits spectral file.  For telescopes with both a blue and red side, two spectra are listed.

\item \texttt{Flags}: Any associated flags with calibration or spectral extraction.  Star: indicates foreground stellar contamination, that a very nearby star (${<}2$ \arcsec) contributed to the emission despite a very small extraction region.  Red: indicates that only the red side is extracted because the Galactic extinction was so high (e.g. $A_V{>}$3) that no source is detected in the blue. Calibration: indicates that the object was observed under poor conditions or the standard star was observed on a different night, so spectral calibration may be more uncertain than usual.  Tellurics: indicates that the spectrum suffers from worse-than-usual telluric correction or that the \molecfit\ correction was unsuccessful. Shortblue: the setup has a shorter than normal blue wavelength coverage due to a reduction issue.

\item \texttt{Date}: UT date of observation.

\item \texttt{Spectral Range}: Range of the spectra in \AA. For telescopes with both a blue and red side, two spectral ranges are listed with the blue side first.

\item \texttt{Grating}: Name of associated grating or grism.  If the instrument had both a blue (shorter wavelength) and red camera (longer wavelength), two gratings are listed with the blue side listed first.

\item \texttt{R} and \texttt{Res FWHM}: Instrumental resolution and FWHM in \AA.  For telescopes with both a blue and red side, two numbers are provided with the blue side first.

\item \texttt{Slit}: Slit width in \arcsec.  For telescopes with both a blue and red side, two widths are listed with the blue side first.

\item \texttt{Slit length}: Extraction length along the slit in \arcsec.  For telescopes with both a blue and red side, two lengths are listed with the blue side first. If multiple exposures were combined with optimal extraction (e.g. Palomar/DBSP), the average value is listed.

\item \texttt{Angle}: Position angle in degrees, measured east of north.  In most cases the sources were observed at parallactic unless a nearby galaxy was observed in the same slit.

\item \texttt{Seeing}: Recorded seeing of observations.  When possible we use the average seeing.  We have not corrected the seeing observations to the observed air mass.

\item \texttt{Exposure}: Total exposure from all combined observations for the individual spectra.

\item \texttt{CDELT}:  Pixel dispersion in \AA per pixel.  For telescopes with both a blue and red side, two widths are listed with the blue side first.  Only included for spectra with linear dispersions (e.g. not the SDSS).

\item \texttt{Airmass}: Average air mass during observation.

\item \texttt{BC}: Barycenter correction \kmps needed for the Earth's motion based on observation time and observatory location.  The computed correction should be added to any observed velocity to determine the final barycentric radial velocity. As this correction is small (e.g. $<30$\,\kmps), it has not been applied to any catalog measurements in the DR2.

\item \texttt{EBV}:  Atmospheric extinction.

\end{enumerate}

We note that all of these observing parameters are not available for every spectrum, including most of the archival sample, but we provide them when possible.

\subsection{Overview of Samples}
\begin{figure*} 
\centering
\includegraphics[width=16cm]{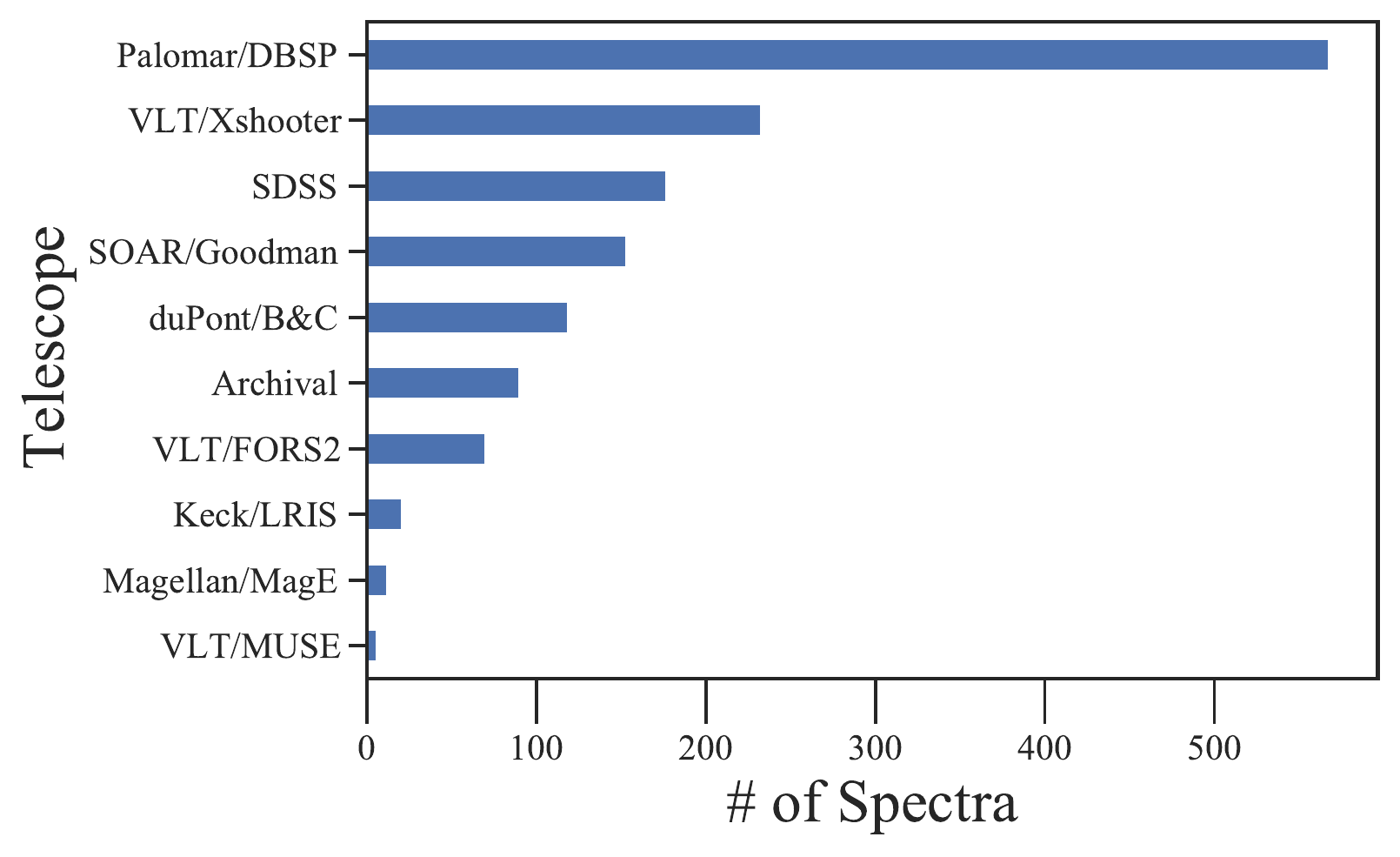}
\caption{Horizontal bar chart showing the number of spectra taken with each telescope, instrument, or survey.  The archival sample is from earlier surveys that were not included in the DR1, including {\em ROSAT} AGN that overlap with BASS in unpublished or published \citep{Grupe:2004:156} works, from the Palermo surveys of Swift BAT AGN \citep[][]{Rojas:2017:A124}, or as part of an atlas of low-redshift AGN \citep{Ho:2009:398}.}
\label{fig:spectra_num}
\end{figure*}

\begin{figure*} 
\centering
\includegraphics[width=8.2cm]{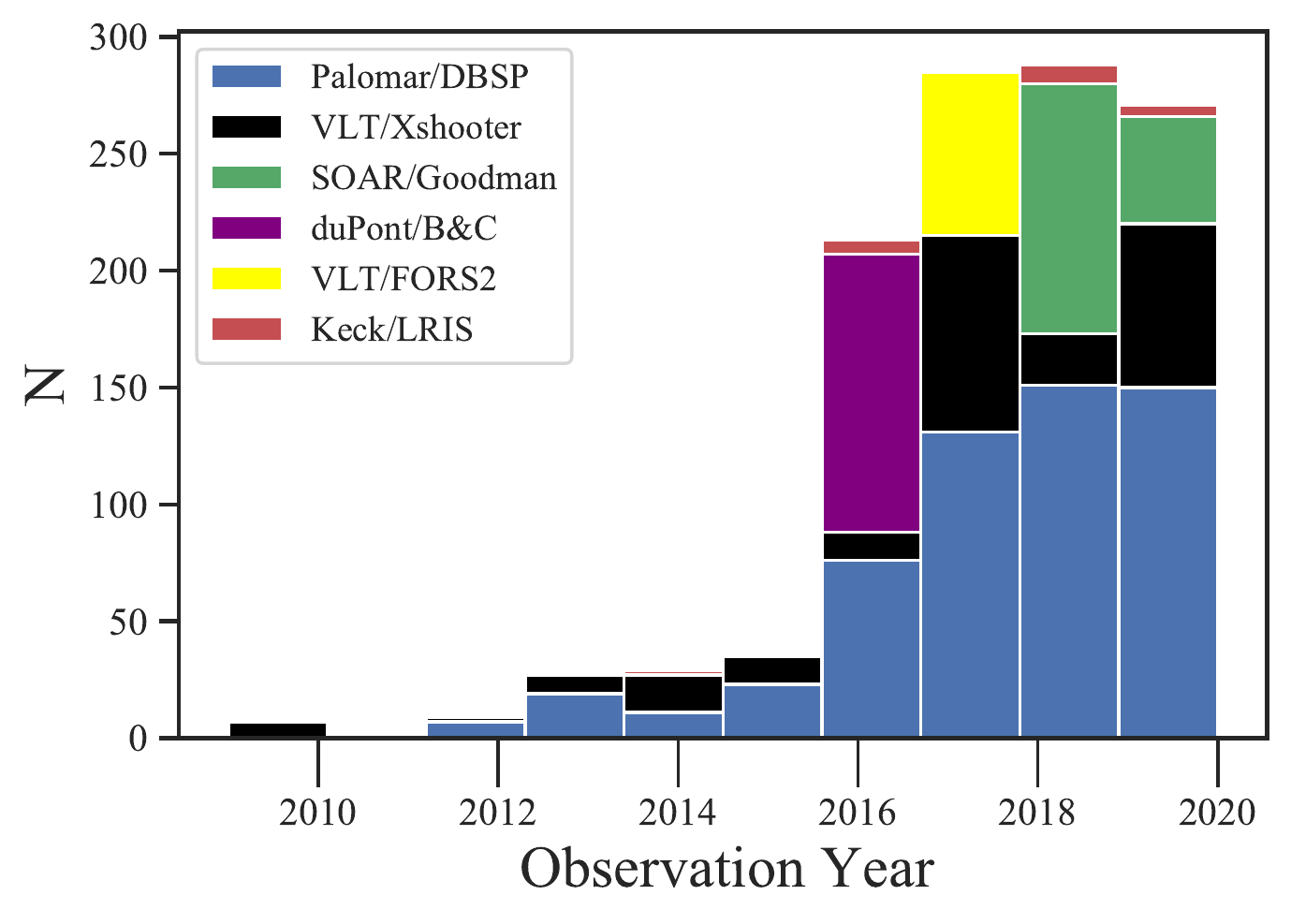}
\includegraphics[width=8.2cm]{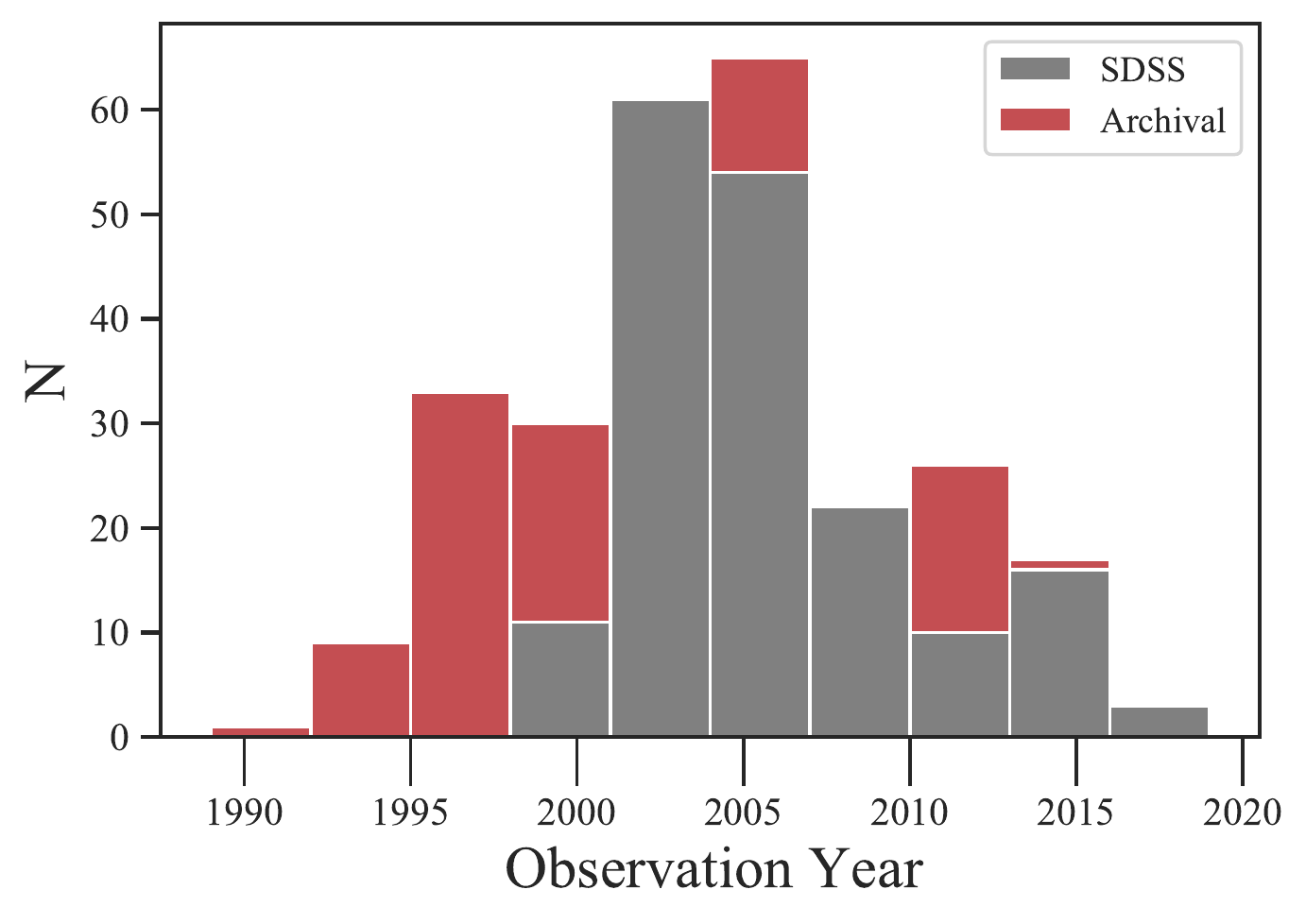}
\includegraphics[width=8.2cm]{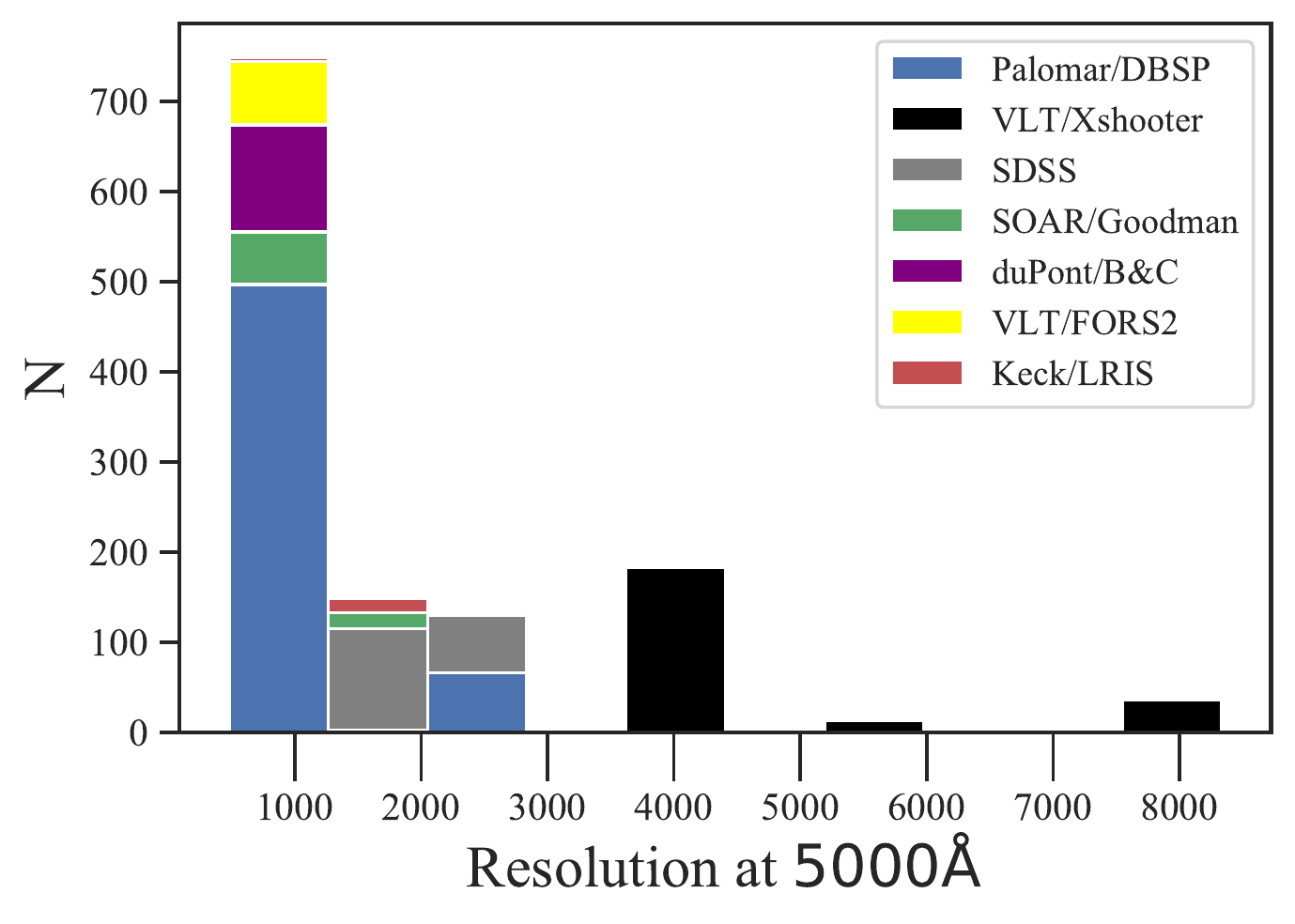}
\includegraphics[width=8.2cm]{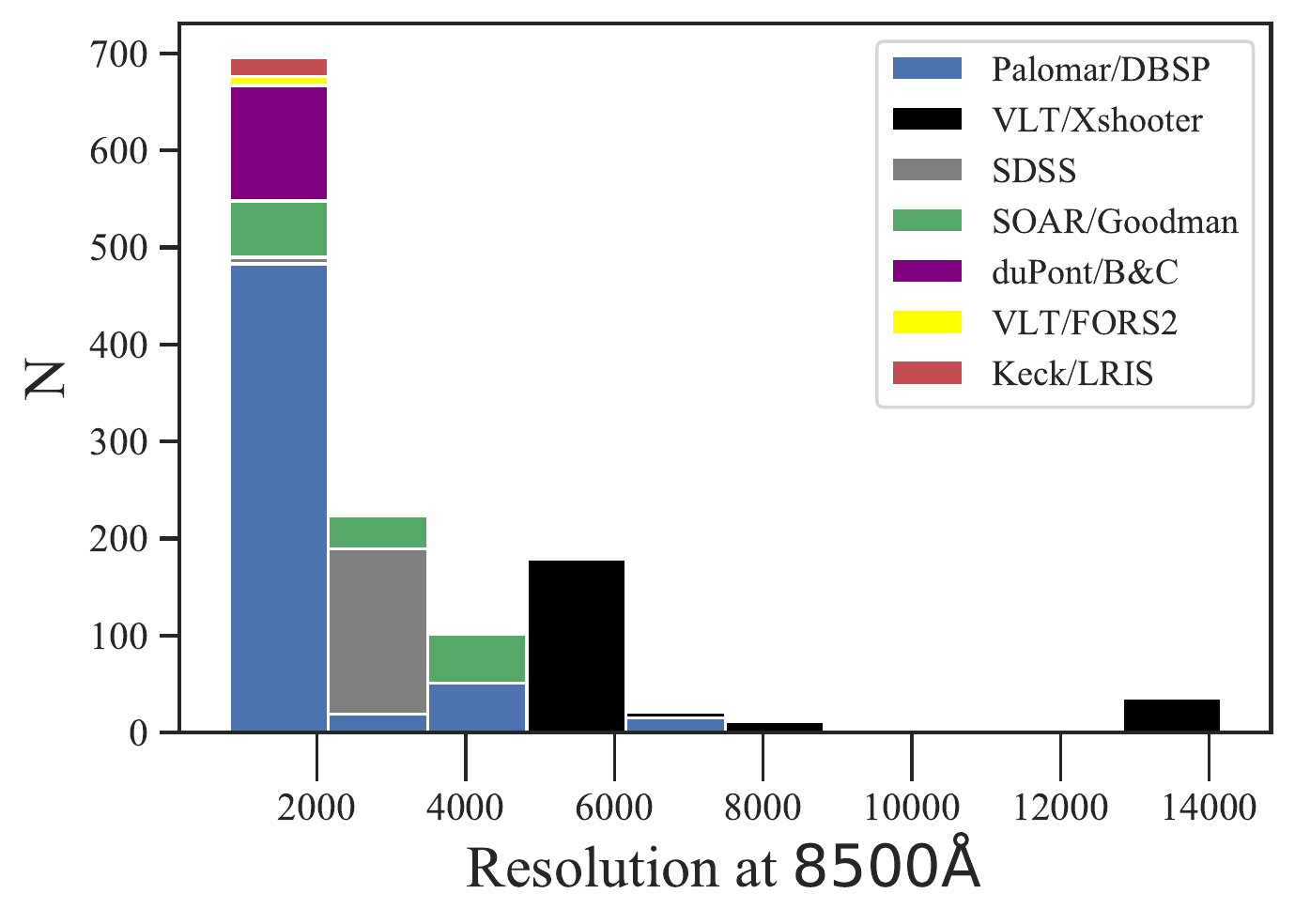}
\includegraphics[width=8.2cm]{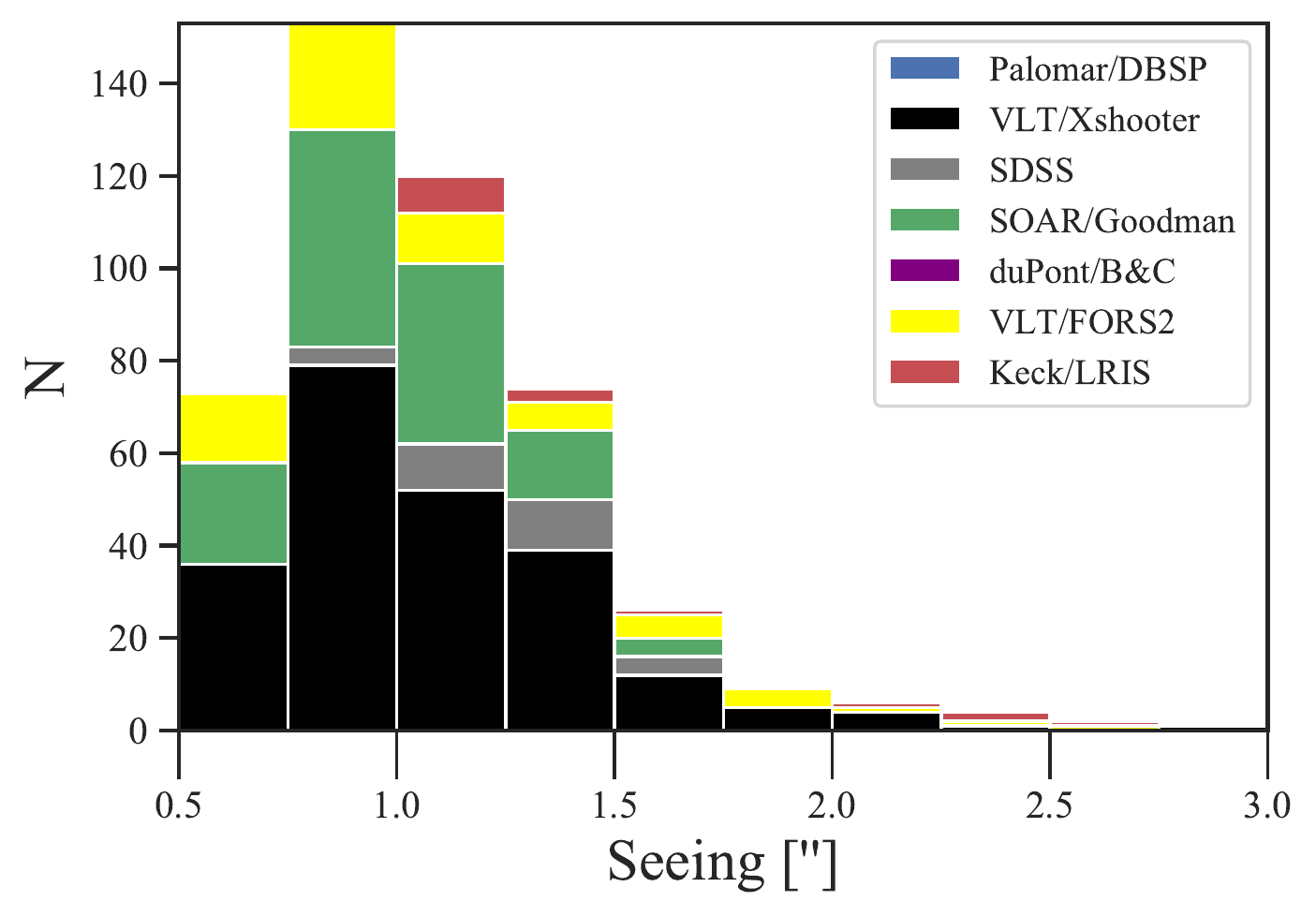}
\includegraphics[width=8.2cm]{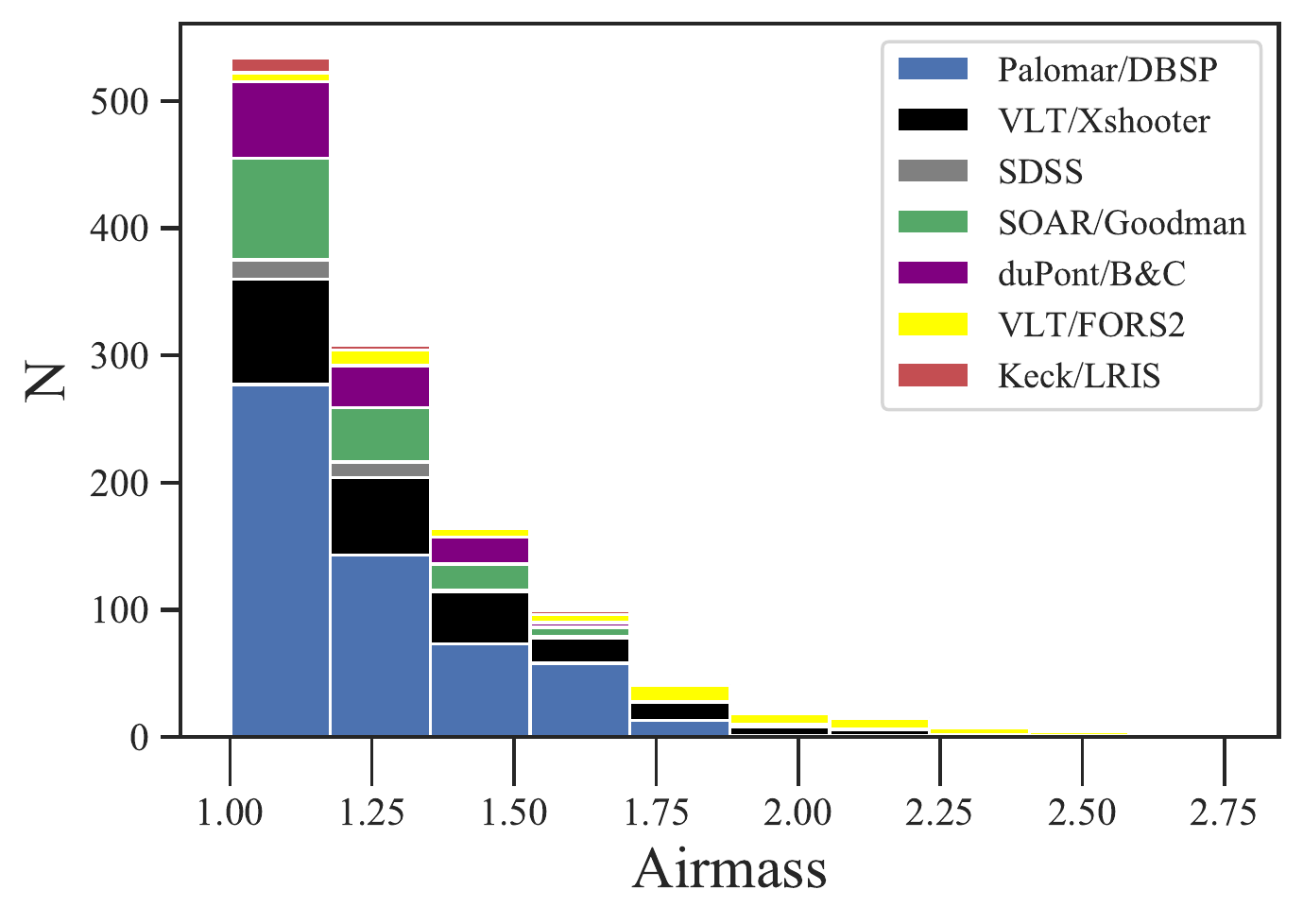}
\caption{Summary of the different observing conditions for DR2 spectra for telescopes with more than 15 spectra. Observing dates are shown for the newly observed sample (top left) and archival sample (top right).  The spectral resolution at 5000\,\AA\ (middle left) and 8500\,\AA\ (middle right) for the different observations.  Finally, the seeing and average air mass of the observations (bottom row).     }
\label{fig:spectra_sum}
\end{figure*}

\begin{figure*} 
\centering
\includegraphics[width=8cm]{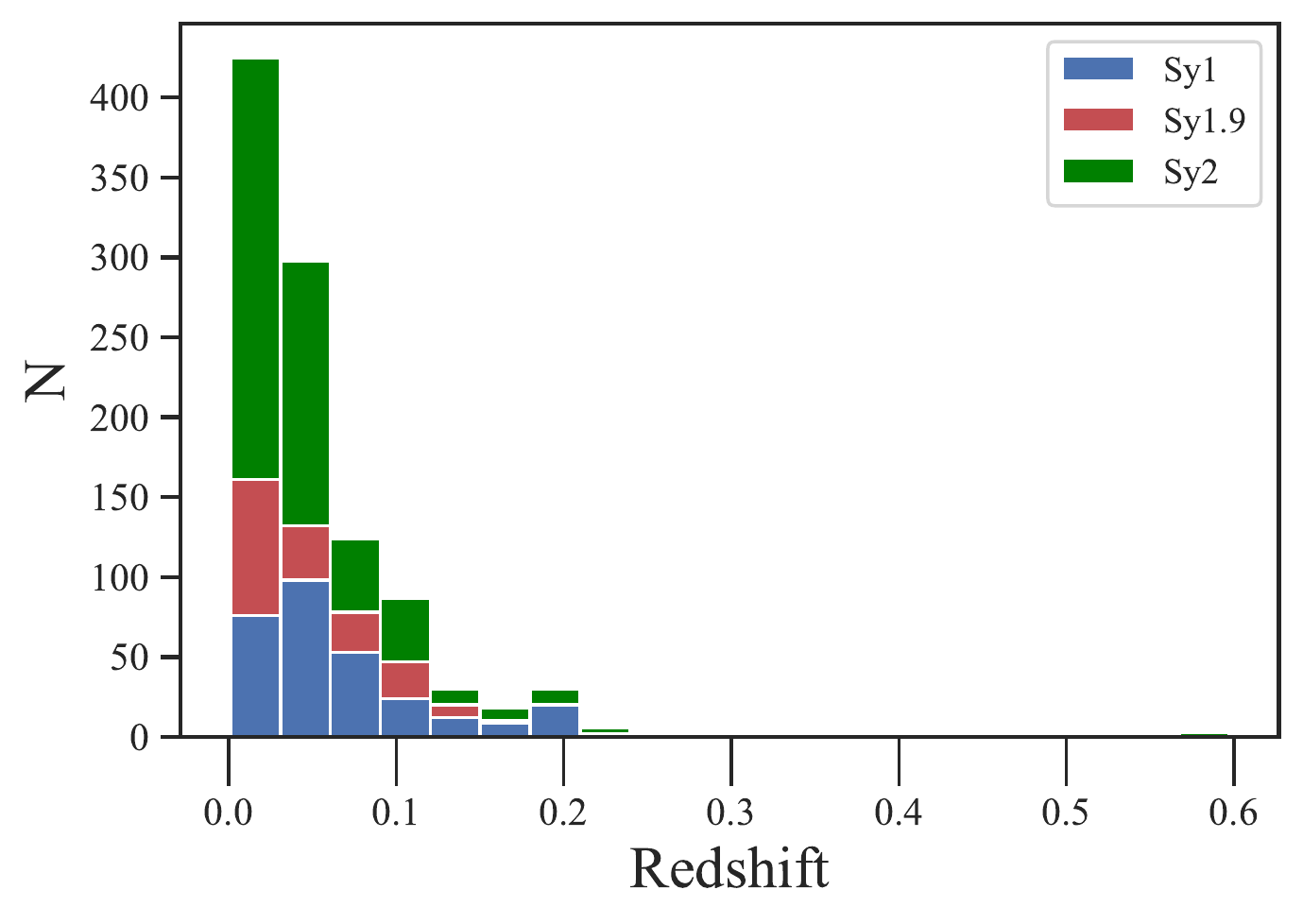}
\includegraphics[width=8cm]{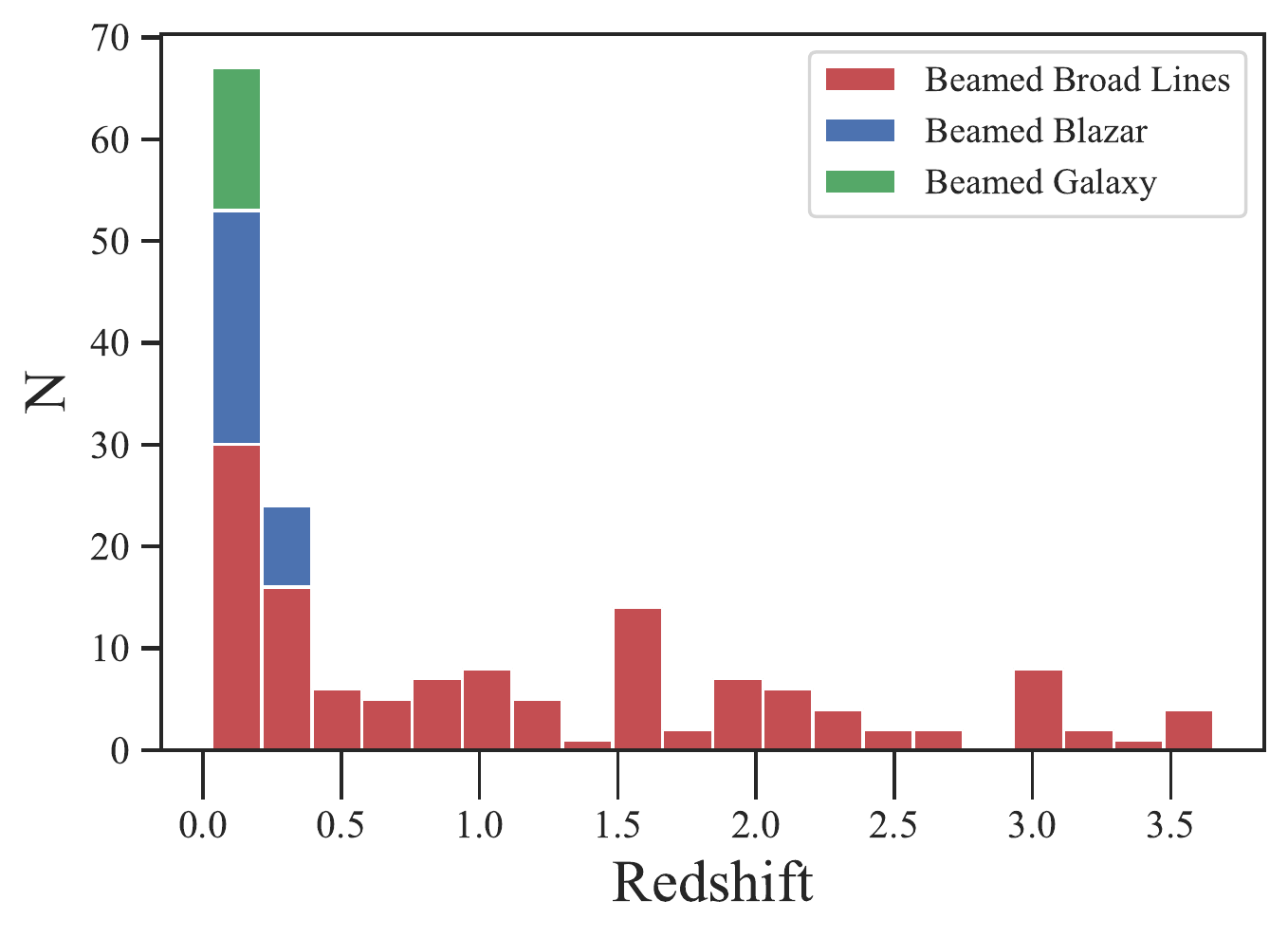}
\includegraphics[width=8cm]{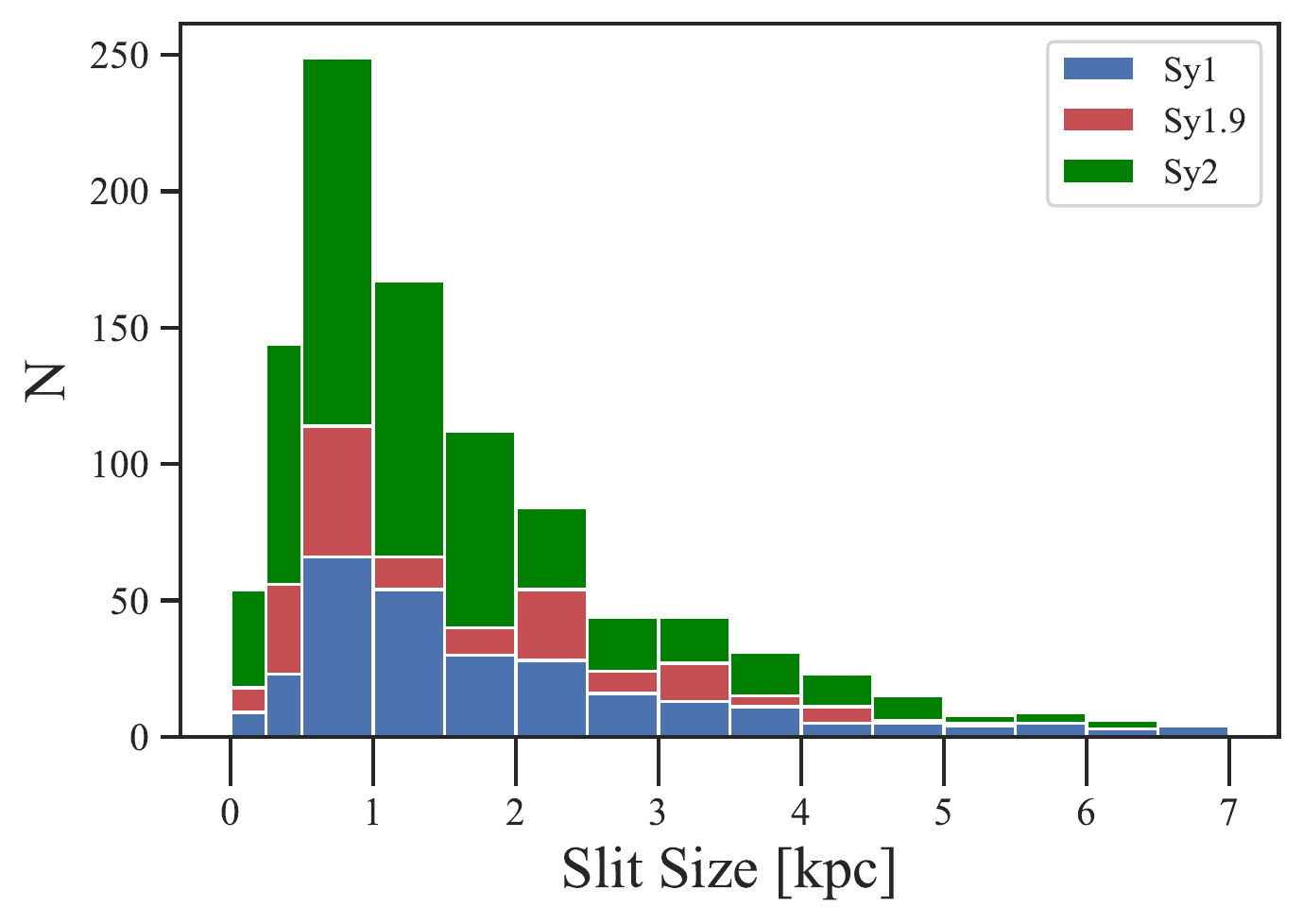}
\caption{Redshift range of all the AGN in the sample, split into unbeamed (upper left) and beamed AGN (upper right).  Additionally, the lower panel shows the slit width size in kpc for all the unbeamed AGN.   }
\label{fig:spectraz}
\end{figure*}

Here we provide a list of the telescopes used and their respective data reductions.  A plot of the number of spectra from each telescope is provided (\autoref{fig:spectra_num})  as well as summary plots of typical observing conditions and resolution (\autoref{fig:spectra_sum}).  The redshift range of observations can be found in \autoref{fig:spectraz}.

\subsubsection{Palomar Telescope}
The largest sample of targeted sources, \NPalomarAGN, was observed with the Palomar Double Spectrograph (DBSP) on the 200-inch Hale telescope.  These AGN were observed as part of a dedicated program on BAT AGN (P.I.\ M. Urry or M.\ Powell) or as part of the NuSTAR program (P.I. F.\ Harrison and D.\ Stern), where they served as backup targets to faint NuSTAR serendipitous sources. The observations were performed between 2012 October and 2020 November.  The majority of observations were taken with the D55 dichroic, with the 600/4000 and 316/7500 gratings using a 1\farcs{5} slit, This setup provided resolutions of $\sim$4.4\,\AA\ and $\sim$5.8\,\AA\ FWHM, over the $\sim$3150--5650\,\AA\ (blue) and $\sim$5400--10500\,\AA\ (red) region, respectively, providing full spectroscopic coverage of the optical region.  The size of the aperture used for extraction along the slit depended on the specific observing run, but was either fixed at 1.5-2\arcsec or matched to the extended profile of the source in the \iraf\ {\tt APALL} task.  We note that there are some flux calibration issues at some of the grating edges due to loss of sensitivity specifically at 5400--5650\,\AA\ on the blue side and 5400--5600\,\AA\ on the red side and at 10000--10500\,\AA\ (see examples in \autoref{spec_example}).  In some cases, the region between 5400 and 5650\,\AA\ was not extracted. Also, we did not force the spectra to be flux-calibrated in the overlapping regions, and differences of 10-20\% in flux may occur between the blue and red sides.

A smaller set of \NPalomarhigh\ narrow-line AGN and Sy1.9 AGN were also observed using the higher-resolution 1200 lines/mm grating for higher spectral resolution velocity dispersion measurements of the 3960--5500\,\AA\ and 8150--9600\,\AA\ regions, respectively.

\subsubsection{Very Large Telescope}

A total of \NXSHOOTERAGN\ were observed with X-shooter, a multiwavelength  (3000--25\,000\,\AA) echelle spectrograph with medium spectral resolution $R = 4000-18\,000$  \citep{Vernet:2011:A105}. Two dichroics are used to split the incoming light into the three arms for efficient observation of all three arms simultaneously.  The UVB arm (3000--5595\,\AA), VIS arm (5595--10240\,\AA) and NIR arm were used (10240--24800\,\AA). In some cases the NIR range only extended to 21010\,\AA\ rather than 24800\,\AA.  More information on the NIR reductions and scientific results can be found in \citet{denBrok_DR2_NIR}.


The majority of observations were observed with 1\farcs{6}, 1\farcs{5} in the UVB and VIS arms, respectively, in the NODDING mode and extracted with a 4\arcsec\ aperture along the slit.  The focus of the program was on obscured AGN (e.g. Sy1.9 and Sy2) to use the broad wavelength coverage and high spectral resolution to measure BH masses from velocity dispersion. These observations were performed as part of a filler program sometimes during bad weather conditions; however, the median seeing was still 1\farcs{02} owing to the inherently good conditions at the VLT.  In the case of VLT/X-shooter the spectra were first reduced using the standard pipeline in the ESO reflex software \citep[v2.3.0;][]{Freudling:2013:A96}.

Another \NllamaTAGN\ sources were observed as part of the LLAMA sample \citep{Davies:2015:127} of low-redshift, luminous BAT AGN ($z{<}0.01$).  These sources were observed in the IFU-offset mode with a field of view (FOV) of $1\farcs8 {\times} 4\arcsec$ with resolution $R\sim$8400 and $R\sim$13200 in the UVB and VIS arms, respectively.  The spectra were reduced using the ESO X-shooter pipeline v2.6.0.  The spectra were corrected for telluric absorption using telluric standard stars.  A more detailed description of the VLT/X-shooter data processing is given in \citet{2021A&A...654A.132B}.

Finally, three spectra were part of the Science Verification data from VLT/X-shooter in 2009 and lack header information necessary to be processed in the standard way using ESO/Reflex.  Two of these spectra, obtained in the SLIT, NODDING mode, were processed using optimal extraction and telluric standard stars following the reductions procedure of \citet{Becker:2019:163}.  The UVB arm was binned to 15\,\kmps\ pixels.  VIS and NIR arms were binned to 10\,\kmps.  A final IFU spectrum of NGC 7319 was extracted with a 4\arcsec\ by 2\arcsec\ region following the LLAMA sample.

There were also \NFORS\ observations with VLT/FORS2 done in a single observing semester in 2017 (099.A-0403A) and were focused on Sy1 or Sy1.9 AGN. For FORS2, the majority were observed with the 600B grating, with 1\arcsec\ slit, covering 3400--6100\,\AA.  A smaller subset of higher-redshift sources ($z{>}0.8$) was done with the 300I grating covering 6100--11\,000\,\AA.  All sources were reduced with v5.3.32 of the pipeline.  Optimal extraction was used with typical extractions along the slit of 9\arcsec.

\subsubsection{Southern Astrophysical Research Telescope}
We observed \NSOAR\ sources at the Southern Astrophysical Research (SOAR) telescope using the Goodman instrument as part of six programs between 2017 and 2020 (P.I. C.\ Ricci).    Observations were acquired in two lower-resolution setups focused on unobscured AGN with the 400 lines mm$^{-1}$ grating and GG455 blocking filter or 600 lines mm$^{-1}$ and GG385 blocking filter.  We performed higher resolution observations of obscured sources focused on the calcium triplet (CaT: 8498, 8542, and 8662\,\AA) using the 930 lines mm$^{-1}$ or 1200 lines mm$^{-1}$ grating.  Nearly all observations were done with a 1\farcs{2} slit with just two sources done with a 0\farcs{45} slit, because of the very low velocity dispersions.  All sources were extracted optimally, with typically slit lengths of 4\farcs{8}.

\subsubsection{du Pont Telescope}
Over 11 nights in 2016, \NduPont\ AGN were observed with the du Pont telescope with the Bollens \& Chivens spectrograph (P.I. C.\ Ricci).  All sources were observed with a 1\arcsec slit, the 300 lines mm$^{-1}$ grating covering 3000--9070\,\AA.  The sources were extracted with an optimal extraction with typical lengths along the slit of 6\farcs{6}.  The sources were typically unobscured AGN, due to the relatively low resolution (FWHM$\sim$8.7\,\AA).

\subsubsection{Keck Telescope}
Some AGN were also observed with the Keck telescopes associated with observations of NuSTAR-observed AGN and mergers \citep[e.g.,][]{Koss:2016:L4}.  A total of \Nkecklris\ observations were carried out using the Low Resolution Imaging Spectrometer  \citep[LRIS;][]{Oke:1995:375} on the Keck I Telescope. The setup used the blue (600 lines $^{-1}$) grism and the red (400 lines mm$^{-1}$ grating, with the D560 dichroic.  The majority of observations were done with a 1\arcsec\ slit with a handful done with a 1\farcs5 slit. 


\subsubsection{Magellan Telescope}
We performed \NMage\ observations with the Magellan Echellete (MagE) spectrograph \citep{Marshall:2008:701454} on the Magellan Clay telescope.  We used a 1\arcsec\ slit for observations with the slit angle set to parallactic.  The data were processed with the MagE pipeline, which is part of the Carnegie Python Distribution ($\mathtt{CarPy}$, v. 1.4.2).  The wavelength coverage was $\sim$3300--10000\,\AA\ though there was significant detector fringing above 8280\,\AA, resulting in strong instrumental issues above this range.  Typical exposure time was 1 hr and targeted higher-redshift type Sy1.9 and Sy2 AGN ($z{>}0.08$) for velocity dispersion measurements.

\subsubsection{SDSS and Archival Data}
We also included additional spectra from archival sources.  We used spectra from the SDSS \cite[][]{York:2000:1579}, with \NSDSSAGN\ sources from data release 16 \citep[DR16;][]{Ahumada:2020:3}, which were observed in the legacy survey with 3\arcsec\ fiber at $\sim$3800--9200\,\AA\ coverage or 2\arcsec\ fiber from BOSS or eBOSS with $\sim$3600--10200\,\AA\ coverage, respectively.

We also include \Narchival\ additional archival spectra of AGN that were acquired after the DR1.  These include AGN observed with earlier surveys of {\em ROSAT} AGN that overlap with BASS of unpublished and published \citep{Grupe:2004:156} sources, from the Palermo surveys of Swift BAT AGN \citep[][]{Rojas:2017:A124}, or as part of an atlas of low-redshift AGN \citep{Ho:2009:398}. 

Finally, we include \Nvltmuse\ archival spectra from VLT/MUSE when our existing BASS spectra were insufficient.  This includes
the nearby Circinus galaxy and NGC 3393, which was too bright to observe using our standard setups and was not part of the DR1 archival sample.  In addition, the dual AGN NGC 6240N and NGC 6240S have a VLT/MUSE spectrum for each AGN obtained in adaptive optics (AO) mode owing to their close separation.    Finally, the relatively distant Sy 2 sources, BAT ID 1204 (z=0.6) and ID 209 (at z=0.09) were included in order to enable velocity dispersion measurements.  We use the processed data from the ESO Science archive. We extract a 2\arcsec-radius aperture at the WISE position, except for NGC 6240N and NGC6240S, for which we use an 1\arcsec\ aperture owing to their closeness.

\subsection{Telluric Absorption Correction}
In DR2 we have implemented the software \texttt{molecfit} \citep{Smette:2015:A77} to correct telluric absorption regions that affect the measurements of emission and absorption lines.  This includes the oxygen bands (the A, B and the weaker $\Gamma$ bands at $\sim7590 < \lambda/$\AA$ < 7720$, $\sim6860 < \lambda/$\AA$ < 6950$, and $\sim6280 < \lambda/\AA < 6340$, respectively) as well as water vapor bands ($\sim8100 < \lambda/$\AA$ < 8300$, $\sim8930 < \lambda/$\AA$ < 9800$).  The performance has been described already in detail in  \citet{Kausch:2015:A78} and \citet{Smette:2015:A77} for ESO instruments.

\begin{figure*}
\centering
\includegraphics[width=8cm]{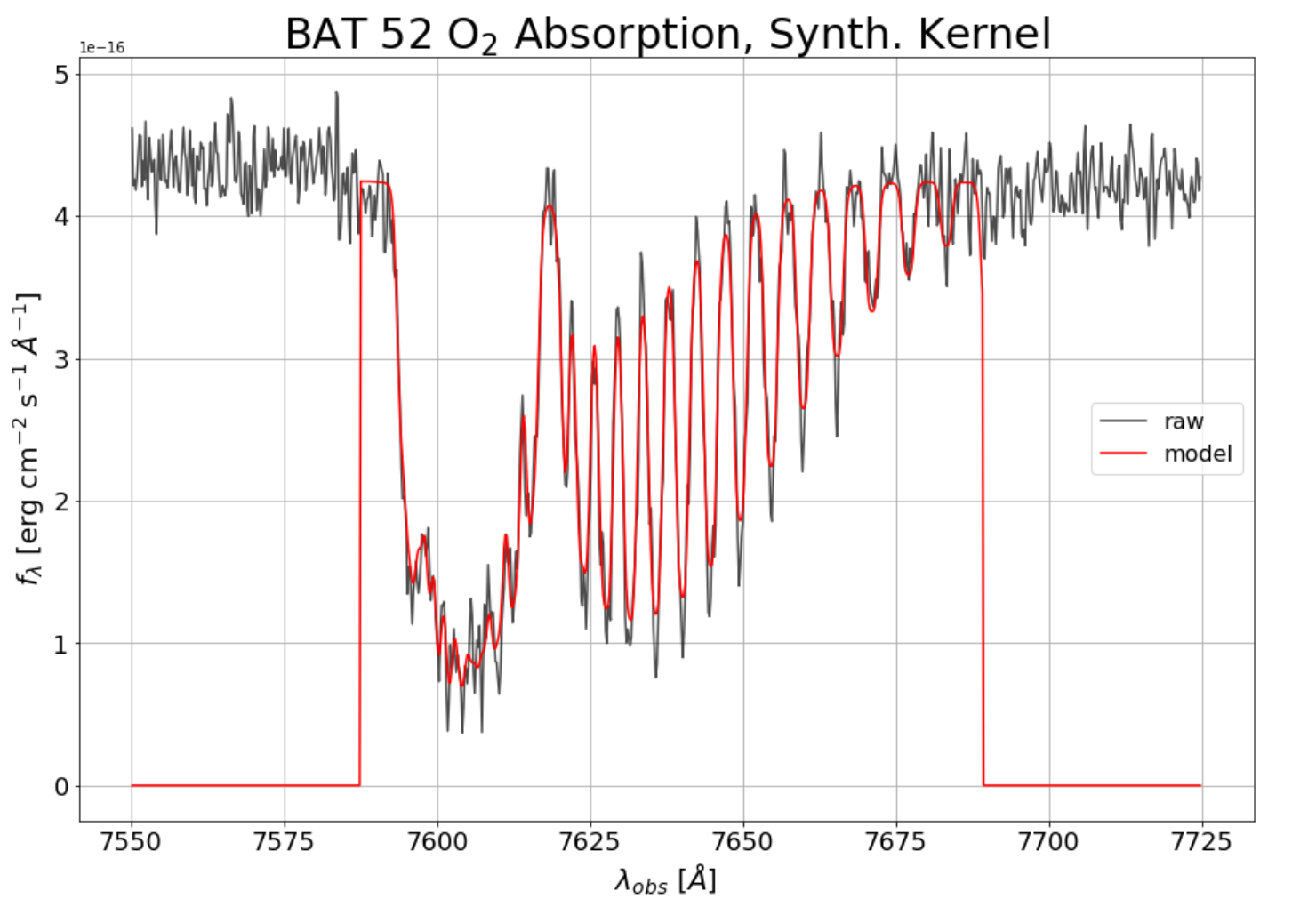}
\caption{Example absorption-line fit of the model spectrum to the observed spectrum using synthetic kernels of Gaussian kernels in the O$_2$ A-band region ($\sim$7580-7680~\AA). }
\label{fig:molecfit_tell}
\end{figure*}

In \texttt{molecfit}, model spectra are fitted to the observed spectra to derive the best-fit atmospheric parameters by iteratively computing transmission curves using a simple Earth atmosphere structure at the time of observation.  Global weather data are combined with local weather data to provide a likely best fit (humidity, pressure, temperature).  We use four regions with strong atmospheric features (O$_2$ at $\sim6800 < \lambda/$\AA$ < 6900$, O$_2$ at $\sim7460 < \lambda/$\AA$ < 7560$, telluric regions at $\sim9100 < \lambda/$\AA$ < 9200$, and telluric regions at $\sim9400 < \lambda/$\AA$ < 9500$) to determine the best-fit atmospheric parameters.  An example is provided in Figure \autoref{fig:molecfit_tell}.   We mask regions with strong AGN emission-line features from the fitting.  We use a Gaussian kernel variable with wavelength (\texttt{varkern}=1).

\begin{figure*}
\centering
\includegraphics[width=16cm]{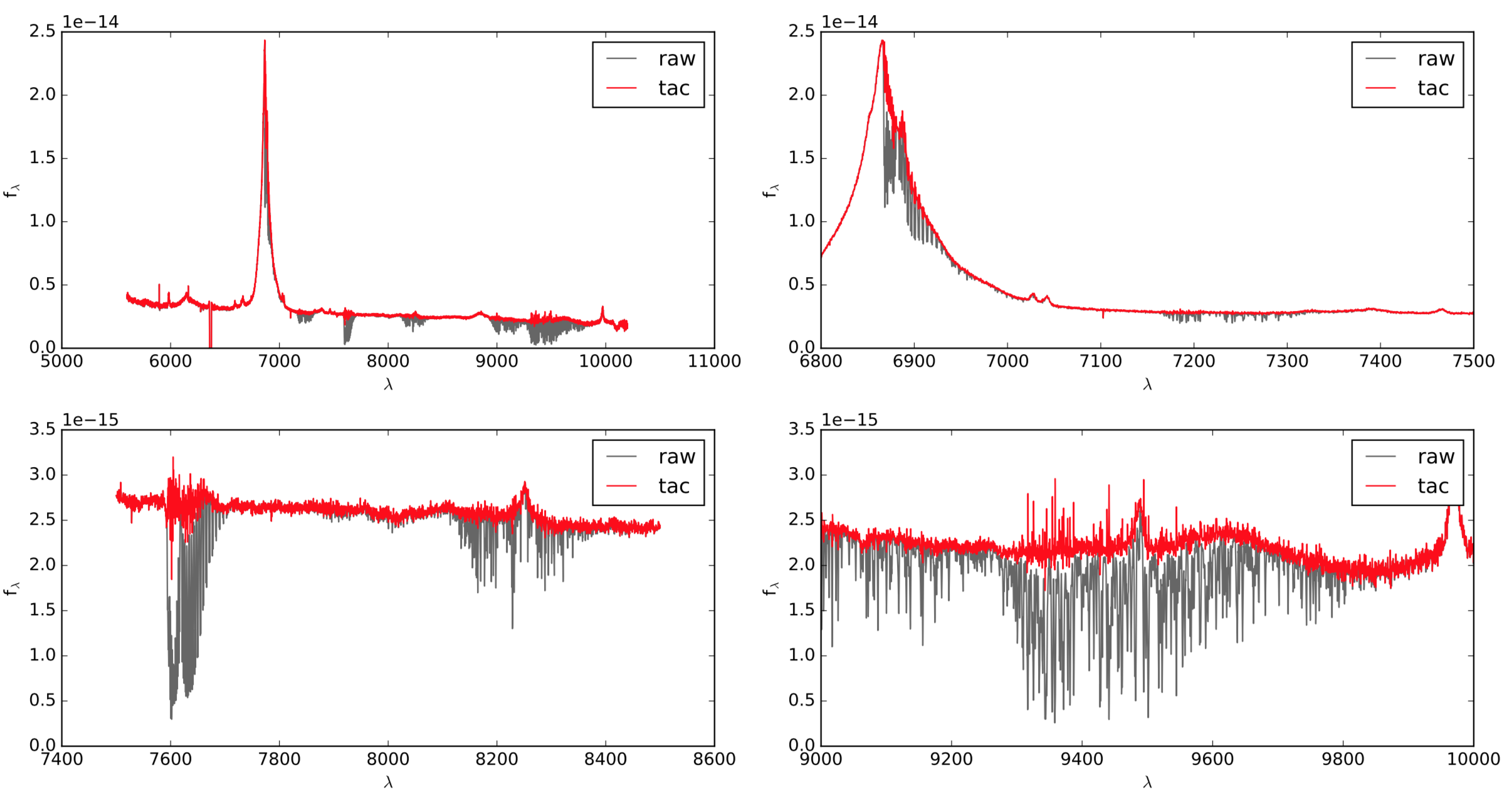}
\includegraphics[width=16cm]{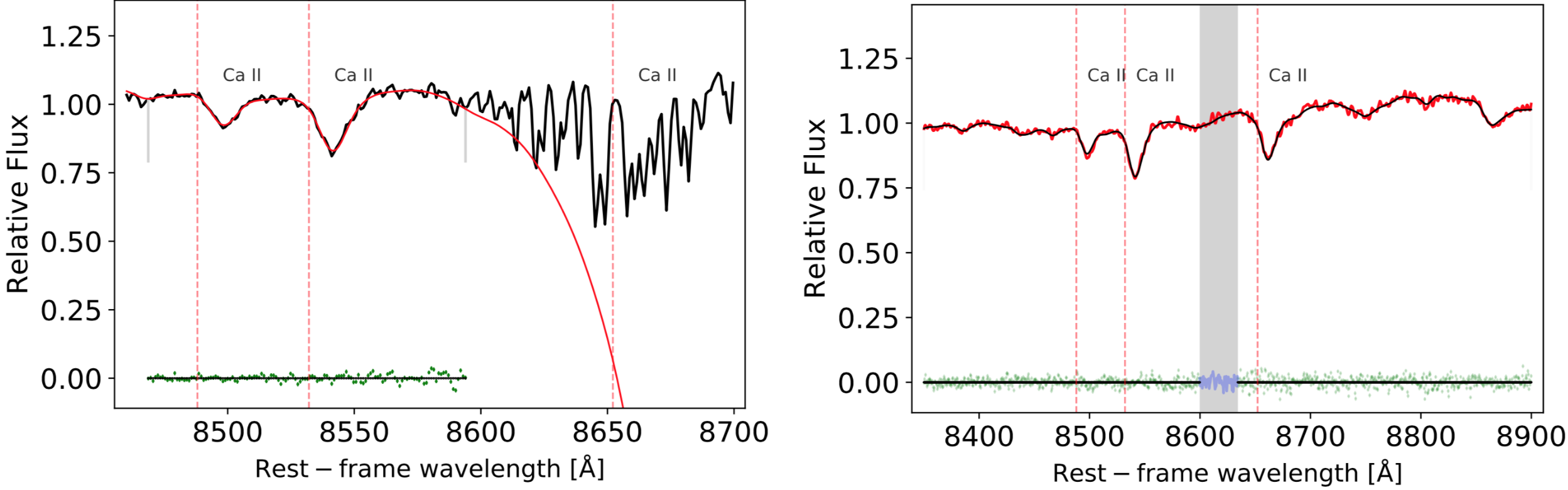}
\caption{Figure showing \molecfit\ correction on a spectrum within the BASS DR2.  Top left: VLT/X-shooter spectra of a broad-line BAT AGN Fairall 9 in the visual arm (5500-10500~\AA).  Top right panel and middle row show zoom-in regions around absorption features.  The red lines are the telluric-absorption-corrected (TAC) spectra, and gray lines are the input raw spectra.  The telluric absorption features cover broad \halpha as well as the prominent narrow emission lines at \HeIIir\ and \SIII\ emission lines.  \molecfit\ recovers these emission lines as shown with the TAC spectra, which are particularly important for accurately measuring \halpha, \HeIIir, and \SIII.  The bottom row shows the CaT region before (left) and after (right) \molecfit\ correction when being fit with a galaxy template for a velocity dispersion measurement \citep{Koss_DR2_sigs}.  The model fit is shown in red, with residuals from the fit shown in green dots below the spectra.  The TAC spectrum is able to recover a larger region of the \caii\ triplet spectral region (8450--8700\,\AA) which is redshifted into telluric features at $z{>}0.04$ with no significant increase in residuals.}
\label{fig:molecfit_example}
\end{figure*}

Telluric corrections with \molecfit\ were applied to all spectra with coverage above 7500\,\AA\ except archival observations (which lacked local weather data), Magellan/MAGE observations (because of significant fringing), and SDSS spectra (which have already had this correction applied).  An example of the telluric correction and its importance for narrow- and broad-line measurements is provided in Figure \autoref{fig:molecfit_example}.  We note that the ability to recover the intrinsic spectra is dependent the ability to measure foreground atmospheric absorption lines in the source, and thus very faint intrinsic spectra, such as those with very high Galactic extinction have little or no correction.  While most spectroscopic regions were adequately corrected, the
9300-9700\AA\ region suffers very high extinction, and emission-line fitting should be approached with caution. 

\subsubsection{Overlap with the DR1}
Initial targeting priority focused on sources without BH mass measurements or spectra from the archival data in the DR1, but this was later expanded to complete the whole sample.   The DR2 includes all of the SDSS spectra in the DR1 as well as 35 early Palomar spectra that were reprocessed to include corrections with \molecfit.  Beyond this there are still \NDRoneonly\ AGN in the DR1, with no new DR2 spectra.

\subsection{Instrumental Resolution \label{sec:instres}}
We determine the instrumental resolution and line-spread function (LSF) FWHM for each spectral setup and provide the best estimate in \autoref{tab:setups}.  A description of how this quantity was measured is given in \autoref{sec:instres}.

For some sources we perform Gaussian fits of night-sky lines including Hg I (e.g. 4046\,\AA, 4358\,\AA, and 5461\,\AA) and OI (5577\,\AA, 6300\,\AA, and 6364\,\AA) for bluer spectral regions.  We used OH lines for redder regions (e.g. 6864\,\AA, 7341\,\AA, 7821\,\AA, and 9872\,\AA).  If this was not feasible owing to spectral coverage, we used arc line spectra. 

In real observations, the slit width is not the only factor in determining the instrumental resolution. If the slit is large, the image quality at the entrance slit can be smaller than the slit width. The observed spectral resolution is then better than what is measured from the slit width for arc or sky lines which fully fill the slit, but is instead determined by the sharpness of the image of the object at the entrance slit. Thus, the spectral resolution in real observations of AGN may be somewhat smaller than when measured using sky lines or arc line spectra that fully fill the slit.  

The measurements from \molecfit, which trace the absorption of telluric lines, can provide an additional estimate of the effective resolution.  We note, however, that the ability to fit the absorption profile is difficult in faint sources and only done for setups that include coverage of telluric features (e.g., ${>}$9000\,\AA).  In several setups without telluric features, only a few AGN were observed, and we use sky or arc lines for resolution measurements.

An independent estimate of the spectral resolution and the line spread was also done with the penalized PiXel-Fitting method \citep[\texttt{pPXF},][]{Cappellari:2004:138, Cappellari:2017:798} by fitting stellar absorption lines to individual Galactic stars that were observed during the observations.  We use two template libraries depending on the resolution.  We use the X-shooter data release 2 library, which was obtained at much higher resolutions than typical observations (e.g. $R\sim10000$).  For spectra obtained at higher resolutions ($R>$3000) based on sky lines, we use the $\mathtt{PHOENIX}$ theoretical spectral library \citep{Husser:2013:A6} as a template which has much higher resolutions ($R\sim500,000$).  We fit the 3880--5500\,\AA\ region and 8350--8730\,\AA\ to determine the LSF in the blue and red ranges, respectively, to target stellar absorption features.

 We find that the spectral resolution as measured from individual Galactic stars during real observations or telluric absorption lines in the AGN galaxies tends to indicate spectral resolutions somewhat sharper ($\sim20\%$) than those of sky lines.  For instance, for the largest sample observed using the 1.5\arcsec\ slit with Palomar/Doublespec, the average of stars observed on different nights is $4.7\pm0.3$\,\AA\ in the 
  CaT, compared to the average of the telluric absorption with \molecfit\ which is $4.9\pm0.6$\,\AA.  The average from fitting sky lines for these observations is 5.8$\pm$0.3\,\AA.  In this case, we use use the \molecfit\ average, though it is statistically similar to that from stars.  For the 3880--5500\,\AA\ region, \molecfit\ was not used because of a lack of telluric features, but the \ppxf\ fits were $4.1\pm0.2$\,\AA\ vs. $4.8\pm0.4$\,\AA\ for the sky lines within each observation, and the average of the \ppxf\ fits was used.  

\section{Survey Measurements} \label{sec:surveytable}
The spectroscopic release provides spectra for \Ndrtwospecper\ (\Ndrtwospec/\NAGN) of DR2 AGN.  When combined with the DR1 spectra, all AGN have spectra (e.g. 100\%, \NAGN/\NAGN), representing a complete census of luminous hard-X-ray-selected AGN over nearly the entire sky outside of a small region on the Galactic plane (94.8\%,$\lvert b\rvert{>}3^{\circ}$).   Here we describe the main survey measurements (e.g. $z$, AGN type, \mbh).  Further derived measurements (broad-line widths, narrow-line widths, velocity dispersions) will be provided in subsequent papers.  

For all the AGN in our sample, we provide the following key parameters when possible for each AGN \autoref{tab:agnprop}:
\begin{enumerate}
\item \texttt{BAT ID:} Catalog ID in the BAT survey.\footnote{https://swift.gsfc.nasa.gov/results/bs70mon/}

\item R.A. \texttt{$\alpha_{\rm J2000}$} and decl. \texttt{$\delta_{\rm J2000}$}: Right ascension and decl. of the optical/IR counterpart of the BAT AGN, in decimal degrees, based on \wise positions.

\item \texttt{DR2 Type}:  AGN type based on optical spectroscopy. Sy1 (with broad \hbeta), Sy1.9 (narrow \hbeta\ and broad \halpha), and Sy 2 (with narrow \hbeta\ and \halpha).  For beamed AGN, the types include those with the presence of broad lines (BZQ), only host galaxy features lacking broad lines (BZG), or traditional continuum-dominated blazars with no emission lines or host galaxy features (BZB).  BZU refers to a beamed AGN where the type is uncertain because of a lack of optical spectroscopy or very low S/N.

\item \texttt{$z$ and $z_\textrm{type}$}: Best DR2 redshift measurement and the line or method used for the measurement.  The majority of fits are done with \OIII\ (88\%, 755/\NAGN). Measurements are from a broad-line fitting code \citep{Mejia_Broadlines} referred to as OIII broad, MgII broad, and CIV broad, respectively, when available for all Sy1 and BZQ sources with broad-line \hbeta.  For narrow-line sources, the redshift is based on emission-line fitting of \OIII, when possible.     For some high Galactic extinction sources, single emission-line fits to other lines are used (in the table as Ha for \Halpha, SIII for \SIII, and HeI for \HeIir). For some high-redshift sources $z{>}1$ without high quality broad-line fitting we report the estimates from single emission-line fits of \CIV\ and \MgII\ (referred to as CIV and MgII). Host galaxy templates (referred to as Gal Temp) are used for some BZB with no emission lines.  For the remaining sources, DR1 fits or those from SIMBAD are used for redshift estimates.    

\item \texttt{Dist} and \texttt{z$_\textrm{ind}$}: Distance assumed based on redshift or redshift independent distance measurements in Mpc.  


\item \texttt{Best MBH} and  \texttt{MBH$_\textrm{Meth}$}.  The best BH mass measurement and the method used for the measurement.  We do not report errors from either broad-line fitting or velocity dispersions as the they are less than 0.1 dex and the errors are dominated by the intrinsic spread of virial and $\sigma_\star$ based BH mass estimates of order 0.5 dex \citep{Ricci_DR2_NIR_Mbh}.

\item \texttt{\Lbol\ and \lledd}:  Measurement of the AGN bolometric luminosity and Eddington ratio based on BH mass.  

\end{enumerate}

\begin{deluxetable}{llrrlrrrrlrr}
\tabletypesize{\scriptsize}
\tablewidth{0pt}
\tablecaption{General AGN Properties \label{tab:agnprop}}
\tablehead{
\colhead{BAT ID} &                   \colhead{Counterpart} &          \colhead{R.A.} &        \colhead{Decl.} &       \colhead{DR2 Type} &     \colhead{$z$} &  \colhead{$z_{\mathrm{type}}$} &     \colhead{Dist} &  \colhead{$M_{\mathrm{BH}}$} &       \colhead{BH meth.} &    \colhead{$\log$ \Lbol} &  \colhead{\lledd}\\
\colhead{} &                   \colhead{} &          \colhead{(\deg)} &        \colhead{(\deg)} &       \colhead{} &     \colhead{} &  \colhead{} &     \colhead{(Mpc)} &      \colhead{(\Msun)} &       \colhead{} &    \colhead{($\log$(\ergps))} &  \colhead{}
}
\startdata
1 &        2MASXJ00004876-0709117 &    0.203234 &  -7.153223 &      Sy1.9 &  0.037496 &        OIII &    165.2 &      7.61 &     Vdisp &   44.45 &      -1.33 \\
     2 &        2MASXJ00014596-7657144 &    0.442035 & -76.953963 &        Sy1 &  0.058505 &  OIII Broad &    261.7 &      7.41 &        Hb &   44.77 &      -0.82 \\
     3 &                       NGC7811 &    0.610105 &   3.351912 &        Sy1 &  0.025457 &  OIII Broad &    111.2 &      6.70 &        Hb &   44.01 &      -0.86 \\
     4 &        2MASXJ00032742+2739173 &    0.864248 &  27.654725 &        Sy2 &  0.039784 &        OIII &    175.6 &      7.89 &     Vdisp &   44.52 &      -1.55 \\
     5 &        2MASXJ00040192+7019185 &    1.008241 &  70.321752 &        Sy2 &  0.095681 &        OIII &    439.1 &      6.20 &    Ha obs &   45.29 &       0.92 \\
     6 &                        Mrk335 &    1.581400 &  20.202951 &        Sy1 &  0.025906 &  OIII Broad &    113.2 &      7.23 &       Lit &   44.24 &      -1.17 \\
     7 &       SDSSJ000911.57-003654.7 &    2.298335 &  -0.615222 &        Sy2 &  0.073345 &        OIII &    331.5 &      8.54 &     Vdisp &   45.05 &      -1.67 \\
     8 &                       Mrk1501 &    2.629175 &  10.974862 &        BZQ &  0.089385 &  OIII Broad &    408.5 &      8.07 &       Lit &   45.63 &      -0.61 \\
     9 &                [HB89]0014+813 &    4.285420 &  81.585596 &        BZQ &  3.377817 &   CIV Broad &  29310.9 &      9.90 &       CIV &   48.90 &       0.82 \\
    10 &                      LEDA1348 &    5.281408 & -19.168191 &      Sy1.9 &  0.095846 &        OIII &    439.9 &      8.94 &     Vdisp &   45.41 &      -1.70 \\
    13 &                    LEDA136991 &    6.385030 &  68.362439 &        Sy2 &  0.012492 &        OIII &     54.0 &      7.68 &     Vdisp &   43.94 &      -1.92 \\
    14 &                    LEDA433346 &    6.669470 & -53.163275 &        Sy1 &  0.063219 &  OIII Broad &    283.7 &      8.44 &        Hb &   44.99 &      -1.63 \\
    16 &                    PG0026+129 &    7.307096 &  13.267761 &        Sy1 &  0.141997 &  OIII Broad &    671.7 &      8.49 &       Lit &   45.68 &      -0.98 \\
    17 &                      ESO112-6 &    7.682626 & -59.007215 &        Sy2 &  0.029004 &        OIII &    127.0 &      7.90 &     Vdisp &   44.44 &      -1.64 \\
    18 &        2MASXJ00331831+6127433 &    8.326442 &  61.462015 &      Sy1.9 &  0.104184 &        OIII &    480.9 &      7.95 &        Ha &   45.42 &      -0.71 \\
    19 &                          RHS3 &    8.570040 & -79.088963 &        Sy1 &  0.074268 &  OIII Broad &    335.9 &      8.02 &        Hb &   44.79 &      -1.41 \\
    20 &        2MASXJ00343284-0424117 &    8.636619 &  -4.403423 &        Sy2 &  0.212982 &        OIII &   1051.6 &      9.22 &     Vdisp &   46.22 &      -1.18 \\
    22 &                       Z535-12 &    9.087263 &  45.664900 &        Sy1 &  0.047608 &  OIII Broad &    211.3 &      7.36 &        Hb &   44.75 &      -0.79 \\
    24 &                        Mrk344 &    9.633812 &  23.613408 &        Sy2 &  0.025246 &        OIII &    110.2 &      7.54 &     Vdisp &   44.29 &      -1.43 \\
    25 &        SWIFT J004039.9+244539 &   10.166178 &  24.760937 &      Sy1.9 &  0.078365 &        OIII &    355.4 &      7.73 &     Vdisp &   44.86 &      -1.05 \\
    28 &                       NGC235A &   10.720042 & -23.541046 &      Sy1.9 &  0.022065 &         DR1 &     96.1 &      8.49 &     Vdisp &   44.61 &      -2.06 \\
    30 &        2MASXJ00423991+3017515 &   10.666287 &  30.297621 &        BZQ &  0.140128 &  OIII Broad &    662.1 &      8.56 &        Hb &   45.41 &      -1.33 \\
\enddata
\tablecomments{\autoref{tab:agnprop} is published in its entirety in the machine-readable format for \NAGN\ AGN.  A portion is shown here for guidance regarding its form and content.  See \autoref{sec:spectable} for a description of each data column.}
\end{deluxetable}

\subsection{AGN Type}
For classification, we first split the sources into \Nunbeamed\ unbeamed AGN and \Nbeamed\ beamed AGN (and 1 lensed AGN) following the spectroscopic classification of the Roma Blazar Catalog catalog \citep[BZCAT;][]{Massaro:2009:691}.  The beamed AGN were split into three categories based on their optical spectral properties, specifically, based on the presence of broad lines (BZQ), only host galaxy features lacking broad lines (BZG), or traditional continuum-dominated blazars with no emission lines or host galaxy features (BZB).  


Overall 72/105 (69\%) of the DR1 beamed AGN maintained the same classification with further optical spectroscopic study.  20/105 (19\%) changed specific beamed AGN classification based on additional DR2 optical spectra (e.g., BZG to BZQ, based on the detection of broad \hbeta\ or any other broad line such as \halpha).  

We provide an unbeamed AGN type based on the presence of broad Balmer lines from visual inspection after fitting with host galaxy templates \citep{Koss_DR2_sigs}.  These include sources with broad \hbeta\ (Sy1), sources with narrow \hbeta, but broad \halpha\ (Sy1.9), and sources with only narrow optical lines (Sy2).  Further classification of broad-line AGN (e.g. Sy 1.2, 1.5, 1.8 etc.) is provided in subsequent studies \citep{Mejia_Broadlines} and also of narrow-line AGN such as LINERs \citep{Oh_DR2_NLR}.  When spectra are not available in DR2 for AGN type, we use the DR1 AGN type.

Overall, there are 168 AGN in DR2 for which we derive a revised or first classification of AGN type based on our measurements. A detailed comparison of DR2 measurements compared to the DR1 is provided in \autoref{dr1comp_appen}.

\subsection{Redshifts} \label{red_explain}
To determine the best redshifts for the sample \autoref{tab:zmeth}, we use the \OIII\ line emission as the primary measurement.  For broad line sources (e.g. Sy1 and BZQ), the \oiii\ redshift is based on the fitting procedure in \citet{Mejia_Broadlines} which is consistent with the procedure used in the DR1.  For more distant beamed AGN, when the \OIII\ line is redshifted out of the spectrum, the \MgII\ ($0.8{<}z{<}2.6$) or \CIV\ ($2.1{<}z{<}3.6$) line is used.  

For the Sy1.9 and Sy2 AGN, or any broad-line AGN where the fitting failed, we fit the \oiii\ emission line in our sample using  {\tt PySpecKit}, an extensive spectroscopic analysis toolkit for astronomy, which uses a Levenberg-Marquardt algorithm for spectral fitting \citep{Ginsburg:2011:1109.001}. We fit the \oiii\ emission line using a single Gaussian.  Finally, for 10 sources that are highly reddened and often in the Galactic plane where an \oiii\ line was not detected we use the \halpha\ or \SII\ or \HeIir\ line.  We note that for NGC6240N and NGC6240S, the sources were too close together to resolve in our spectra, so we provide a single measurement.

There are 30 beamed AGN that are strongly continuum-dominated blazars (BZB), with only weak stellar features, and host-galaxy-dominated blazars (BZG), where no \OIII\ emission lines are measured.  For 25 of these sources, we can measure redshifts using \ppxf\ with the Ca H+K lines or \caii\ absorption features.  Further details, including a full list of redshifts from stellar absorption features from galaxy template fits for Sy1.9 and Sy2 AGN can be found in \citet{Koss_DR2_sigs}. 

We used the BASS DR1 data for the  33 AGN in DR2, which did not have a new spectrum at all or were missing one that covered the \OIII\ line.

For the remaining sources we rely on NED or SIMBAD or past publications for redshift measurements.    Of the remaining six BZB sources we could not detect strong host galaxy features for a redshift, there are five sources with an existing redshift from NED or SIMBAD, which we use as the redshift measurement. There is one very high-extinction AGN ($A_{V}$=9.5) 2MASS J10445192-6025115, which was found to have a redshift of $z{=}0.047$ \citep{Fortin:2018:A150} based on \HeIir\ line.  The beamed and lensed AGN PKS 1830-21, was measured using  \halpha\ in the NIR at $z{=}2.507$ \citep{Lidman:1999:L57} due to it's high extinction ($A_V$) from being in the Galactic plane.

\begin{figure*} 
\centering
\includegraphics[width=16cm]{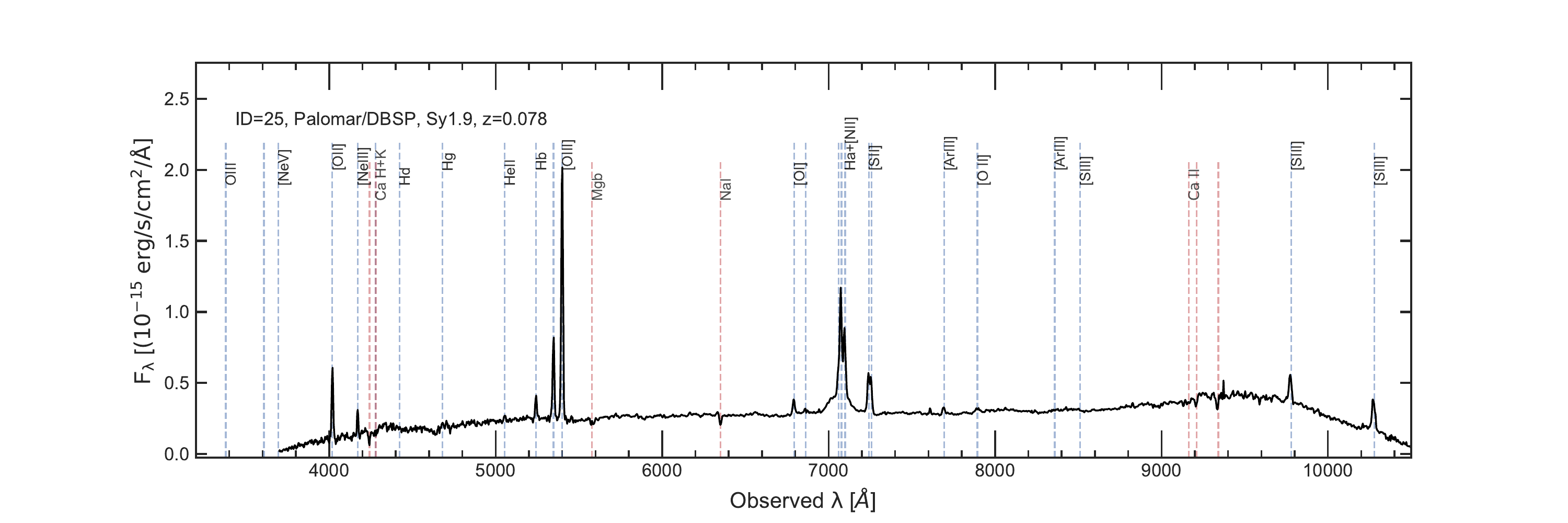}
\includegraphics[width=16cm]{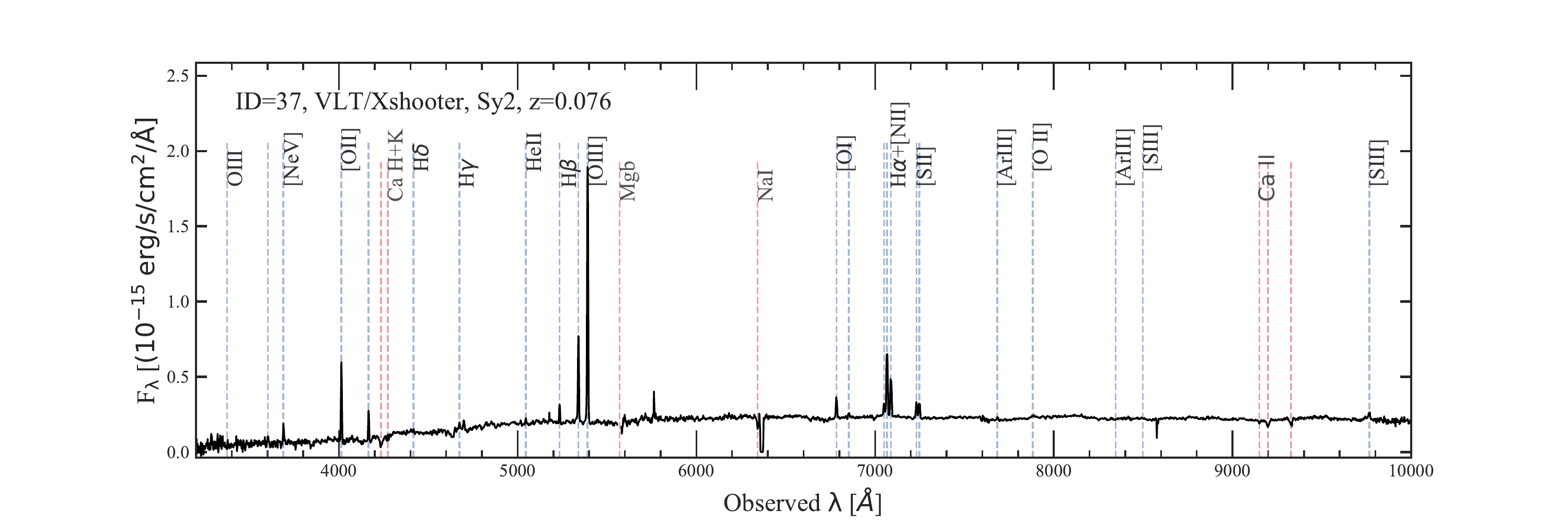}
\includegraphics[width=16cm]{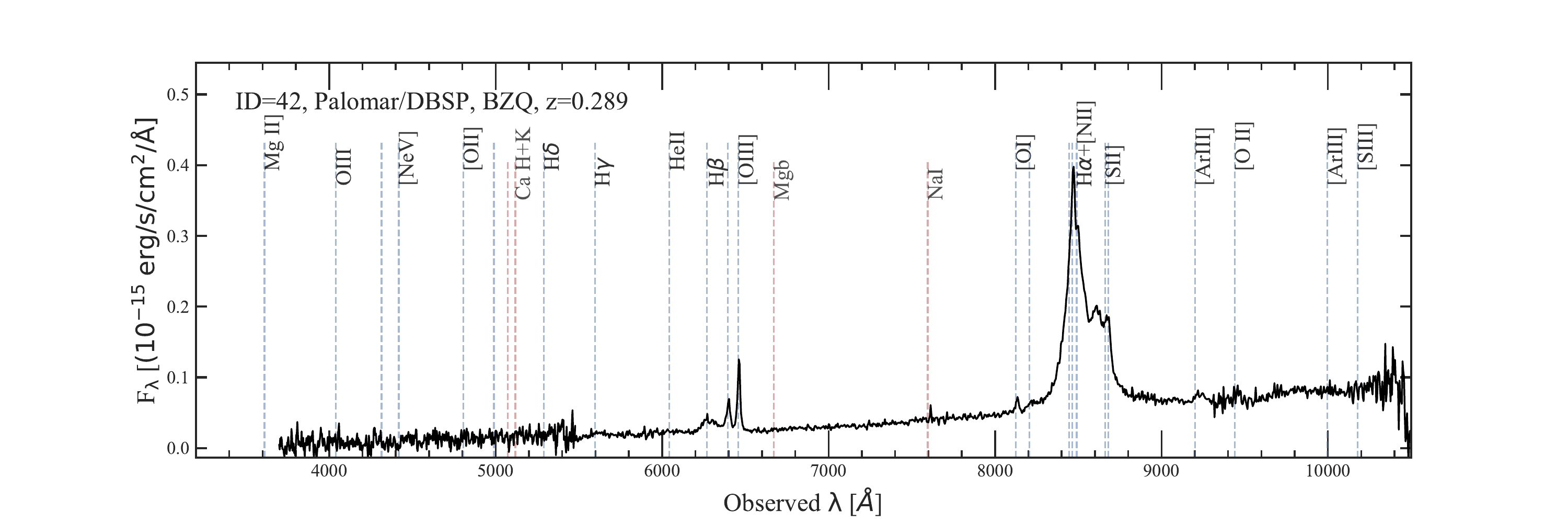}
\includegraphics[width=16cm]{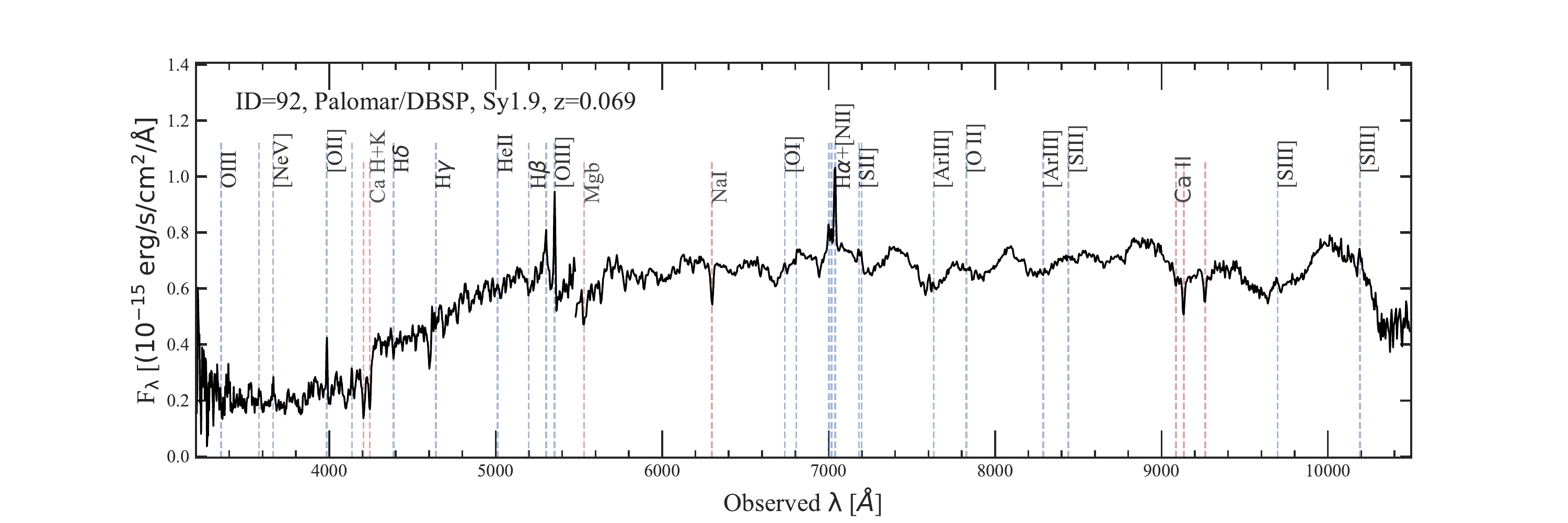}
\caption{Examples of \Nnewz\ sources with newly identified redshifts.}
\label{fig:example_newz}
\end{figure*}

Overall the redshift completion is extremely high, 99.8\% (857/\NAGN) for the full sample.  Of these AGN redshifts, \Nnewz\ (\autoref{tab:newz}) are found for the first time (\autoref{fig:example_newz}). The only AGN without a redshift is a continuum-dominated blazar (BZB). The blazar, B3 0133+388, was first discovered in the third Bologna sky survey of 408 MHz radio objects \citep{Ficarra:1985:255} and is also shows bright gamma-ray emission above 1 GeV in Fermi.  The source shows faint Ca H+K lines at redshift zero in two different Palomar spectra (and also in a Keck/LRIS spectra shown in \citet{Aliu:2012:102}).  However, given the radio and Fermi detection, the source is unlikely to be Galactic but maybe a blazar with a foreground star.

\begin{deluxetable}{ll}
\tablecaption{Summary of Redshift Measurements \label{tab:zmeth}}
\tablehead{
\colhead{Redshift Method } & \colhead{Count}}
\startdata
\OIII       &    393 \\
\OIII\ Broad &    364 \\
DR1        &     34 \\
Gal Temp   &     22 \\
\CIV\ Broad  &     13 \\
\MgII\ Broad &     10 \\
\Halpha         &      7 \\
SIMBAD     &      6 \\
\CIV        &      3 \\
\SIII       &      3 \\
\HeIir        &      1 \\
NED        &      1 \\
None       &      1 \\
\enddata
\tablecomments{The \oiii\ broad redshift indicates that a broad line fitting procedure for \hbeta\ was used as described in \citep{Mejia_Broadlines}.  Otherwise a single emission-line fit or a galaxy template fit was performed.  See \autoref{red_explain} for details.}
\end{deluxetable}

\begin{deluxetable}{cccccccc}
\tabletypesize{\footnotesize}
\tablewidth{0pt}
\tablecaption{Newly Identified Redshifts
\label{tab:newz}}
\tablehead{
\colhead{BAT Index}& \colhead{Swift Name}& \colhead{Counterpart}& \colhead{DR2 Type}& \colhead{Telescope}& \colhead{$z$}& \colhead{Line}& \colhead{$A_V$}}
\startdata
25&SWIFT J0041.0+2444&SWIFT J004039.9+244539&Sy1.9&Palomar/DBSP&0.078365&\OIII&0.1\\
42&SWIFT J0057.0+6405&NVSS J005712+635942&BZQ&Palomar/DBSP&0.289627&\oiii&4.5\\
65&SWIFT J0116.5-1235&2MASX J01163118-1236171&Sy1.9&VLT/X-shooter&0.142447&\oiii&0.1\\
92&SWIFT J0149.2+2153A&LEDA 1656658&Sy1.9&Palomar/DBSP&0.069397&\oiii&0.3\\
154&SWIFT J0252.1-6758&2MASX J02513173-6803059&Sy1.9&VLT/X-shooter&0.18263&\oiii&0.1\\
161&SWIFT J0259.9+4419&2MASX J02593756+4417180&Sy1&Palomar/DBSP&0.031291&\oiii&0.7\\
210&SWIFT J0413.3+1659&MG1J041325+1659&BZQ&Palomar/DBSP&0.211541&\oiii&2.1\\
240&SWIFT J0449.6-5515&2MASX J04500193-5512404&Sy2&VLT/X-shooter&0.021576&\oiii&0\\
250&SWIFT J0459.7+3502&LEDA168924&Sy2&Palomar/DBSP&0.044383&\oiii&2.8\\
257&SWIFT J0505.6-6735&2MASX J05052442-6734358&Sy2&VLT/X-shooter&0.046517&\oiii&0.8\\
323&SWIFT J0612.2-4645&PMN J0612-4647&BZQ&VLT/X-shooter&0.317767&\oiii&0.2\\
333&SWIFT J0626.6+0729&LEDA136513&Sy1&duPont/BC&0.042447&\oiii&2\\
343&SWIFT J0640.0-4737&SWIFT J064013.50-474132.9&Sy2&VLT/X-shooter&0.057242&\oiii&0.4\\
353&SWIFT J0659.3+2406&2MASX J06591070+2401400&Sy2&Keck/LRIS&0.090822&\oiii&0.2\\
359&SWIFT J0709.3-1527&PKS 0706-15&BZB&Palomar/DBSP&0.142277&Gal. Temp.&2\\
367&SWIFT J0723.8-0804&1RXS J072352.4-080623&Sy1&Palomar/DBSP&0.144926&\oiii&0.9\\
380&SWIFT J0741.4-5447&2MASX J07410919-5447461&Sy2&VLT/X-shooter&0.106098&\oiii&0.5\\
396&SWIFT J0755.4+8402&2MASS J07581638+8356362&Sy1&Palomar/DBSP&0.133952&\oiii&0.15\\
433&SWIFT J0854.3-0827&SWIFT J085429.35-082428.6&Sy2&VLT/X-shooter&0.188435&\oiii&0.1\\
487&SWIFT J1007.4+6534&1RXSJ100712.6+653511&Sy1&Palomar/DBSP&0.126589&\oiii&0.2\\
494&SWIFT J1020.5-0237A&SDSS J102103.08-023642.6&Sy2&VLT/X-shooter&0.293645&\oiii&0.13\\
510&SWIFT J1041.4-1740&2MASX J10410120-1734300&Sy2&VLT/X-shooter&0.080844&\oiii&0.2\\
660&SWIFT J1310.9-5553&IGR J13109-5552&BZQ&VLT/X-shooter&1.55906&\MgII&1.1\\
745&SWIFT J1449.5+8602&2MASX J14545815+8554589&Sy2&Palomar/DBSP&0.111951&\oiii&0.5\\
747&SWIFT J1451.0-5540B&LEDA 3085605&Sy2&VLT/X-shooter&0.018663&\Halpha&2.5\\
756&SWIFT J1508.6-4953&PMNJ1508-4953&BZQ&VLT/X-shooter&1.520201&\MgII&1.2\\
761&SWIFT J1512.2-1053A&NVSS J151148-105023&BZQ&Palomar/DBSP&0.94672&\oiii&0.34\\
762&SWIFT J1512.2-1053B&2MASX J15120505-1046356&Sy2&VLT/X-shooter&0.165799&\oiii&0.3\\
780&SWIFT J1548.1-6406&SWIFT J1548.1-6406&BZQ&VLT/X-shooter&1.693124&\oiii&0.6\\
792&SWIFT J1605.9-7250&LEDA 259433&Sy2&VLT/X-shooter&0.069271&\oiii&0.3\\
894&SWIFT J1733.3+3635&2MASX J17333689+3631319&Sy1.9&Palomar/DBSP&0.043661&\oiii&0.1\\
897&SWIFT J1737.7-5956A&1RXS J173751.2-600408&BZQ&VLT/X-shooter&3.656025&\oiii&0.2\\
906&SWIFT J1742.1-6054&PKS 1737-60&Sy1&SOAR/GM&0.152012&\oiii&0.2\\
974&SWIFT J1825.7+7215&LEDA 61865&Sy2&Palomar/DBSP&0.110806&\oiii&0.2\\
1000&SWIFT J1852.2+8424A&SWIFT J185024.2+842240&Sy1&Palomar/DBSP&0.183122&\oiii&0.3\\
1001&SWIFT J1852.2+8424B&1RXS J184642.2+842506&Sy1&Palomar/DBSP&0.225381&\oiii&0.3\\
1007&SWIFT J1852.8+3002&GALEXASC J185249.68+300425.8&Sy1.9&Palomar/DBSP&0.057301&\oiii&0.6\\
1066&SWIFT J2010.6-2521&1RXS J201020.0-252356&BZQ&VLT/X-shooter&0.824924&\oiii&0.5\\
1075&SWIFT J2024.0-0246&1RXS J202400.8-024527&Sy1.9&VLT/X-shooter&0.137523&\oiii&0.2\\
1078&SWIFT J2029.4-6149&2MASX J20293125-6149087&Sy2&VLT/X-shooter&0.124274&\oiii&0.2\\
1083&SWIFT J2034.0-0943&2MASX J20341926-0945586&Sy2&VLT/X-shooter&0.081551&\oiii&0.2\\
1091&SWIFT J2048.4+3815&1RXS J204826.8+381120&Sy1&Palomar/DBSP&0.105394&\oiii&2.8\\
1096&SWIFT J2059.6+4301B&SWIFT J210001.06+430209.6&Sy2&Palomar/DBSP&0.066023&\oiii&4.1\\
1105&SWIFT J2117.5+5139&2MASX J21174741+5138523&BZQ&Palomar/DBSP&0.053392&\SIII&9.7\\
1130&SWIFT J2156.2+1724&2MASX J21561518+1722525&Sy1.8&VLT/X-shooter&0.03417&\oiii&0.3\\
1164&SWIFT J2243.2-4539&2MASX J22422135-4539093&Sy1.9&VLT/X-shooter&0.120675&\oiii&0\\
1208&SWIFT J2352.6-1707&2MASX J23525143-1704370&Sy1&duPont/BC&0.054695&\oiii&0.1\\
\enddata
\tablecomments{Column descriptions are the same as in \autoref{tab:newagn}, unless otherwise noted.  A detailed description of this table's contents is given in \autoref{sec:surveytable}.}
\end{deluxetable}

\subsection{Distance and Luminosity} \label{dist_desc}
The AGN host galaxies span a large range of redshifts down to very nearby (${<}50$ Mpc) systems.  These AGN can have substantial peculiar velocities compared to velocities of the Hubble flow  where a simple assumption of uniform expansion would lead to large errors.  High-quality redshift-independent distances to nearby galaxies such as through using the tip of the red giant branch (TRGB) are now available \citep[e.g.,][]{McQuinn:2017:51}.  Further compilations such as the Extragalactic Distance Database \citep[EDD,][]{Tully:2009:323} or the Cosmicflows-3 project \citep{Courtois:2017:L6} have now compiled motions of many thousands of local galaxies.

We follow the approach of \citet{Leroy:2019:24}, which performed a careful analysis of which compilations to adopt at different distances based on statistical uncertainties.  Specifically, we limit our search to ${<}$50 Mpc (or 3500\,\kmps) galaxies to adopt redshift-independent distances as beyond this the typical uncertainties are larger than those in the Hubble flow. We focus on using EDD, Cosmicflows, and NED for adopting distances.  We adopt a TRGB and "quality" distances from EDD whenever available.    When this is not available, we take the Cosmicflows-3 value.  If none of these are available, we use the most recent redshift-independent estimate from NED.  This results in redshift-independent measurements for all 59 galaxies (\autoref{tab:zind}) below 3500\,\kmps in our survey.

\begin{deluxetable}{llllll}
\tabletypesize{\scriptsize}
\tablecaption{Redshift Independent Distances \label{tab:zind}}
\tablehead{
\colhead{BAT ID} &                   \colhead{Counterpart}  &     \colhead{$z$} & \colhead{Dist. Ind. [Mpc] \tablenotemark{a}} &\colhead{Dist. $z$ [Mpc]}& \colhead{Method\tablenotemark{b}}}
\startdata
655&NGC4945&0.002259&3.72&9.7&TRGB\\
671&CenA&0.001877&3.66&8.1&TRGB\\
477&M81&-0.000113&3.61&0&TRGB\\
711&CircinusGalaxy&0.001495&4.21&6.4&NED\\
616&NGC4395&0.001106&4.76&4.7&TRGB\\
609&NGC4258&0.001692&7.31&7.3&TRGB\\
585&NGC4051&0.002043&11.02&8.8&CF3\\
838&NGC6221&0.004108&11.86&17.6&CF3\\
731&NGC5643&0.004042&12.68&17.4&TRGB\\
875&NGC6300&0.003131&13.18&13.4&CF3\\
593&NGC4138&0.003187&13.7&13.7&SBF\\
144&NGC1068&0.003484&14.4&15&NED\\
579&NGC3998&0.003574&14.19&15.3&SBF\\
686&NGC5273&0.003606&16.6&15.5&SBF\\
1157&NGC7314&0.004607&16.75&19.8&CF3\\
548&NGC3718&0.003278&17.1&14.1&NED\\
216&NGC1566&0.00474&17.9&20.4&TRGB\\
615&NGC4388&0.008344&18.11&36&CF3\\
595&NGC4151&0.003152&19&13.5&NED\\
665&NGC5033&0.002763&19.05&11.9&CF3\\
140&NGC1052&0.004519&19.23&19.4&SBF\\
590&NGC4102&0.002365&19.5&10.1&CF3\\
184&NGC1365&0.005091&19.57&21.9&TRGB\\
653&NGC4941&0.003881&20.45&16.7&NED\\
484&NGC3079&0.003505&20.61&15&CF3\\
1142&NGC7213&0.004767&22&20.5&NED\\
1188&NGC7582&0.005249&22.49&22.6&CF3\\
1046&NGC6814&0.005792&22.8&24.9&NED\\
497&NGC3227&0.003277&22.95&14.1&NED\\
436&NGC2655&0.004854&24.4&20.9&NED\\
712&NGC5506&0.005979&26.4&25.7&NED\\
607&NGC4235&0.007934&26.6&34.2&SBF\\
1180&NGC7465&0.006343&27.2&27.3&NED\\
319&ESO5-4&0.006162&28.18&26.5&CF3\\
437&NGC2712&0.006754&31.19&29.1&CF3\\
480&NGC3081&0.008071&32.5&34.8&NED\\
1135&NGC7172&0.008511&33.9&36.7&NED\\
823&ESO137-34&0.008759&34.1&37.8&NED\\
308&NGC2110&0.0075&34.3&32.3&NED\\
93&NGC678&0.009485&34.5&40.9&NED\\
688&NGC5290&0.008566&34.51&36.9&CF3\\
621&NGC4500&0.010357&34.51&44.7&CF3\\
1184&NGC7479&0.007105&36.81&30.6&CF3\\
631&NGC4593&0.00832&37.2&35.9&NED\\
739&NGC5728&0.010321&37.5&44.6&CF3\\
471&NGC2992&0.007675&38&33.1&NED\\
558&NGC3783&0.008958&38.5&38.6&NED\\
103&LEDA89913&0.011836&38.8&51.2&NED\\
530&NGC3516&0.008718&38.9&37.6&NED\\
654&NGC4939&0.010543&42.07&45.5&CF3\\
599&NGC4180&0.006532&43.05&28.1&CF3\\
560&NGC3786&0.008916&43.9&38.4&NED\\
766&NGC5899&0.008597&45.08&37.1&CF3\\
1092&IC5063&0.011267&45.9&48.7&NED\\
237&LEDA86269&0.010529&46.13&45.5&CF3\\
62&IC1657&0.011688&48.31&50.5&CF3\\
58&NGC424&0.010885&51.05&47&CF3\\
684&NGC5283&0.010365&51.52&44.7&CF3\\
451&IC2461&0.007535&58.88&32.5&CF3\\
\enddata
\tablenotetext{a}{Best redshift-independent distance.  See \autoref{red_explain} for a detailed description of the redshift-independent measurements.}
\tablenotetext{b}{TRGB: tip of the red giant branch \citep[e.g.,][]{McQuinn:2017:51}, CF3: Cosmicflows-3 project \citep{Courtois:2017:L6}, NED: NASA Extragalactic Database.}
\end{deluxetable}

\subsection{Black Hole Mass, Bolometric Luminosity, and Eddington Ratios}
We also provide the best BH mass measurements for each AGN in our catalog outside of continuum-dominated blazars (BZB).  A small number of sources have direct (or higher-quality) measurements of BH masses, either from reverberation mapping ($N$=\Nrever), OH megamasers ($N$=\Nmegamaser), or high-quality IFU observations of gas or stars ($N$=\NIFU) which we have adopted and tabulated when available.

Some AGN may have multiple BH mass measurements from broad lines and velocity dispersions, so we select the best measurement (\autoref{tab:mbhmeth}) based on the following ordered scheme:
\begin{enumerate}
    \item Literature measurements with megamasers, reverberation mapping, or stellar and gas dynamics.
    \item Broad-line \hbeta\ if \NH${<}10^{22}$ \nhunit from \citet{Mejia_Broadlines}.  The conversion of broad-line measurements to BH masses is given by \citet{Trakhtenbrot:2012:3081}.  
    \item Broad-line \halpha\ from \citet{Mejia_Broadlines}  if \NH${<}10^{22}$ \nhunit\ and broad-line \hbeta\ is present but not measurable because of instrumental or telluric issues.  They use \citet{Greene:2005:122}, but adjusted by 4/3 (or +0.125 dex) for a consistent virial factor of 1 across all broad-line measurements.
    \item Broad-line \MgII\ followed by \CIV\ for high-redshift sources ($z{>}0.8$) from \citet{Mejia_Broadlines} for those without broad-line \hbeta\ using the relation of \citet{Mejia-Restrepo:2016:187}.
    \item Stellar velocity dispersion measurements for all Sy1.9 and Sy2 AGN from \citet{Koss_DR2_sigs}.  We calculated the BH mass, \mbh, using the $\mbh-\sigs$ relation of equation 13 from \citet[][]{Kormendy:2013:511}.
    \item If \NH${>}10^{22}$ \nhunit, and if no velocity dispersion measurement is possible, broad-line \halpha\ followed by broad-line \hbeta\ is used from \citet{Mejia_Broadlines}.  These final measurements should be used with caution because of the tendency to underestimate \mbh\ due to obscuration \citep[e.g.,][]{Mejia_Broadlines,Ricci_DR2_NIR_Mbh,denBrok_DR2_NIR}
\end{enumerate}

\begin{figure*} 
\centering
\includegraphics[width=8cm]{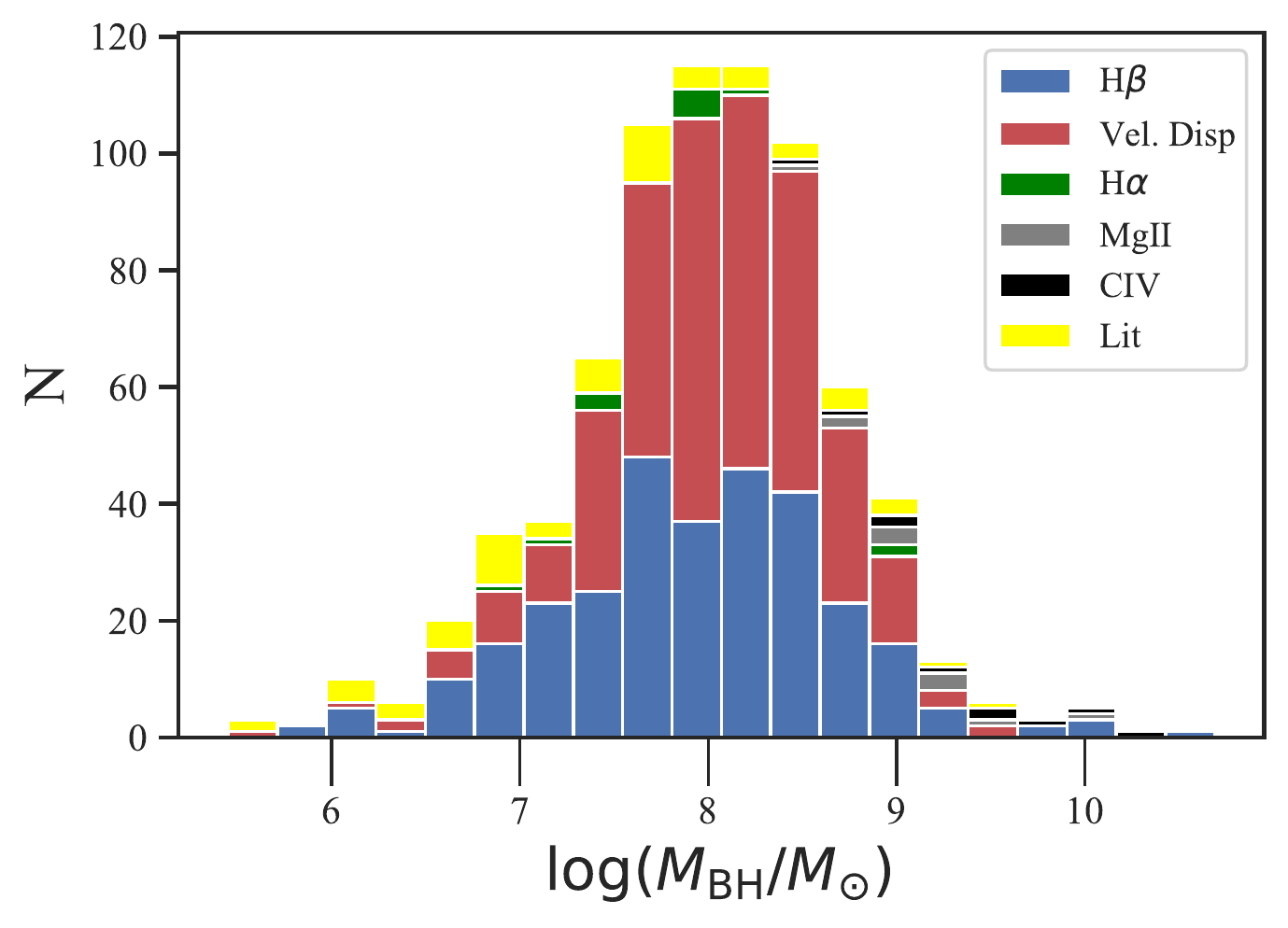}
\includegraphics[width=8cm]{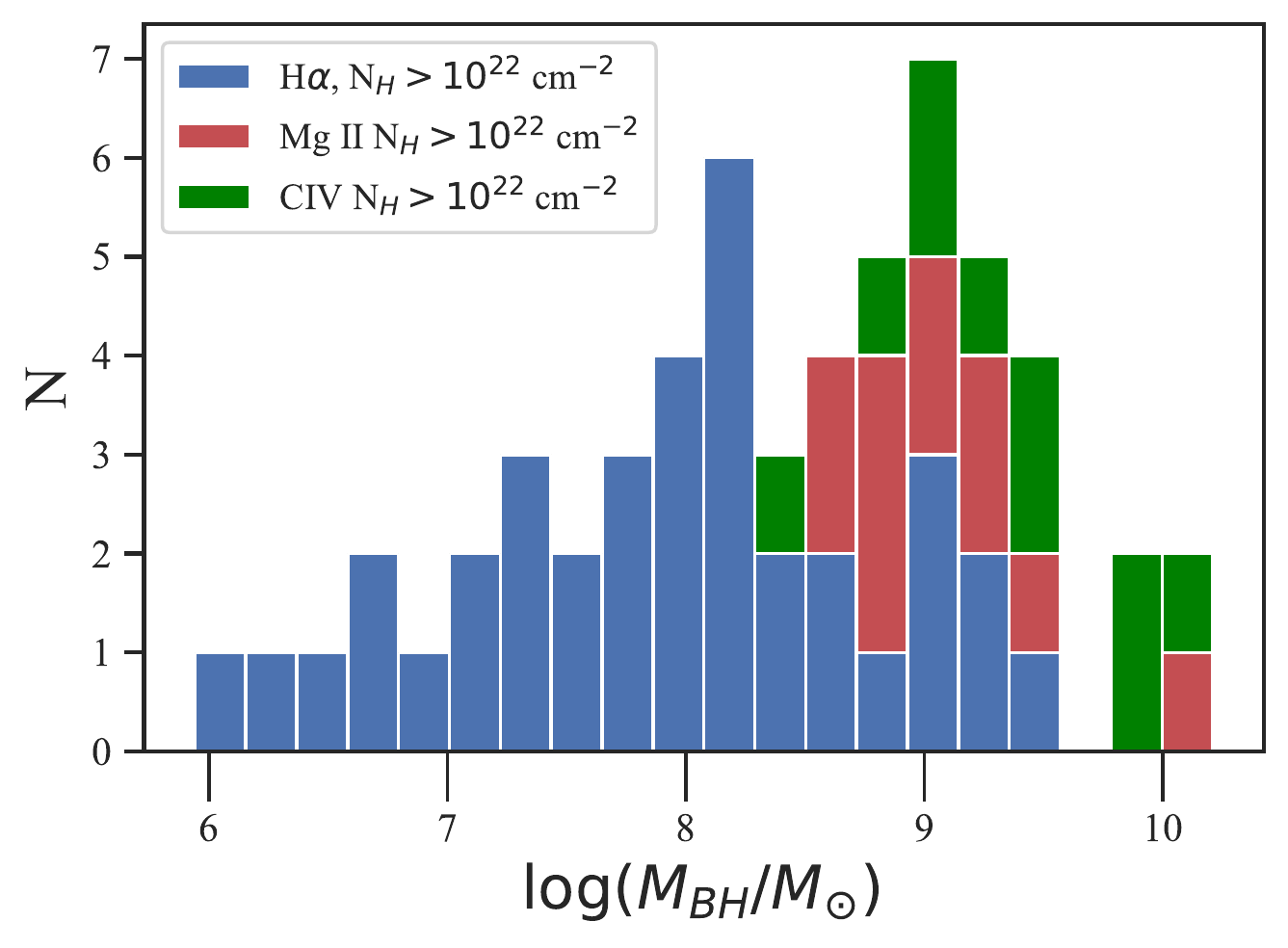}
\caption{Left:  histogram of the BH mass estimates excluding broad-line measurements in X-ray-obscured AGN (\NH${>}10^{22}$ \nhunit).  Right:  Histogram of the BH mass estimates in the small number of X-ray-obscured AGN with only broad-line measurements.  While many of these AGN do have BH masses consistent with the other unobscured distribution (e.g. $>10^7$\,\Msun) there is a larger fraction with low masses ($<10^7$\,\Msun), which may be significantly underestimated.
}
\label{fig:mbh_dist_meas}
\end{figure*}

A summary of the number of best BH mass measurements is found in \autoref{fig:mbh_dist_meas}.  The majority of measurements come from either velocity dispersion measurements or broad \hbeta\ with a smaller number from literature measurements, broad \halpha, broad \halpha\ in X-ray-obscured AGN (\NH${>}10^{22}$ \nhunit), and \MgII\ or \CIV\ broad lines for distant beamed AGN ($z{>}1$). See \citet{Mejia_Broadlines} and \citet{Koss_DR2_sigs} for more details about the individual observations, calculations, and methodologies.


The bolometric luminosity is calculated from the intrinsic luminosity in the 14--150\,keV range as shown in \citep[see Table 12,][]{Ricci:2017:17}.  This analysis was done using the 0.3--150\,keV range by combining the 70-month average Swift BAT spectra with data below 10\,keV from Swift XRT, XMM-Newton, Chandra, Suzaku, and ASCA using detailed spectral models.  Here we calculated the bolometric luminosity using a 14--150\,keV bolometric correction of 8 based on the factor of 20 for the 2--10\,keV range \citep{Vasudevan:2009:1124} and assuming $\Gamma$=1.8.  We prefer this rather than using the direct calculation from the 2--10\,keV range because the corrections are less dependent on \NH\ for CT AGN. The 14--150\,keV emission is also integrated over 70 months, so it is more likely to be representative of the average value.  

For Eddington ratios we assume an Eddington luminosity consistent with solar metallicity:
\begin{equation}
L_{\rm Edd}=1.5\cdot 10^{46}\,\text{erg s}^{-1} \frac{M_{BH}}{10^8 M_\odot}
\end{equation}

We note that more complicated procedures than a simple bolometric corrections from the intrinsic X-ray flux such as in terms of the Eddington ratio (\lledd) are sometimes used \citep[e.g.,][]{Marconi:2004:169, Lusso:2011:A110}.  However, we prefer this simple approach, similar to what was done in the BASS DR1, that can be reliably applied to all AGN. 

We do not list individual errors for each \mbh, \lbol, and \lledd\ measurement, as they are dominated by the systematic uncertainties in the scaling relations rather than the emission-line fitting or velocity dispersion measurements which are typically $<$0.1 dex.  Errors in \mbh\ are of order 0.4-0.5 dex owing to systematic uncertainties in  virial and $\sigma_*$-based scaling relations \citep[e.g.,][]{McLure:2002:795,Vestergaard:2006:689,Ricci_DR2_NIR_Mbh}.  For \lbol, the scatter between BAT 14--195\,keV luminosity and \Lop\ was 0.46 dex in the DR1 \citep{Koss:2017:74}

Efforts are currently underway for future BASS surveys to better calibrate the bolometric correction with AGN source properties and estimate its intrinsic reliability.  A large HST program (${>}$100) AGN is currently underway obtaining high spatial resolution near UV (${<}$3000\,\AA) imaging of the AGN emission, combined with simultaneous measurement of the AGN emission in the X-rays and UV/optical from Swift, with ground based imaging in $griz$.

A summary of the survey completeness in BH mass measurements for unbeamed AGN is provided in \autoref{tab:median_unbeamedagn} separated by AGN type.  Overall the completeness is slightly higher for Sy1 and Sy1.9 (${>}96\%$) than for velocity dispersion measurements (${>}93\%$).  Outside of the Galactic plane ($\lvert b\rvert{>}10^{\circ}$) the survey completeness rises to $98\%$ for all unbeamed AGN because of the typically lower extinction in these regions.  Finally, for beamed AGN with broad lines (BZQ) the measured BH masses (\autoref{tab:median_unbeamedagn}) and completeness are somewhat lower, but still the majority (91\%).

A summary of the typical BH masses, bolometric luminosities, Eddington ratios, and X-ray column densities is provided in \autoref{tab:median_unbeamedagn} and \autoref{tab:median_beamedagn} for unbeamed and beamed AGN, respectively.

\begin{deluxetable}{ll}
\tablecaption{Summary of Best Black Hole Mass Method \label{tab:mbhmeth}}
\tablehead{
\colhead{Best BH Mass Method } & \colhead{Count}}
\startdata
$\sigma_\star$&344\\
H$\beta$&305\\
Lit.&62\\
H$\alpha$ (\NH$>10^{22}$\,\nhunit)&37\\
H$\alpha$&13\\
\MgII&11\\
\CIV&10\\
\CIV\ (\NH$>10^{22}$\,\nhunit)&4\\
\MgII\ (\NH$>10^{22}$\,\nhunit)&3\\
H$\beta$ (\NH$>10^{22}$\,\nhunit)&1\\
\hline
Total&790\\
\enddata
\tablecomments{The best BH mass measurement provided in the catalog.  See \autoref{red_explain} for details.}
\end{deluxetable}

\section{Summary}
We have presented an overview of the BASS DR2 survey with \Nspec\ optical spectra,  of which \Nspecnew\ are released for the first time, for the \NAGN\ hard-X-ray-selected AGN in the Swift BAT 70-month sample.   With this first DR2 catalog release we provide the following:

\begin{itemize}
    \item A revised catalog based on optical and NIR spectroscopic follow-up that identifies all of \NAGN\ among unknown sources above ($\lvert b\rvert{>}3^{\circ}$) or below $A_{\mathrm{V}}$=5 mag excluding only 7 unknown sources deep within the Galactic plane at high extinction.  We have included new identifications of 17 Galactic sources.  
    \item We have further classified our sources by AGN type based on the presence of broad lines (e.g. Sy1, Sy1.9, Sy2) as well as beamed and lensed AGN. We have further provided important catalogs for population studies including dual AGN, weakly associated AGN, and multiple weak confused sources within the BAT beam.
    \item A full master catalog summary of the \Nspec, instrumental settings, their reductions, and observing conditions.  With this we have provided a master catalog of redshifts, distances, bolometric luminosities, and BH masses.   
    \item Overall the completion for the survey is 99.9\% in redshift outside the extreme regions in the Galactic plane ($\lvert b\rvert{>}3^{\circ}$).  In BH mass, the survey is 98\% complete  using broad lines and velocity dispersions for unbeamed AGN outside the Galactic plane ($\lvert b\rvert{>}10^{\circ}$).  The final catalog contains \Nnewz\ new redshift measurements and \Nmbh\ BH mass measurements.
\end{itemize}

\begin{deluxetable}{ccccccccccccc}
\tabletypesize{\footnotesize}
\tablewidth{0pt}
\tablecaption{Summary of Unbeamed AGN Properties \label{tab:median_unbeamedagn}}
\tablehead{
\colhead{Type}& \colhead{$N$}&\colhead{$N$}& \colhead{$z$}& \colhead{N$_{\mbh}$}& \colhead{N$_{\mbh}$}& \colhead{\% Meas \mbh}& \colhead{\% Meas}& \colhead{$\log$ \mbh}& \colhead{$\log$ \Lbol}& \colhead{$\log$ \lledd}& \colhead{$\log$ \NH}\\
\colhead{}& \colhead{}&\colhead{$\lvert b\rvert>10^{\circ}$}& \colhead{}& \colhead{}& \colhead{$\lvert b\rvert>10^{\circ}$}& \colhead{}& \colhead{$\lvert b\rvert>10^{\circ}$}& \colhead{(\Msun)}& \colhead{(\ergps)}& \colhead{}& \colhead{(\nhunit)}
\\
\colhead{(1)}& \colhead{(2)}& \colhead{(3)}& \colhead{(4)}&\colhead{(5)}& \colhead{(6)}& \colhead{(7)}&
\colhead{(8)}&
\colhead{(9)}&
\colhead{(10)}&
\colhead{(11)}&
\colhead{(12)}
}
\startdata
Sy1&359&318&0.050$\pm$0.003&350&311&97&98&7.81$\pm$0.04&44.87$\pm$0.04&-1.17$\pm$0.03&20.0$\pm$0.05\\
Sy1.9&101&86&0.030$\pm$0.004&97&84&96&98&7.98$\pm$0.06&44.59$\pm$0.08&-1.61$\pm$0.09&22.28$\pm$0.13\\
Sy2&292&259&0.029$\pm$0.003&275&253&94&98&8.06$\pm$0.04&44.50$\pm$0.04&-1.71$\pm$0.04&23.27$\pm$0.05\\
\hline
Total&752&663&0.038$\pm0.002$&722&648&96&98&7.96$\pm$0.03&44.67$\pm$0.03&-1.42$\pm$0.03&21.98$\pm$0.06
\enddata
\tablecomments{Summary of the medians and standard error of the median for different populations of unbeamed AGN.  Column (1): AGN optical type based on the presence of broad \hbeta\ and \halpha.  Column (2): Total for the whole sample. Column (3): total excluding the Galactic plane region $\lvert b\rvert<10^{\circ}$ where high optical extinction makes measurements more difficult. Column (4): median redshift from optical lines.  Columns (5--8): number of unique AGN with \mbh\ measurements and excluding the Galactic plane region $\lvert b\rvert<10^{\circ}$ where high optical extinction makes measurements more difficult.  Also listed as percentages.  Columns (9-12): median \mbh, \lbol, \lledd, and, \logNH\ for the sample.  }
\end{deluxetable}

\begin{deluxetable}{ccccccccccc}
\tabletypesize{\small}
\tablewidth{0pt}
\tablecaption{Summary of Beamed AGN Properties \label{tab:median_beamedagn}}
\tablehead{
\colhead{Type}& \colhead{$N$}&\colhead{$N$, $\lvert b\rvert>10^{\circ}$}& \colhead{$z$}& \colhead{$N_{\mbh}$}& \colhead{\% Meas \mbh}&  \colhead{$\log$ \mbh}& \colhead{$\log$ \Lbol}& \colhead{$\log$ \lledd}& \colhead{$\log$ \NH}
\\
\colhead{(1)}& \colhead{(2)}& \colhead{(3)}& \colhead{(4)}&\colhead{(5)}& \colhead{(6)}& \colhead{(7)}&
\colhead{(8)}&
\colhead{(9)}&
\colhead{(10)}
}
\startdata
BZQ&74&63&0.88$\pm$0.12&67&91&8.83$\pm$0.09&47.66$\pm$0.16&0.38$\pm$0.12&20$\pm$0.11\\
BZB&22&18&0.13$\pm$0.02&&&&45.81$\pm$0.12&&20.57$\pm$0.12\\
BZG&8&6&0.07$\pm$0.02&&&&45.11$\pm$0.20&&20.81$\pm$0.11\\
Sy1/Lense&1&1&0.65&1&100&8.79&47.18&0.21&20\\
BZQ/Lense&1&0&2.51&&&&49.49&&22.77\\
\hline
Total&106&88&0.33$\pm$0.10&68&&8.83$\pm$0.09&46.53$\pm$0.14&0.38$\pm$0.12&20.54$\pm$0.08\\
\enddata
\tablecomments{Summary of the medians and standard error of the median for different populations of beamed and/or lensed AGN.  (1) AGN optical type based on presence of broad lines (BZQ), only host galaxy features lacking broad lines (BZG), or traditional continuum-dominated blazars with no emission lines (BZB), or lensing. (2) Total for the whole sample and excluding the Galactic plane region $\lvert b\rvert<10^{\circ}$ where high optical extinction makes measurements more difficult. (4) Median redshift from optical lines.  (5--6) Number of unique AGN with \mbh\ measurements and percentages.  (7--10) Median \mbh, \lbol, \lledd, and, \logNH\ for the sample. }
\end{deluxetable}

\section{Acknowledgment} 

We thank the reviewer for the constructive comments that helped us improve the quality of this paper.

BASS/DR2 was made possible through the coordinated efforts of a large team of astronomers, supported by various funding institutions, and using a variety of facilities. 

We acknowledge support from NASA through ADAP award NNH16CT03C (M.K.); 
the Israel Science Foundation through grant number 1849/19 (B.T.);
the European Research Council (ERC) under the European Union's Horizon 2020 research and innovation program, through grant agreement number 950533 (B.T.);
FONDECYT Postdoctorado 3180506 (F.R.) and 3210157 (A.R.); 
FONDECYT Regular 1190818 (E.T., F.E.B.) and 1200495 (E.T.,F.E.B);
ANID grants CATA-Basal AFB-170002, ACE210002, and FB210003 (C.R., F.R., E.T., F.E.B., and E.C.); 
ANID Anillo ACT172033 (E.T.); and Millennium Science Initiative Program  – ICN12\_009 (FEB); an ESO fellowship (M.H., J.M.); Fondecyt Iniciacion grant 11190831 (C.R.). K.O. acknowledges support from the National Research Foundation of Korea (NRF-2020R1C1C1005462) and the Japan Society for the Promotion of Science (JSPS, ID: 17321); M.B. acknowledges support from the YCAA Prize Postdoctoral Fellowship;   R.R. thanks CNPq (311223/2020-6), CAPES and FAPERGS (16/2551-0000251-7 and 19/2551-0001750-2). This work was performed in part at Aspen Center for Physics, which is supported by National Science Foundation grant PHY-1607611. 

We also acknowledge the following people who assisted in acquiring the Palomar observations presented herein: Dave Alexander, Roberto Assef, Rosamaria Carraro, Alison Dugas, Peter Eisenhardt, Clarke Esmerian, Carla Fuentes, Felix Fuerst, Maya Fuller, Daniel Gawerc, David Girou, Ana Glidden, Matthew Graham, Claire Greenwell, Brian Grefenstette, Marianne Heida, Hyunsung Jun, Peter Kosec, Stephanie LaMassa, Jeff Maggio, Alejandra Melo, Catalina-Ana Miritescu, Wenli Mo, Eric Mukherjee, Gael Noirot, Antonija Oklop\v{c}i\'{c}, Sean Pike, Joseph Simon, Tawny Sit, Navin Sridhar, Aaron Stemo, Becky Tang, Thomas Venville, Jingyi Wang, and Dominika Wylezalek.    We acknowledge financial contribution from Joanna Wall Muir and the California Institute of Technology's Student Faculty Program to assist in the Palomar observations for Tea Freedman-Susskind, Feiyang Liu, Milan Robertson, Yerong Xu, and Emily Zhang.  We acknowledge the work that Swift BAT team has done to make this project possible.

We acknowledge the various telescopes used in this paper.  We are tremendously thankful to all the observing and support staff in all the observatories, and their headquarters, for their great assistance in planning and conducting the observations that made BASS/DR2 possible. 
Specifically, BASS/DR2 is based on the following:\\
%
(1) Observations collected at the European Organisation for Astronomical Research in the Southern Hemisphere under 34 ESO programs: 
60.A-9024(A), 60.A-9421(A), 062.H-0612(A), 086.B-0135(A),
089.B-0951(A), 089.B-0951(B),
090.A-0830(A), 090.D-0828(A),
091.B-0900(B), 091.C-0934(B),
092.B-0083(A), 093.A-0766(A), 094.B-0321(A),
095.B-0059(A), 098.A-0062,
098.A-0635(B), 098.B-0551(A),
099.A-0403(A+B), 099.A-0442(A),
099.B-0785(A), 
0101.A-0765(A), 0101.B-0456(B), 0101.B-0739(A), 
0102.A-0433(A), 0102.B-0048,
0103.A-0521(A), 0103.B-0566(A), 0103.A-0777(A),
0104.A-0353(A), 0104.B-0959(A), 0106.A-0521(A), 385.B-1035(A), 2100.B-5018(B), and 60.A-9100(J).

(2) Data obtained with the Hale 200-inch telescope at Palomar observatory, including five dedicated Yale programs: 
2017B/Y04, 2018A/Y03, 2018B/Y04, 2019A/Y03, and 2019B/Y04 (PI: M. Powell). Additional data were obtained as part of Caltech runs including 2014B/P05, 2015A/P04, 2015B/P06, 2016A/P06, 2016B/P04, 2017A/P13, 2017B/P01, 2018A/P15, 2018B/P02, 2019A/P18, 2019B/P11, 2020A/P14, 2020B/P19 (PI: F. Harrison) and also JPL runs including 2016B/J12, 2017A/J07, 2017B/J15 2019B/J09, 2019A/J06, 2019B/J19, 2020A/J11, 2020B/J04 (PI: D. Stern).

(3) Observations obtained at the Southern Astrophysical Research (SOAR) telescope, which is a joint project of the Minist\'{e}rio da Ci\^{e}ncia, Tecnologia e Inova\c{c}\~{o}es (MCTI/LNA) do Brasil, the US National Science Foundation’s NOIRLab, the University of North Carolina at Chapel Hill (UNC), and Michigan State University (MSU).  These include from CNTAC (CN2018A-104, CN2018B-83) and NOIRLab program 2012A-0463 (PI M. Trippe).

(4) Observations at Kitt Peak National Observatory at NSF’s NOIRLab (NOIRLab Prop. ID 52, 2946; PI: F. Bauer), which is managed by the Association of Universities for Research in Astronomy (AURA) under a cooperative agreement with the National Science Foundation.   We are honored to be permitted to conduct astronomical research on Iolkam Du’ag (Kitt Peak), a mountain with particular significance to the Tohono O’odham.
%

%
(5) Data gathered with the 6.5 m Magellan Telescopes (CN2019A-70 and CN2019B-77) and the 2.5 m Dupont Telescope (CN2016A-80) located at Las Campanas Observatory, Chile.  Support for the design and construction of the Magellan Echellette Spectrograph was received from the Observatories of the Carnegie Institution of Washington, the School of Science of the Massachusetts Institute of Technology, and the National Science Foundation in the form of a collaborative Major Research Instrument grant to Carnegie and MIT (AST0215989).  

(6) Data obtained at the W.M. Keck Observatory, which is operated as a scientific partnership among the California Institute of Technology, the University of California, and the National Aeronautics and Space Administration. The Observatory was made possible by the generous financial support of the W. M. Keck Foundation.  The authors wish to recognize and acknowledge the very significant cultural role and reverence that the summit of Maunakea has always had within the indigenous Hawaiian community.  We are most fortunate to have the opportunity to conduct observations from this mountain.

%
(7) Data obtained as part of the various stages of the Sloan Digital Sky Survey (SDSS-I/II, III, and IV).
Funding for the SDSS has been provided by the Alfred P. Sloan Foundation, the US Department of Energy Office of Science, and the Participating Institutions. 
SDSS acknowledges support and resources from the Center for High 
Performance Computing  at the University of Utah. 
The SDSS website is www.sdss.org.
The most recent generation of the SDSS we benefited from, SDSS-IV, is managed by the Astrophysical Research Consortium for the Participating Institutions of the SDSS Collaboration including the Brazilian Participation Group, the Carnegie Institution for Science, Carnegie Mellon University, Center for Astrophysics | Harvard \& Smithsonian, the Chilean Participation Group, the French Participation Group, Instituto de Astrof\'isica de Canarias, The Johns Hopkins University, Kavli Institute for the Physics and Mathematics of the Universe (IPMU) / University of Tokyo, the Korean Participation Group, Lawrence Berkeley National Laboratory, Leibniz Institut f\"ur Astrophysik Potsdam (AIP),  Max-Planck-Institut f\"ur Astronomie (MPIA Heidelberg), Max-Planck-Institut f\"ur Astrophysik (MPA Garching), Max-Planck-Institut f\"ur Extraterrestrische Physik (MPE), National Astronomical Observatories of China, New Mexico State University, New York University, University of Notre Dame, Observat\'ario Nacional / MCTI, The Ohio State University, Pennsylvania State University, Shanghai Astronomical Observatory, United Kingdom Participation Group, Universidad Nacional Aut\'onoma de M\'exico, University of Arizona, University of Colorado Boulder, University of Oxford, University of Portsmouth, University of Utah, University of Virginia, University of Washington, University of Wisconsin, Vanderbilt University, and Yale University.

This research has made use of the NASA/IPAC Extragalactic Database (NED), which is operated by the Jet Propulsion Laboratory, California Institute of Technology, under contract with the National Aeronautics and Space Administration. This research has made use of the SIMBAD database, operated at CDS, Strasbourg, France.
 
A significant part of the BASS observations and work took place during the COVID-19 crisis. We thank the health care experts in communities around the world, for their tireless efforts to keep us all as safe and healthy as possible.

\appendix





\section{Comparison to DR1 and Past Surveys}\label{dr1comp_appen} 
Here we provide a comparison to the 641 optical spectroscopic measurements from the BASS DR1 including redshifts, AGN classification, and BH mass measurements as well as past measurements in the literature.
,
\subsection{Beamed AGN Changes in DR2} \label{appen_beamed}
\begin{figure*} 
\centering
\includegraphics[width=18cm]{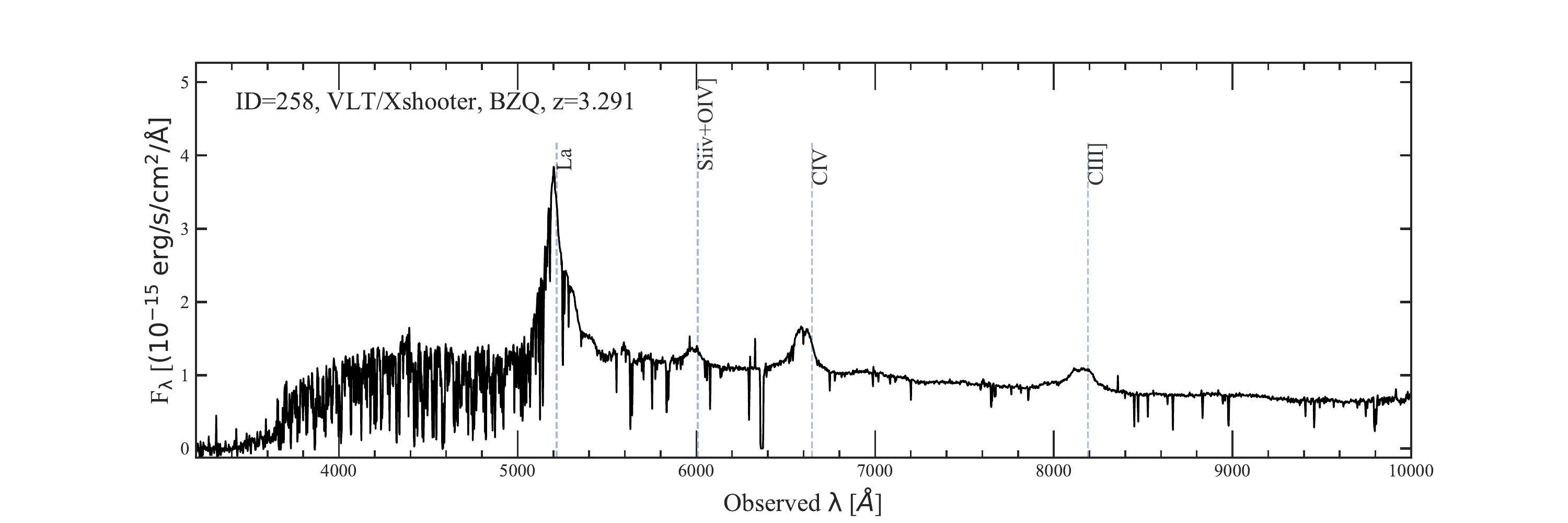}
\includegraphics[width=18cm]{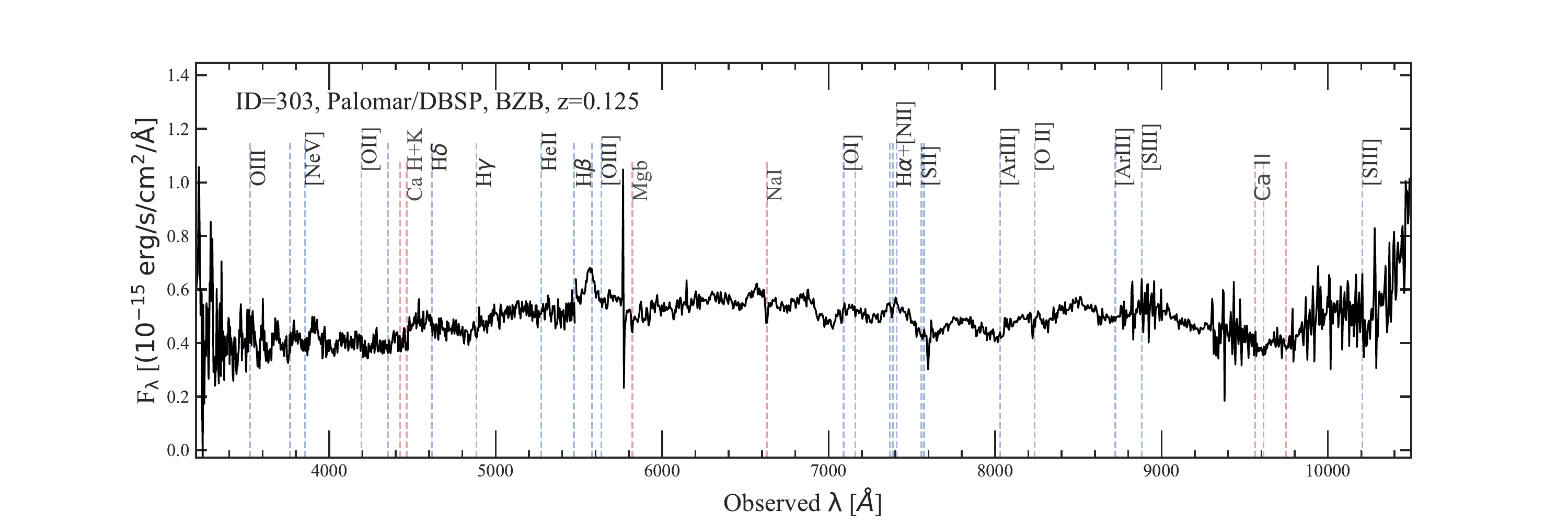}
\includegraphics[width=18cm]{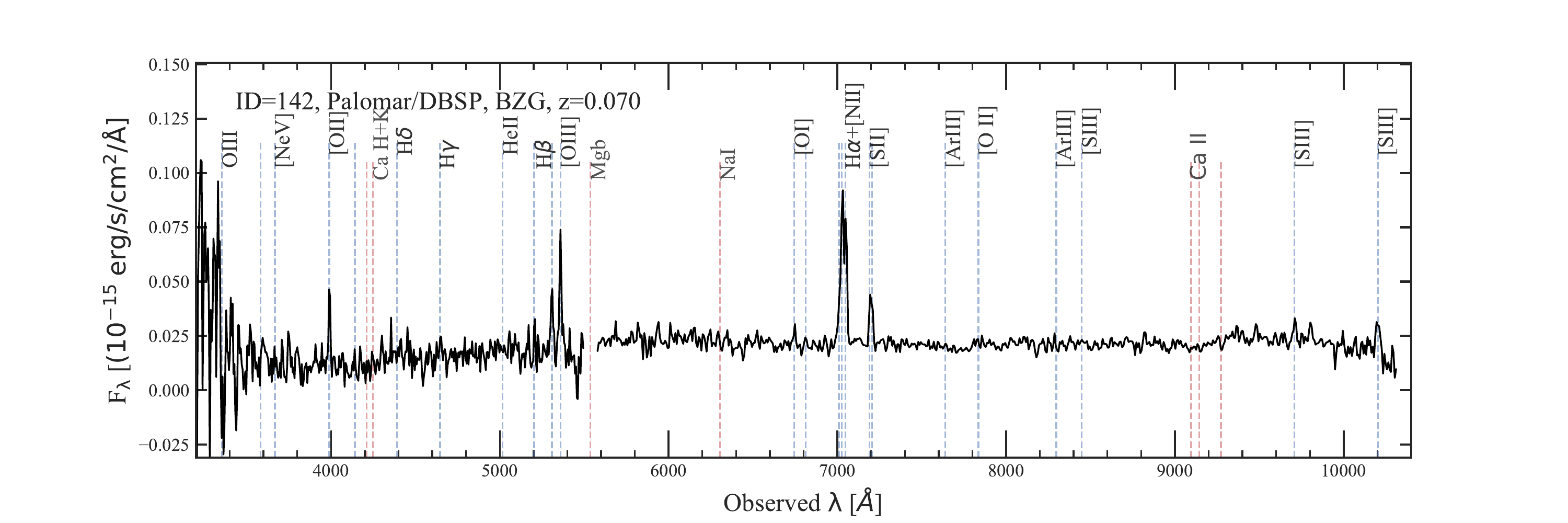}
\caption{Figure showing the spectral classification of beamed sources based on the presence of broad lines (BZQ, top), traditional continuum-dominated blazars with no emission lines or host galaxy features (BZB, middle), or only host galaxy features lacking broad lines (BZG, bottom). }
\label{fig:type_example_beamed}
\end{figure*}

Here we provide a list of all beamed AGN that are newly identified or changed in DR2 in \autoref{tab:beamedchange} compared to the DR1.  Example spectra for the various classes (e.g., BZB, BZG, BZQ) are found in  \autoref{fig:type_example_beamed}.

\begin{deluxetable}{ccccc}
\tabletypesize{\footnotesize}
\tablewidth{0pt}
\tablecaption{Beamed AGN Changes in DR2 \label{tab:beamedchange}}
\tablehead{
 \colhead{BAT ID} & \colhead{Swift Name} & \colhead{DR1 Class\tablenotemark{a}} & \colhead{DR2 Type}& \colhead{Ref.}}
\startdata
30&SWIFT J0042.9+3016B&&BZQ&P19\\
33&SWIFT J0048.8+3155&BZB&Sy2&P19\\
59&SWIFT J0113.8+2515&&BZQ&P19\\
140&SWIFT J0241.3-0816&BZU&Sy2&P19\\
170&SWIFT J0312.9+4121&BZU&Sy1&T78\\
173&SWIFT J0319.7+4132&BZU&Sy2&P19\\
178&SWIFT J0326.0-5633&&BZG&P19\\
226&SWIFT J0433.0+0521&BZU&Sy1&P19\\
273&SWIFT J0519.5-3140&BZU&Sy1&A19\\
277&SWIFT J0525.3-4600&BZQ&Sy2&P19\\
323&SWIFT J0225.0+1847&&BZQ&P19\\
359&SWIFT J1959.6+6507&&BZB&P19\\
377&SWIFT J0733.9+5156&&BZG&P19\\
660&SWIFT J1310.9-5553&&BZQ&P19\\
671&SWIFT J1325.4-4301&BZU&Sy2&P19\\
690&SWIFT J1347.1+7325&&BZQ&P19\\
761&SWIFT J1943.5+2120&&BZQ&P19\\
780&SWIFT J1145.6-6956&&BZQ&P19\\
787&SWIFT J1557.8-7913&BZU&Sy2&P19\\
876&SWIFT J1719.7+4900&BZU&Sy1.9&P19\\
897&SWIFT J1458.9+7143&&BZQ&P19\\
906&SWIFT J1742.1-6054&BZU&Sy1&B16\\
1082&SWIFT J2033.4+2147&&BZQ&P19\\
1105&SWIFT J2117.5+5139&&BZQ&P19\\
1142&SWIFT J2209.4-4711&BZU&Sy1.9&P19\\
1181&SWIFT J2303.1-1837&BZU&Sy1&P19\\
\enddata
\tablecomments{Column descriptions are the same as in \autoref{tab:newagn}, unless otherwise noted.}
\tablenotetext{a}{Beamed AGN type in BASS DR1 \citep{Koss:2017:74}.}
\tablerefs{A19: \citet{Angioni:2019:A148}; B16: \citet{Bassani:2016:3165}; P19: \citet{Paliya:2019:154}; T78: \citet{Tzanetakis:1978:63P}.}
\end{deluxetable}

\subsection{AGN classification}
Overall, there are 168 DR2 AGN that we provided revised or the first classification of AGN type based on our measurements.  For comparison, we first look at the 641 DR1 AGN types compared to overlapping DR2 AGN types.  There are 10\% changes (64/641). This includes 52 reclassifications from Sy2 to Sy1.9 or Sy1, or from Sy1.9 to Sy1 based on the detection of broad lines that were not detected in DR1 spectra.  Conversely, 12 spectra change from Sy1 to Sy1.9 or Sy2 or from Sy1.9 to Sy2.  This shift largely reflects the higher-quality spectra in terms of resolution and signal-to-noise ratio compared to the archival DR1, which used much smaller telescopes and lower spectral resolutions rather than "bona fide" AGN that have undergone changes.  The "bona fide" changing optical type AGN are part of a future study (Temple, M. et al. 2022, in preparation).  Among the 216 AGN which were not part of the DR1 release, roughly half have their first or revised classifications  (48\%, 103/216).  This is compared to the most recent 105-month survey \citep{Oh:2018:4} which includes updates from SIMBAD and NED.   From these 103, 72 measurements are to previously unknown AGN without available optical spectroscopy or classification.

\subsection{Redshifts}
We compare the redshifts from the 599 DR1 to the revised measurements from spectroscopy in DR2.  Among the low-redshift sample ($z{<}0.3$) the agreement is excellent,  with no  differences larger than 1000\,\kmps.  At higher redshifts ($0.3{>}z{>}1$) the median offset increases ($\lvert z_{DR2}-z_{DR1}\rvert{=}$ 157\,\kmps), and finally increases to 1180\,\kmps at $z{>}1$ owing to the use of the intrinsically broad lines of \MgII\ and \CIV\ for the derivation of the redshift.

 
 \clearpage
\section{Example Spectra} \label{spec_example}
Here we provide example spectra of various classes of AGN and telescope setups.  Example spectra for each AGN will be provided at the BASS website.{\footnote{\href{https://www.bass-survey.com/}{https://www.bass-survey.com/}}}

\begin{figure*} 
\centering
\includegraphics[width=16cm]{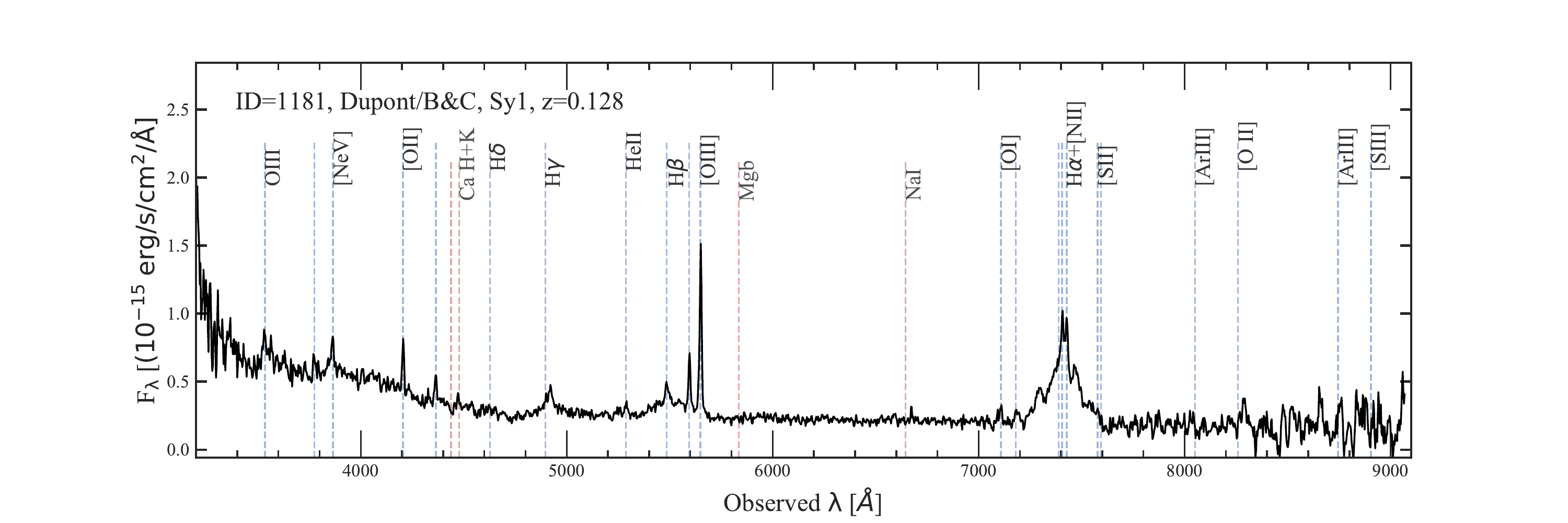}
\includegraphics[width=16cm]{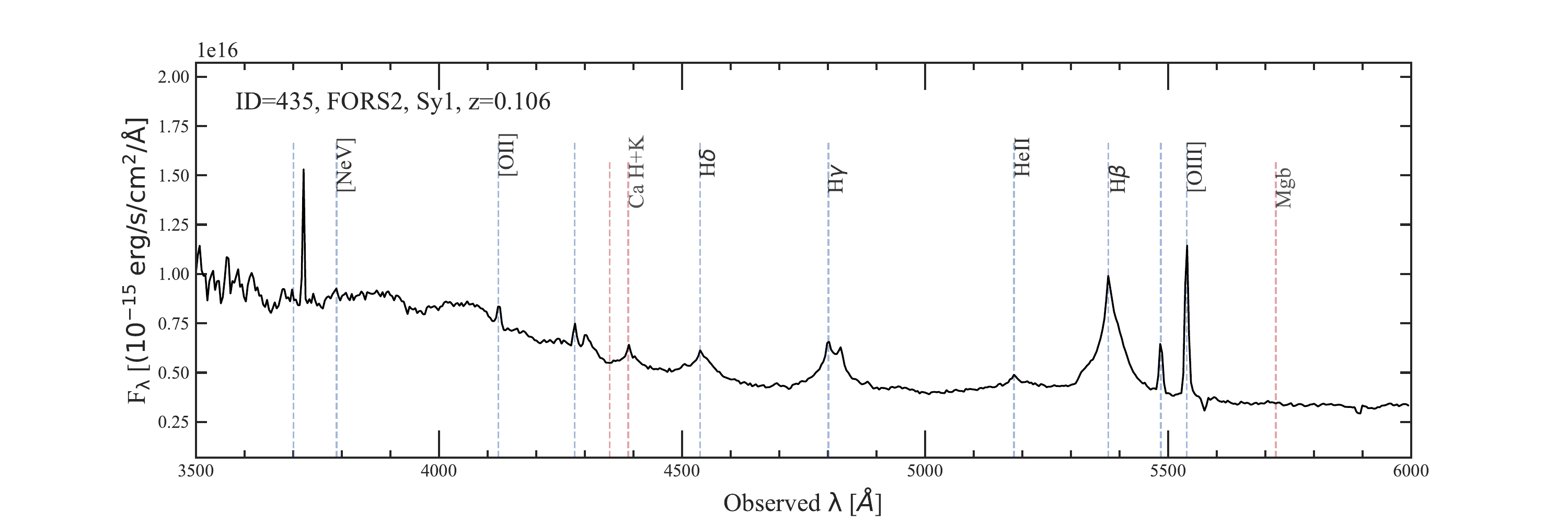}
\includegraphics[width=16cm]{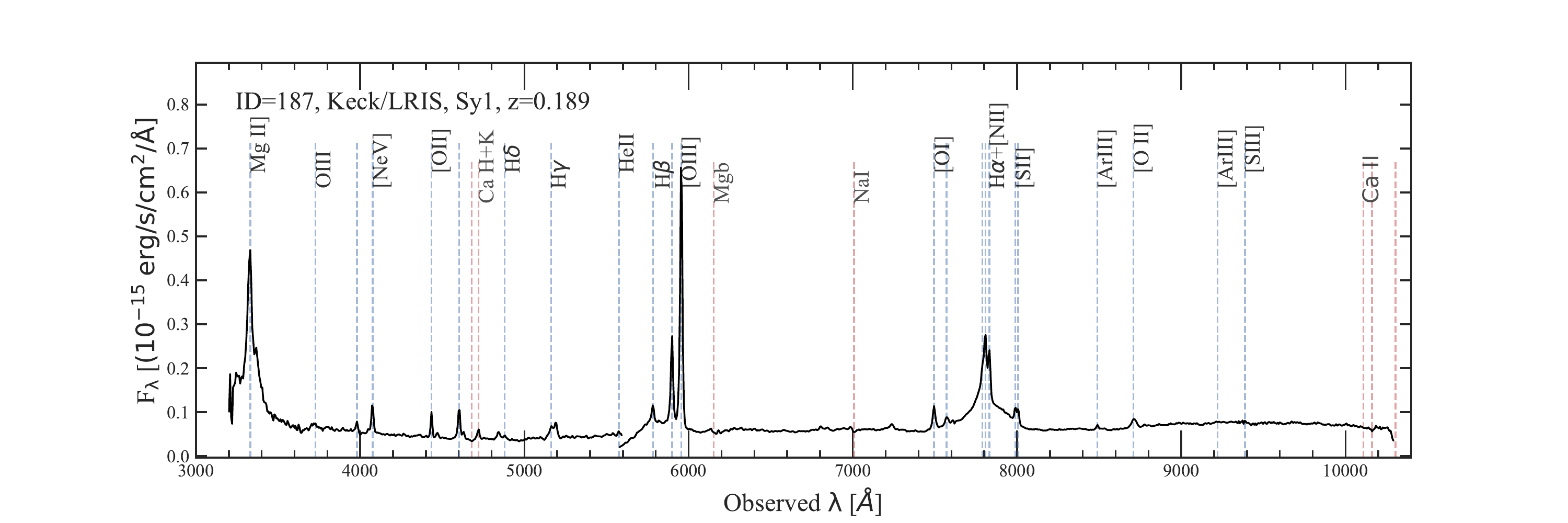}
\caption{Figure showing examples of spectral setups not shown earlier in the text excluding the archival observations.  A spectrum from the du Pont telescope with the Bollens \& Chivens spectrograph with the 300 lines/mm grating is shown in the top panel, a spectrum from the VLT with the FORS2 instrument is shown in the middle panel, and a spectrum with Keck using the LRIS instrument is shown in the bottom panel.}
\label{fig:example_speclowres}
\end{figure*}

\begin{figure*} 
\centering
\includegraphics[width=16cm]{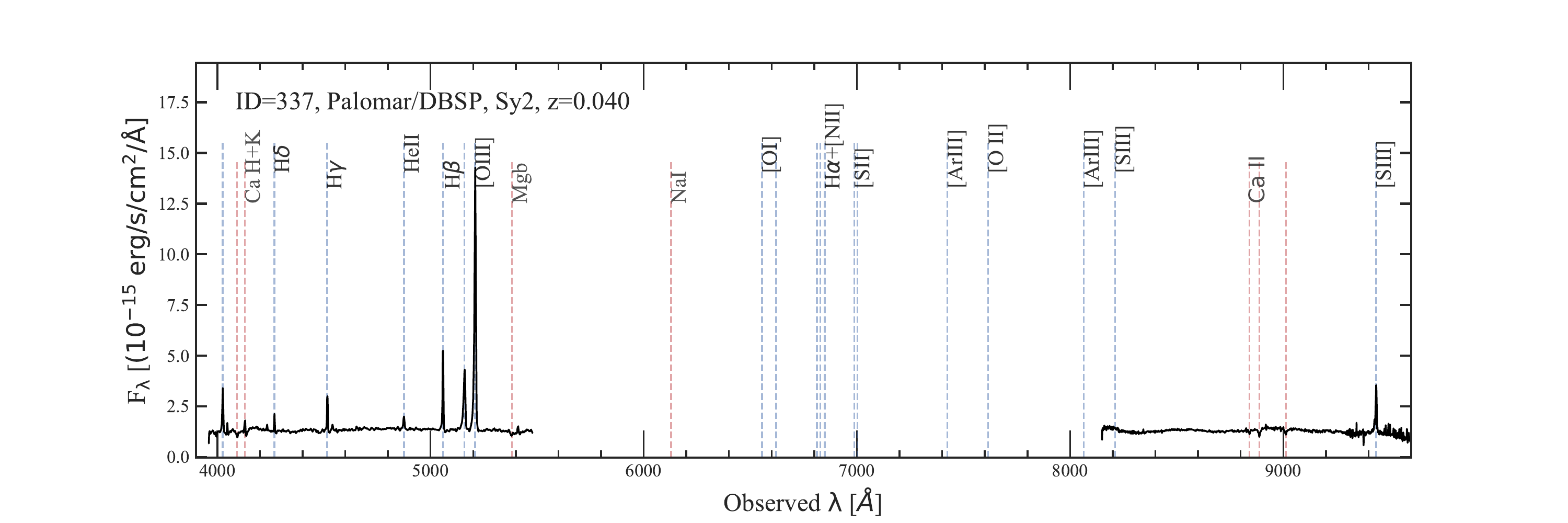}
\includegraphics[width=16cm]{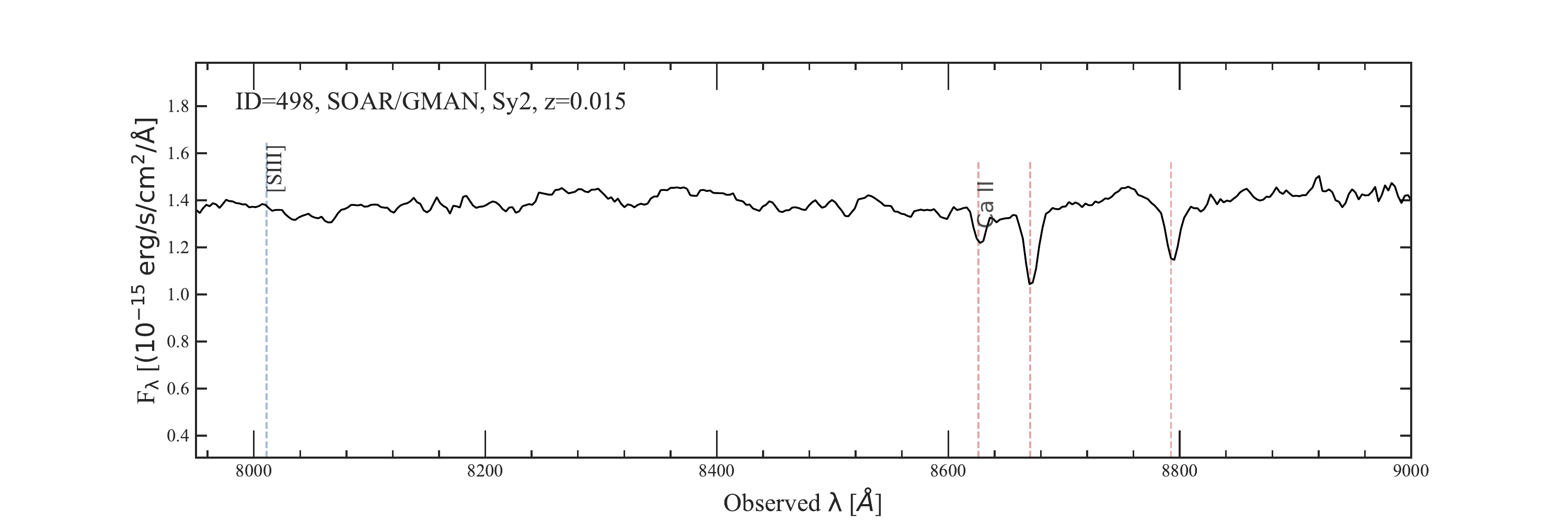}
\includegraphics[width=16cm]{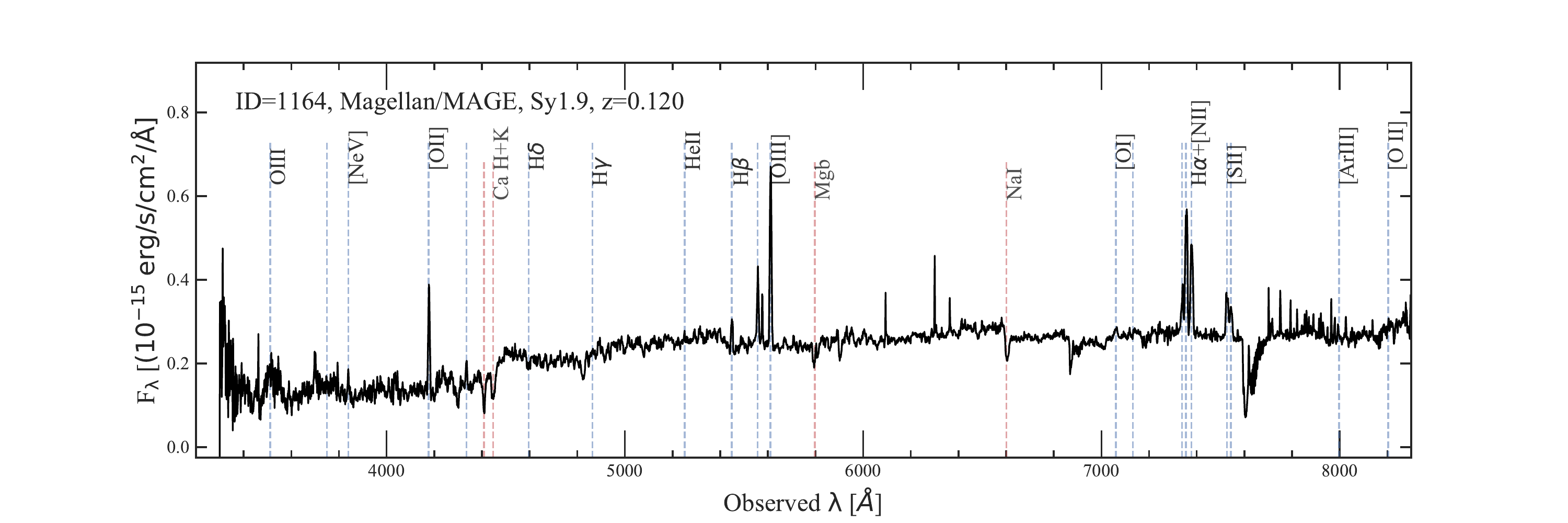}
\caption{Figure showing examples of high-resolution spectral setups not shown earlier in the text excluding the archival observations.  A spectrum taken at the Palomar telescope with the Doublespec instrument with the 1200 lines/mm grating is shown on top, with a spectrum from SOAR with 1200 lines/mm grating using the Goodman instrument in middle, and finally an echelle spectrum from Magellan using the MAGE instrument.}
\label{fig:example_spechighres}
\end{figure*}


\vspace{5mm}
\facilities{Du Pont (Boller \& Chivens spectrograph), Keck:I (LRIS), Magellan:Clay, Hale (Doublespec), NuSTAR, Swift (XRT and BAT), VLT:Kueyen (X-Shooter), VLT:Antu (FORS2), SOAR (Goodman)}

\software{Astropy \citep{Collaboration:2013:A33}, ESO Reflex \citep{Freudling:2013:A96},
           IRAF \citep{Observatories:1999:ascl:9911.002}, Matplotlib \citep{Hunter:2007:90}, \texttt{molecfit} \citep{Smette:2015:A77},
          Numpy \citep{vanderWalt:2011:22}, Pandas (\url{https://doi.org/10.5281/zenodo.3630805}), {\tt PySpecKit} \citep{Ginsburg:2011:1109.001}}

\bibliography{bibfinal,bib_dr2papers,bib_add_here}{}
\bibliographystyle{aasjournal}

\end{document}